\documentstyle{houches}

\begin{document}

\author[E. Bertschinger]{Edmund Bertschinger}

\address{Department of Physics\\
	 MIT, Room 6-207\\
	 Cambridge, MA 02139, USA.}

\chapter{Cosmological Dynamics}

\setcounter{section}{-1}
\section{Preface}

The theory of large-scale structure is presently one of the most
active research areas in cosmology.  The important questions being
studied include: Did structure form by gravitational instability?
What are the nature and amount of dark matter?  What is the background
cosmological model?  What were the initial conditions for structure
formation?  It is exciting that we can ask these questions seriously,
knowing that observational tests are rapidly improving.

Numerous papers and reviews discuss specific theoretical models of
large-scale structure, or specific theoretical techniques for constructing
and analyzing models.  However, there are few coherent presentations of the
basic physical theory of the dynamics of matter and spacetime in cosmology.
Although there are now several textbooks in this area, I think there is
still room for further pedagogical development.  My aim in these lecture
notes is to provide a detailed yet readable introduction to cosmological
dynamics.

Although I gave an evening seminar on N-body techniques for simulating
large-scale structure, for reasons of length I have excluded that
subject from these notes.  The subject is presented elsewhere
(e.g., Hockney \& Eastwood 1981, Efstathiou et al. 1985, Bertschinger
\& Gelb 1991, and S. White's notes in this volume).  Otherwise, these
notes generally follow the lectures I gave in Les Houches, except that
my lecture on Lagrangian fluid dynamics has been subsumed into the
section on relativistic perturbation theory.  The former subject is
still evolving, and does not seem to be as fundamental as the subjects
of my other lectures.

I would like to thank Andrew Hamilton, Lam Hui, Bhuvnesh Jain,
Chung-Pei Ma, Dominik Schwarz, Uro\v s Seljak, and Simon White for
useful comments and discussion, and Rennan Bar-Kana, Chung-Pei Ma,
Nick Gnedin, and Marie Machacek for correcting several errors in
early drafts.  I am grateful to the organizers and
students of the Les Houches Summer School for providing the opportunity
to present this material.  I appreciate the hospitality of John Bahcall
and the Institute for Advanced Study, where much of the writing was done.
This work was supported by NASA grants NAGW-2807 and NAG5-2816.

\section{Elementary mechanics}

This lecture applies elementary mechanics to an expanding universe.
Attention is given to puzzles such as the role of boundary conditions
and conservation laws.

\subsection{Newtonian dynamics in cosmology}

For a finite, self-gravitating set of mass points with positions
$\vec r_i(t)$ in an otherwise empty universe, Newton's laws
(assuming nonrelativistic motions and no non-gravitational forces) are
\begin{equation}
  {d^2\vec r_i\over dt^2}=\vec g_i\ ,\quad
  \vec g_i=-\sum_{j\ne i}Gm_j{(\vec r_i-\vec r_j)\over\vert\vec r_i-
    \vec r_j\vert^3}\ .
\label{Newton}
\end{equation}
In the limit of infinitely many particles each with infinitesimal mass
$\rho d^3r$, we can also obtain $\vec g_i=\vec g(\vec r_i,t)$ as the
irrotational solution to the Poisson equation,
\begin{equation}
  \vec\nabla\cdot\vec g=-4\pi G\rho(\vec r,t)\ ,\quad
  \vec\nabla\times\vec g=0\ ,
\label{Poisson}
\end{equation}
which may be written
\begin{equation}
  \vec g(\vec r,t)=-\int G\rho(\vec r',t){(\vec r-\vec r')\over
    \vert\vec r-\vec r'\vert^3}\,d^3r'\ .
\label{g-int}
\end{equation}
The Newtonian potential $\phi$, defined so that $\vec g=-\partial\phi/
\partial\vec r$ (using partial derivatives to indicate the gradient with
respect to $\vec r$), obeys $\nabla^2\phi=4\pi G\rho$.

If the mass density $\rho$ is finite and nonzero only in a finite
volume, then $\vec g$ (and also $\phi$) generally converges to a finite
value everywhere, with $g\to0$ as $r\to\infty$.  If, however, $\rho$
remains finite as $r\to\infty$, then $\phi$ diverges and $\vec g$ depends
on boundary conditions at infinity.

Consider the dilemma faced by Newton in his correspondence with
Bentley concerning the gravitational field in cosmology (Munitz 1957).
What is $\vec g$ in an infinite homogeneous medium?  If we consider
first a bounded sphere of radius $R$, Gauss' theorem quickly gives us
$\vec g=-(4\pi/3)G\rho\vec r$ for $r<R$.  This result is unchanged as
$R\to\infty$, so we might conclude that $\vec g$ is well-defined at
any finite $r$.  Suppose, however, that the surface bounding the mass
is a spheroid (a flattened or elongated sphere, whose cross-section is
an ellipse) of eccentricity $e>0$.  In this case the gravity field is
nonradial (see Binney \& Tremaine 1987, \S 2.3, for expressions).  The
only difference in the mass distribution is in the shell between the
spheroid and its circumscribed sphere, yet the gravity field is changed
everywhere except at $\vec r=0$.  An inhomogeneous density field further
changes $\vec g$.  Thus, the gravity field in cosmology depends
on boundary conditions at infinity.

There is an additional paradox of Newtonian gravity in an infinite
homogeneous medium: $\vec g=0$ at one point but is nonzero elsewhere
(at least in the spherical and spheroidal examples given above),
in apparent violation of the Newtonian relativity of absolute space.
Newton avoided this problem (incorrectly, in hindsight) by assuming
that gravitational forces due to mass at infinity cancel everywhere
so that a static solution exists.

These problems are resolved in general relativity (GR), which forces
us to complicate the treatment of Newtonian gravity in absolute space.
First, in GR distant matter curves spacetime so that $(\vec r,t)$ do
not provide good coordinates in cosmology.  Second, in GR we must specify
a global spacetime geometry explicitly taking into account distant boundary
conditions.

What coordinates shall we take in cosmology?  First note that a homogeneous
self-gravitating mass distribution cannot remain static (unless non-Newtonian
physics such as a fine-tuned cosmological constant is added to the model,
as was proposed by Einstein in 1917).  The observed mass distribution is
(on average) expanding on large scales.  For a uniform expansion, all
separations scale in proportion with a cosmic scale factor $a(t)$.  Even
though the expansion is not perfectly uniform, it is perfectly reasonable
to factor out the mean expansion to account for the dominant motions at
large distances as in Figure 1.  We do this by defining comoving coordinates
$\vec x$ and conformal time $\tau$ as follows:
\begin{equation}
  \vec x=\vec r/a(t)\ ,\quad d\tau=dt/a(t)\ \ \hbox{or}\ \
    \tau=\int_0^t {dt'\over a(t')}\ .
\label{comoving}
\end{equation}
The starting time for the expansion is $\tau=0$ and $t=0$ when $a=0$;
if this time was nonexistent (or ill-defined in classical terms) then
we can set the lower limit of integration for $\tau(t)$ to any convenient
value.  Although the units of $a$ are arbitrary, I follow the standard
convention of Peebles (1980) in setting $a=1$ today when $t=t_0$ and
$\tau=\tau_0$.  A radiation source emitting radiation at $\tau<\tau_0$
has redshift $\Delta\lambda/\lambda_0=z=-1+a^{-1}$ where $\lambda_0$ is
the rest wavelength.

\begin{figure}
  \vskip 3.3truein \hskip -1.8truein
  \includegraphics{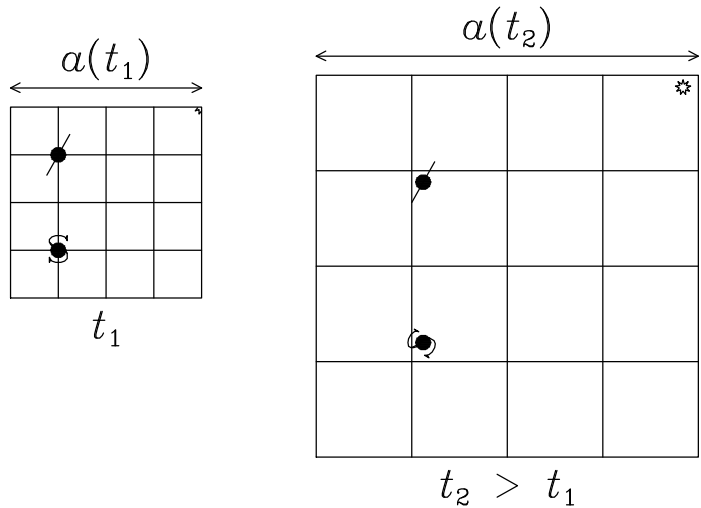}
  \vskip -0.3truein
  \caption{Perturbed Hubble expansion.}
\label{fig1}
\end{figure}

For a perfectly uniform expansion, the comoving position vectors $\vec x$
remain fixed for all particles.  For a perturbed expansion, each particle
follows a trajectory $\vec x\,(\tau)$ [or $\vec x\,(t)$].   The comoving
coordinate velocity, known also as the peculiar velocity, is
\begin{equation}
  \vec v\equiv{d\vec x\over d\tau}={d\vec r\over dt}-H(t)\vec r\ ,
\label{v-pec}
\end{equation}
where $H(t)=d\ln a/dt=a^{-2}da/d\tau$ is the Hubble parameter.  Note
that $\vec v$ is the proper velocity measured by a comoving observer
at $\vec x$, i.e., one whose comoving position is fixed.

[The distinction between ``proper'' and ``comoving'' quantities is
important.  Proper quantities are physical observables, and they do
not change if the expansion factor is multiplied by a constant.
Thus, $\vec v=d\vec x/d\tau=(ad\vec x)/(adt)$ is a proper quantity,
while $d\vec x/dt$ is not.  This is why I prefer $\tau$ rather than $t$
as the independent variable.]

We shall assume that peculiar velocities are of the same order at all
distances and in all directions, consistent with the choice of a
homogeneous and isotropic mean expansion scale factor.  These assumptions
are consistent with the {\bf Cosmological Principle}, which states that
the universe is approximately homogeneous and isotropic when averaged
over large volumes.  In general relativity theory, the Cosmological
Principle is applied by assuming that we live in a perturbed
Robertson-Walker spacetime.  Locally, the GR description is equivalent to
Newtonian cosmology plus the boundary conditions that the mass distribution
is (to sufficient accuracy) homogeneous and isotropic at infinity.

Unless otherwise stated, in this and the following lectures (until
section 4) I shall use 3-vectors for spatial vectors assuming an
orthonormal basis. Thus, $\vec A\cdot\vec B=A_iB_i=A^iB_i=A^iB^i$
with summation implied from $i=1$ to 3.  Note that $A_i=A^i$ are
Cartesian components, whether comoving or proper, and they are to
be regarded (in this Newtonian treatment) as 3-vectors, not the spatial
parts of 4-vectors.  (If we were to use 4-vectors, then $A_i=g_{ij}A^j=
a^2A^i$ in a Robertson-Walker spacetime.  Because we are not using
4-vectors, there is no factor of $a^2$ distinguishing covariant and
contravariant components.) This treatment requires space to be Euclidean,
which is believed to be an excellent approximation everywhere except
very near relativistic compact objects such as black holes and, possibly,
on scales comparable to or larger than the Hubble distance $c/H$.  (In
section 4 the restrictions to Cartesian components and Euclidean space
will be dropped.)  Also, gradients and time derivatives will be taken
with respect to the comoving coordinates: $\vec\nabla\equiv\partial/
\partial\vec x$, $\dot{}\equiv\partial/\partial\tau$.

Before proceeding further we must derive the laws governing the mean
expansion.  Consider a spherical uniform mass distribution with mass
density $\bar\rho$ and radius $r=xa(t)$ with $x=\hbox{constant}$.
Newtonian energy conservation states
\begin{displaymath}
  {1\over2}\left(dr\over dt\right)^2-{GM\over r}=E\ ,
\end{displaymath}
\noindent
implying
\begin{equation}
  \left(d\ln a\over d\tau\right)^2=(aH)^2={8\pi\over3}Ga^2\bar\rho-K\ ,
  \quad K\equiv -2Ex^{-2}\ .
\label{Friedmann}
\end{equation}
This result, known as the Friedmann equation, is valid (from GR) even
if $\bar\rho$ includes relativistic particles or vacuum energy density
$\rho_{\rm vac}=\Lambda/(8\pi G)$ (where $\Lambda$ is the cosmological
constant).  The cosmic density parameter is $\Omega\equiv8\pi G\bar\rho/
(3H^2)$, so the Friedmann equation may also be written $K=(\Omega-1)
(aH)^2$.  Homogeneous expansion, with $a=a(\tau)$ independent of
$\vec x$, requires $K=\hbox{constant}$ in addition to $\vec\nabla
\bar\rho=0$.  In GR one finds that $K$ is related to the curvature
of space (i.e., of hypersurfaces of constant $\tau$).  The solutions
of eq. (\ref{Friedmann}) for zero-pressure (Friedmann) models,
two-component models with nonrelativistic matter and radiation, and
other simple equations of state may be found in textbooks (e.g.,
Padmanabhan 1993, Peebles 1993) or derived as good practice for the
student.

At last we are ready to describe the motion of a nonuniform medium
in Newtonian cosmology with mass density $\rho(\vec x,\tau)=
\bar\rho(\tau)+\delta\rho(\vec x,\tau)$.  We start from Newton's law
in proper coordinates, $d^2\vec r/dt^2=\vec g$, and transform to comoving
coordinates and conformal time:
\begin{displaymath}
  {d^2\vec x\over d\tau^2}+{\dot a\over a}\,{d\vec x\over d\tau}
    +\vec x{d\over d\tau}\left(\dot a\over a\right)=-Ga^2
    \int(\bar\rho+\delta\rho){(\vec x-\vec x')\over
    \vert\vec x-\vec x'\vert^3}\,d^3x'\ .
\end{displaymath}
\noindent
We eliminate the homogeneous terms (those present in a homogeneous
universe) as follows.  First, assuming that the universe is, on average,
spherically symmetric at large distance, the first term on the right-hand
side becomes (from Gauss' theorem) $-(4\pi/3)Ga^2\bar\rho\vec x$.  (This
is where the boundary conditions at infinity explicitly are used.)
To get the term proportional to $\vec x$ on the left-hand side,
differentiate the Friedmann equation: $(\dot a/a)d(\dot a/a)/d\tau=
(4\pi G/3)d(\bar\rho a^2)/d\tau$.  For nonrelativistic matter,
$\bar\rho\propto a^{-3}$, implying $d(\bar\rho a^2)/d\tau=-\dot a\bar
\rho a$, so $d(\dot a/a)/d\tau=-(4\pi/3)Ga^2\bar\rho$.  (If $\bar\rho$
includes relativistic matter, not only is $d\rho/d\tau$ changed,
so is the gravitational field.  Our derivation gives essentially the
correct final result in this case, but its justification requires GR.)
We conclude that the homogeneous terms cancel, so that the equation of
motion becomes
\begin{displaymath}
{d^2\vec x\over d\tau^2}+{\dot a\over a}\,{d\vec x\over d\tau}=-Ga^2
  \int\delta\rho(\vec x',\tau){(\vec x-\vec x')
  \over\vert\vec x-\vec x'\vert^3}\,d^3x'
  \equiv-\vec\nabla\phi'\ ,
\end{displaymath}
\noindent
where
\begin{displaymath}
  \phi'(\vec x,\tau)=-Ga^2\int{\delta\rho(\vec x',\tau)\,
    d^3x'\over\vert\vec x-\vec x'\vert}\ .
\end{displaymath}
\noindent
Note that $\phi'$ is a proper quantity: $a^2d^3x'/\vert\vec x-\vec
x'\vert\sim d^3r/\vert\vec r-\vec r'\vert$.

If $\int\delta\rho\,d^3x\to0$ when the integral is taken over all space
--- as happens if the density field approaches homogeneity and isotropy
on large scales, with $\bar\rho$ being the volume-averaged density ---
then $\phi'$ is finite and well-defined (except, of course, on top of
point masses, which we ignore by treating the density field as being
continuous).   Newton's dilemma is then resolved: we have no ambiguity
in the equation of motion for $\vec x(\tau)$.  We conclude that
$\phi'$, sometimes called the ``peculiar'' gravitational potential,
is the correct Newtonian potential in cosmology provided we work in
comoving coordinates.  Therefore we shall drop the prime and the
quaint historical adjective ``peculiar.''  In summary, the equations
of motion become
\begin{equation}
  {d^2\vec x\over d\tau^2}+{\dot a\over a}\,{d\vec x\over d\tau}=
  -\vec\nabla\phi\ ,\quad\nabla^2\phi=4\pi Ga^2\delta\rho(\vec x,\tau)\ .
\label{Newton-eom}
\end{equation}
As we shall see in section 4, the same equations follow in the weak-field
($\vert\phi\vert\ll c^2$), slow-motion ($v^2\ll c^2$) limit of GR for a
perturbed Robertson-Walker spacetime.  If Newton had pondered more carefully
the role of boundary conditions at infinity, he might have invented modern
theoretical cosmology!

\subsection{Lagrangian and Hamiltonian formulations}

The equations of Newtonian cosmology may be derived from Lagrangian
and Hamiltonian formulations.  The latter is particularly useful for
treatments of phase space.

In the Lagrangian approach, one considers the trajectories $\vec x(\tau)$
and the action $S[\vec x(\tau)]$.  From elementary mechanics (with proper
coordinates and no cosmology, yet), $S=\int L\,dt$ with Lagrangian
$L=T-W={1\over2}mv^2-m\phi$ for a particle moving in a potential $\phi$
($T$ is the kinetic energy and $W$ is the gravitational energy).  We now
write a similar expression in comoving coordinates,  bearing in mind that
the action must be a proper quantity:
\begin{equation}
  S=\int L(\vec x,\dot\vec x,\tau)\,d\tau\ ,\quad
  L=a\left({1\over2}mv^2-m\phi\right)\ ,
\label{action-1}
\end{equation}
where $\dot\vec x=\vec v$ is the peculiar velocity.  We will show that
eq. (\ref{action-1}) is the correct Lagrangian by showing that it
leads to the correct equations of motion.

Equations of motion for the trajectories follow from Hamilton's principle:
the action must be stationary under small variations of the trajectories
with fixed endpoints.  Thus, we write $\vec x(\tau)\to\vec x(\tau)+
\delta\vec x(\tau)$, $d\vec x/d\tau\to d\vec x/d\tau+(d/d\tau)\delta
\vec x(\tau)$.  The change in the action is
\begin{eqnarray*}
  \delta S&=&\int_{\tau_1}^{\tau_2}\left[{\partial L\over\partial\vec x}\cdot
    \delta\vec x+{\partial L\over\partial\dot\vec x}\cdot{d\over d\tau}
    (\delta\vec x)\,\right]\,d\tau\\
  &=&\int_{\tau_1}^{\tau_2}\left[{\partial L\over\partial\vec x}-{d\over
     d\tau}\left({\partial L\over\partial\dot\vec x}\right)\,\right]\cdot
     \delta\vec x(\tau)\,d\tau\ ,
\end{eqnarray*}
where we have integrated by parts assuming $(\partial L/\partial\dot\vec
x)\cdot\delta\vec x=0$ at $\tau=\tau_1$ and $\tau_2$.  Applying
Hamilton's principle, $\delta S=0$, we obtain the Euler-Lagrange equation
(it works in cosmology, too!):
\begin{equation}
  {d\over d\tau}\left({\partial L\over\partial\dot\vec x}\right)-
    {\partial L\over\partial\vec x}=0\ .
\label{E-L}
\end{equation}
The reader may verify that substituting $L$ from eq. (\ref{action-1})
yields the correct equation of motion (\ref{Newton-eom}).

It is straightforward to extend this derivation to a system of self-gravitating
particles filling the universe.  The Lagrangian is
\begin{equation}
  L=a\left(\sum_i{1\over2}m_iv_i^2-W\right)\ ,
\label{lag-1}
\end{equation}
where the total gravitational energy excludes the part arising from the
mean density:
\begin{eqnarray}
  W&=&{1\over2}\left(\sum_im_i\phi_i-a^3\bar\rho\int\phi\,d^3x\right)\ ,
    \nonumber\\
  \phi_i&=&-\left(\sum_{j\ne i}{Gm_j\over a\vert\vec x_i-\vec x_j\vert}-
    Ga^2\bar\rho\int{d^3x'\over\vert\vec x-\vec x'\vert}\right)\ ,
\label{egrav-1}
\end{eqnarray}
where the factor ${1\over2}$ is introduced to avoid double-counting
pairs of particles.  For a continuous mass distribution we obtain
\begin{equation}
\!\!\!W={1\over2}\int\!\phi\,\delta\rho\,a^3d^3x=-{1\over2}Ga^5\int\!\!d^3x_1
 \int\!\!d^3x_2\,{\delta\rho(\vec x_1,\tau)\,\delta\rho(\vec x_2,\tau)\over
 \vert\vec x_1-\vec x_2\vert}\ .
\label{egrav-2}
\end{equation}

In the Hamiltonian approach one considers the trajectories in the
single-particle (6-dimensional) phase space, $\{\vec x(\tau),
\vec p(\tau)\}$.  The aim is to obtain coupled first-order equations
of motion for $\vec x(\tau)$ and $\vec p(\tau)$, known as Hamilton's
equations, instead of a single second-order equation for $\vec x(\tau)$.

The derivation of Hamilton's equations has several steps.
First we need the canonical momentum conjugate to $\vec x$:\
\begin{equation}
  \vec p\equiv{\partial L\over\partial\dot\vec x}=am\vec v=am
    {d\vec x\over d\tau}\ .
\label{p-conj}
\end{equation}
Note that $\vec p$ is {\it not} the proper momentum measured by a
comoving observer: $m\vec v$ is.  In Hamiltonian mechanics, one must
use the conjugate momentum and not the proper momentum.

The next step is to eliminate $d\vec x/d\tau$ from the Lagrangian
in favor of $\vec p$.  We then transform from the Lagrangian to a new
quantity called the Hamiltonian, using a Legendre transformation:
\begin{equation}
  L(\vec x,\dot\vec x,\tau)\to H(\vec x,\vec p,\tau)\equiv\vec p\cdot
  \dot\vec x-L\ .
\label{legendre}
\end{equation}
Notice that we transform $L$ to $H$ and $\dot\vec x$ to $\vec p$ (the
latter through eq. \ref{p-conj}).  Why do we perform these
transformations?  The answer is that now Hamilton's principle gives
the desired equations of motion for the phase-space trajectory
$\{\vec x(\tau),\vec p(\tau)\}$.  In phase space, Hamilton's principle says
that the  action $S=\int L\,d\tau=\int(\vec p\cdot\dot\vec x-H)\,d\tau$
must be stationary under independent variations of all phase space coordinates:
$\vec x(\tau)\to\vec x(\tau)+\delta\vec x(\tau)$ and $\vec p(\tau)\to
\vec p(\tau)+\delta\vec p(\tau)$.  As an exercise, the reader can show,
using a method similar to the derivation of the Euler-Lagrange equation
above,
 \begin{equation}
   {d\vec x\over d\tau}={\partial H\over\partial\vec p}\ ,\quad
   {d\vec p\over d\tau}=-{\partial H\over\partial\vec x}\ ,
 \label{hamilton}
 \end{equation}
provided that $\vec p\cdot\delta\vec x=0$ at the endpoints of $\tau$.

In our case, $H=p^2/(2am)+am\phi$ (getting the $a$'s right requires
using the Legendre transformation), yielding
\begin{equation}
  {d\vec x\over d\tau}={\vec p\over am}\ ,\quad
  {d\vec p\over d\tau}=-am\vec\nabla\phi\ .
\label{hamilton-c}
\end{equation}
These equations could be combined to yield eq. (\ref{Newton-eom}),
but in the Hamiltonian approach we prefer to think of two coupled
evolution equations.  This is particularly useful when studying the evolution
of a system in phase space, as we shall do in section 3 with hot dark matter.

\subsection{Conservation of momentum and energy?}

Are total momentum and energy conserved in cosmology?  This is a nontrivial
question because the canonical momentum and Hamiltonian differ from
the proper momentum and energy.

Consider first the momentum of a particle in an unperturbed Robertson-Walker
universe.  With no perturbations, $\phi=0$ so that Hamilton's equation for
$\vec p$ becomes $d\vec p/d\tau=-am\vec\nabla\phi=0$, implying that the
canonical momentum $\vec p$ is conserved.  But, the {\it proper} momentum
$m\vec v=a^{-1}\vec p$  measured by a comoving observer decreases as $a$
increases.  What happened to momentum conservation?

The key point is that $\vec v=d\vec x/d\tau$ is measured using a non-inertial
(expanding) coordinate system.  Suppose, instead, that we choose $\vec v$
to be a proper velocity measured relative to some fixed origin.  Momentum
conservation then implies $\vec v=\hbox{constant}$ (if $\vec\nabla\phi=0$, as
we assumed above).  At $\tau=\tau_1$ and $\tau_2$, the particle is at
$\vec x_1$ and $\vec x_2$, respectively.  Because $d\vec x/d\tau$ gives
the proper velocity relative to a {\it comoving} observer at the
particle's position, at $\tau_1$ we have $d\vec x/d\tau=\vec v-(\dot
a/a)_1\vec x_1$, while at $\tau_2$, $d\vec x/d\tau=\vec v-(\dot a/a)_2
\vec x_2$.  (The proper velocity relative to the fixed origin is $\vec
v$ in both cases, but the Hubble velocity at the particle's position
--- the velocity of a comoving observer --- changes because the
particle's position has changed.) Combining these, we find $[\dot\vec x
(\tau_2)-\dot\vec x(\tau_1)]/(\tau_2-\tau_1)\approx-(\dot a/a)[\vec x
(\tau_2)-\vec x(\tau_1)]/(\tau_2-\tau_1)+O(\tau_2-\tau_1)$ or, in the
limit $\tau_2-\tau_1\to0$, $d^2\vec x/d\tau^2=-(\dot a/a)d\vec x/d\tau$.
This is precisely our comoving equation of motion in the case $\vec
\nabla\phi=0$.  Thus, the ``Hubble drag'' term $(\dot a/a)d\vec x/d\tau$
is merely a ``fictitious force'' arising from the use of non-inertial
coordinates.  Stated more physically, the particle appears to slow down
because it is continually overtaking faster moving observers.

Energy conservation is more interesting.  Let us  check whether the
Hamiltonian $H(\vec x,\vec p,\tau)$ is conserved.  Using Hamilton's
equations for a single particle, we get
\begin{equation}
  {dH\over d\tau}={\partial H\over\partial\vec x}\cdot{d\vec x\over d\tau}+
    {\partial H\over\partial\vec p}\cdot{d\vec p\over d\tau}+{\partial H\over
    \partial\tau}={\partial H\over\partial\tau}\ .
\label{Ham-dot}
\end{equation}
Using $H=p^2/(2am)+am\phi$, we obtain $dH/d\tau=-(\dot a/a)(p^2/2am)
+md(a\phi)/d\tau$ which is nonzero even if $d\phi/d\tau=0$.  Is this lack
of energy conservation due to the use of non-inertial coordinates?  While
the appearance of a Hubble-drag term may suggest this is the case, if we
wish to obtain the total Hamiltonian (or energy) for a system of particles
filling all of space, we have no choice but to use comoving coordinates.

Perhaps the Hamiltonian is not conserved because it is not the proper
energy.  To examine this possibility, we use the Hamiltonian for a system
of particles in comoving coordinates, with $H=a(T+W)$.  The proper
kinetic energy (with momenta measured relative to comoving
observers) is
\begin{equation}
T=\sum_i{1\over2}m_iv_i^2=\sum_i{1\over2}{p_i^2\over a^2m_i}\ ,
\label{kinetic-e}
\end{equation}
while the gravitational energy $W$ is given in eq. (\ref{egrav-1}).
Holding fixed the momenta, we see that $a^2T$ is a constant, implying
$\partial(aT)/\partial\tau=-\dot aT$.  Similarly, holding fixed the
particle positions, we find that $a\phi$ is a constant, implying
$\partial(aW)/\partial\tau=0$.  We thus obtain the Layzer-Irvine
equation (Layzer 1963, Irvine 1965)
\begin{equation}
{d\over d\tau}(T+W)=-{\dot a\over a}(2T+W)\ .
\label{layzer-irvine}
\end{equation}

Total energy (expressed in comoving coordinates) is not conserved in
Newtonian cosmology.  (This is also the case in GR --- indeed, there is
generally no unique scalar for the total energy in GR.)  However, if
almost all of the mass is in virialized systems obeying the classical
virial theorem $2T+W\approx0$, we recover approximate total energy
conservation.

\section{Eulerian fluid dynamics}
\subsection{Cosmological fluid equations}

A fluid is a dense set of particles treated as a continuum.  If particle
collisions are rapid enough to establish a local thermal equilibrium
(e.g., Maxwell-Boltzmann velocity distribution), the fluid is an ideal
collisional gas.  If collisions do not occur (e.g., a gas of dark matter
particles), the gas is called collisionless.  (I exclude incompressible
fluids, i.e., liquids, from consideration because the gases considered
in cosmology are generally very dilute and compressible.)  The fluid
equations discussed in this lecture apply only for a collisional gas
(or a pressureless collisionless gas).  They apply, for example, to
baryons (hydrogen and helium gas) after recombination, to cold dark
matter before trajectories intersect (``cold dust''), and (with
relativistic corrections) to the coupled photon-baryon fluid before
recombination.

I shall assume a nonrelativistic gas and ignore bulk electric and magnetic
forces.  These are not difficult to add, but the essential physics of
cosmological fluid dynamics does not require them.

The fluid equations consist of mass and momentum conservation laws and
an equation of state.  Mass conservation is represented by the
{\bf continuity equation}.  In proper coordinates $(\vec r,t)$ this is
\setcounter{equation}{0}
\begin{equation}
  {\partial\rho\over\partial t}+{\partial\over\partial\vec r}\cdot
    (\rho\vec v)=0\ ,\quad \vec v={d\vec r\over dt}\ .
\label{cont-1}
\end{equation}
We convert to comoving coordinates $\tau=\int dt/a(t)$, $\vec x=\vec
r/a(t)$, being careful to transform the partial derivatives as follows:
$\partial/\partial t=(\partial\tau/\partial t)\partial/\partial\tau
+(\partial\vec x/\partial t)\cdot\partial/\partial\vec x$, $\partial/
\partial\vec r=a^{-1}\partial/\partial\vec x\equiv a^{-1}\vec\nabla$.
We also rewrite the density and velocity by factoring out the mean
behavior:
\begin{equation}
  \rho=\bar\rho(1+\delta)\ ,\quad {d\vec r\over dt}=H\vec r+\vec v
\label{rho,v}
\end{equation}
where $\vec v=d\vec x/d\tau$ is now the peculiar velocity.  The reader
may easily show that eq. (\ref{cont-1}) becomes
\begin{equation}
  {\partial\delta\over\partial\tau}+\vec\nabla\cdot\left[(1+\delta)
    \vec v\,\right]=0\ .
\label{cont-2}
\end{equation}

Momentum conservation for an ideal fluid is represented by the
{\bf Euler equation} (Landau \& Lifshitz 1959).  It is most simply obtained
by adding the pressure-gradient force to the equation of motion for a
freely-falling mass element, eq. (\ref{Newton-eom}).  In comoving
coordinates, we find
\begin{equation}
  {d\vec v\over d\tau}+{\dot a\over a}\,\vec v=-\vec\nabla\phi-
   {1\over\rho}\vec\nabla p\ .
\label{euler-1}
\end{equation}
The time derivative is taken along the fluid streamline and is known as
the convective or Lagrangian time derivative:
\begin{equation}
  {d\over d\tau}={\partial\over\partial\tau}+\vec v\cdot\vec\nabla\ .
\label{tderiv-lag}
\end{equation}

Closing the fluid equations requires an evolution equation for the pressure
or some other thermodynamic variable.  Perhaps the most natural is the
entropy.  For a collisional gas, thermodynamics implies an {\bf equation
of state} $p=p(\rho,S)$ where $S$ is the specific entropy.  For example,
for an ideal nonrelativistic monatomic gas, for reversible changes we have
\begin{equation}
  TdS=d\left({3\over2}{p\over\rho}\right)+pd\left(1\over\rho\right)\ ,
\label{thermo}
\end{equation}
which says that the heat input to a fluid element equals the change in
thermal energy plus the pressure work done by the element, i.e., energy
is conserved.  Combining this with the ideal gas law $p=\rho k_{\rm B}
T/\mu$ where $\mu$ is the mean molecular mass and $k_{\rm B}$ is the
Boltzmann constant, we obtain
\begin{equation}
  p(\rho,S)=\rho^{5/3}\exp\left({2\over3}{\mu\over k_{\rm B}}S\right)\ .
\label{eos-adiab}
\end{equation}
The equation of state must be supplemented by an evolution equation for
the specific entropy.  Outside of shock waves, the entropy evolution equation
is
\begin{equation}
  T{dS\over d\tau}=a(\Gamma-\Lambda)\ ,
\label{entropy}
\end{equation}
where $\Gamma$ and $\Lambda$ are, respectively, the proper specific
heating and cooling rates (in erg g$^{-1}$ s$^{-1}$).  They are determined
by microphysical processes such as radiative emission and absorption,
cosmic ray heating, Compton processes, etc.  For the simplest case, adiabatic
evolution, $\Gamma=\Lambda=0$.  For a realistic non-ideal gas, it may be
necessary to evolve the radiation field, the ionization fraction, and other
variables specifying the equation of state.

The fluid equations are much harder to solve than Newton's laws for particles
falling under gravity, for several reasons.  First, they are nonlinear
partial differential equations rather than a set of coupled ordinary
differential equations.  Second, shock waves (discontinuities in $\rho$,
$p$, $S$, and $\vec v$) prevent intersection of fluid elements.  These
discontinuities must be resolved (on a computational mesh or otherwise)
and followed stably and accurately.  Finally, heating and cooling for
realistic gases are complicated and can lead to large temperature or entropy
gradients that are difficult to resolve.  An example of the latter is the
sun, whose temperature changes by about 15 million K in a distance that
is minuscule compared with cosmological distance scales.

Computational fluid dynamics is a difficult art but is important for galaxy
formation.  I shall not summarize the numerical methods here but refer
the reader instead to the literature (e.g., Sod 1985, Leveque 1992,
Monaghan 1992, Bryan et al. 1994, Kang et al. 1994).

Some of the most important effects of gas pressure can be gleaned from
linear perturbation theory, in which we linearize the fluid equations about
the uniform solution for an unperturbed Robertson-Walker spacetime.
This technique is useful for checking for gravitational and other linear
instabilities.  Moreover, the linearized fluid equations may provide
a reasonable description of large-scale, small-amplitude fluctuations in
the (dark+luminous) matter, even if structure is nonlinear on small scales.
This is a common assumption in large-scale structure theory.  It is supported
reasonably well by numerical simulations.

Linearizing the continuity and Euler equations gives
\begin{equation}
  \dot\delta+\vec\nabla\cdot\vec v\approx0\ ,\quad
  \dot\vec v+{\dot a\over a}\vec v\approx-\vec\nabla\phi
    -{1\over\bar\rho}\vec\nabla p\ ,
\label{linear-cons}
\end{equation}
where an overdot denotes $\partial/\partial\tau$.  The pressure gradient
may be obtained from the equation of state $p=p(\rho,S)$.  For an ideal
nonrelativistic monatomic gas,
\begin{equation}
  {1\over\bar\rho}\vec\nabla p=c_{\rm s}^2\vec\nabla\delta+{2\over3}T\vec\nabla
S
    \ ,\quad c_{\rm s}^2={5\over3}{p\over\bar\rho}\ .
\label{p-grad}
\end{equation}
Finally, we must linearize the entropy evolution equation.  If the time scale
for entropy changes is long compared with the acoustic or gravitational time
scales, eq. (\ref{entropy}) becomes $dS/d\tau\approx0$.  For the small
peculiar velocities of linear perturbation theory this reduces to
$\dot S\approx0$.

There are five fluid variables ($\rho$, $S$, and three components of $\vec v$),
hence five linearly independent modes.  The general linear perturbation is a
linear combination of these, which we now proceed to examine.

\subsection{Linear instability 1: isentropic fluctuations and Jeans criterion}

We begin with some nomenclature from thermodynamics.  {\bf Isentropic}
means $\vec\nabla S=0$: the same entropy everywhere.  {\bf Adiabatic}
means $dS/d\tau=0$: the entropy of a given fluid element does not change.
The two concepts are distinct.  It is common in cosmology to say ``adiabatic''
when one means ``isentropic.''  This usage is confusing and I shall adopt
instead the standard terminology from thermodynamics.

Isentropic fluctuations are the natural outcome of quantum fluctuations
during inflation followed by reheating: rapid particle interactions in
thermal equilibrium eliminate entropy gradients.  If $\vec\nabla S=0$,
the linearized fluid and gravitational field equations are
\begin{equation}
  \dot\delta+\vec\nabla\cdot\vec v=0\ ,\quad
  \dot\vec v+{\dot a\over a}\vec v=-\vec\nabla\phi-
    c_{\rm s}^2\vec\nabla\delta\ ,\quad
  \nabla^2\phi=4\pi G\bar\rho a^2\delta\ .
\label{linfl-1}
\end{equation}
Combining these gives a damped, driven acoustic wave equation for $\delta$:
\begin{equation}
  \ddot\delta+{\dot a\over a}\dot\delta=4\pi G\bar\rho a^2\delta+
    c_{\rm s}^2\nabla^2\delta\ .
\label{acoust-1}
\end{equation}
Aside from the Hubble damping and gravitational source terms, this
equation is identical to what one would get for linear acoustic waves
in a static medium.

To eliminate the spatial Laplacian we Fourier transform the wave
equation.  For one plane wave, $\delta(\vec x,\tau)\to\delta(\vec k,
\tau)\exp(i\vec k\cdot\vec x)$.  The wave equation becomes
\begin{equation}
  \ddot\delta+{\dot a\over a}\dot\delta=\left(4\pi G\bar\rho a^2-
    k^2c_{\rm s}^2\right)\delta\equiv\left(k_{\rm J}^2-k^2\right)
    c_{\rm s}^2\,\delta\ ,
\label{acoust-2}
\end{equation}
where we have defined the comoving Jeans wavenumber,
\begin{equation}
  k_{\rm J}\equiv\left(4\pi G\bar\rho a^2\over c_{\rm s}^2\right)^{1/2}\ .
\label{Jeans-k}
\end{equation}

Neglecting Hubble damping (by setting $a=1$), the time dependence
of the solution to eq. (\ref{acoust-2}) would be $\delta\propto
\exp(-i\omega\tau)$, yielding a dispersion relation very similar to that
for high-frequency waves in a plasma, but with an important sign difference
because gravity is attractive:
\begin{equation}
  \omega^2=\omega_{\rm p}^2+k^2c^2\quad\to\quad\omega^2=-\omega_{\rm J}^2+
    k^2c_{\rm s}^2\ .
\label{omega-p}
\end{equation}
The plasma frequency is $\omega_{\rm p}=(4\pi n_ee^2/m_e)^{1/2}$ while the
Jeans frequency is $\omega_{\rm J}=k_{\rm J}c_{\rm s}=(4\pi G\bar\rho)^{1/2}$.
Whereas electromagnetic waves with $\omega^2<\omega_{\rm p}^2$ do not propagate
($k^2<0$ implies they are evanescent, e.g., they reflect off the Earth's
ionosphere), gravitational modes with $k<k_{\rm J}$ are {\it unstable}
($\omega^2<0$), as was first noted by Jeans (1902).  In physical terms,
pressure forces cannot prevent gravitational collapse when the
sound-crossing time $\lambda/c_{\rm s}$ is longer than the gravitational
dynamical time $(G\rho)^{-1/2}$ for a perturbation of proper wavelength
$\lambda=2\pi a/k$.

Including the Hubble damping term slows the growth of the Jeans instability
from exponential to a power of time for $k\ll k_{\rm J}$.  In general there
is one growing and one decaying solution for $\delta(k,\tau)$; these are
denoted $\delta_\pm(k,\tau)$.  For $c_{\rm s}^2=0$ and an Einstein-de Sitter
(flat, matter-dominated) background with $a(\tau)\propto\tau^2$, $\delta_+
\propto\tau^2$ and $\delta_-\propto\tau^{-3}$.  For $k\gg k_{\rm J}$, we
obtain acoustic oscillations.  In a static universe the acoustic
amplitude for an adiabatic plane wave remains constant; in the expanding
case it damps in general.  An important exception is oscillations in the
photon-baryon fluid in the radiation-dominated era; the amplitude of
these oscillations is constant.  (Showing this requires generalizing
the fluid equations to a relativistic gas, a good exercise for the
student.)  In any case, acoustic oscillations suppress the growth
relative to the long-wavelength limit.

It is interesting to write the linear wave equation in terms of $\phi$
rather than $\delta$ using $\nabla^2\phi=4\pi Ga^2\bar\rho\delta
\propto a^{-1}\delta$ for nonrelativistic matter (with $c_{\rm s}^2\ll c^2$):
\begin{equation}
  \ddot\phi+3{\dot a\over a}\dot\phi+\left({\ddot a\over a}-{1\over2}
    {\dot a^2\over a^2}-{3\over2}K\right)\phi+k^2c_{\rm s}^2\phi=0\ ,
\label{acoust-3}
\end{equation}
where we used the Friedmann equation (\ref{Friedmann}); recall that
$K=(\Omega-1)(aH)^2$ is the spatial curvature constant.  In a
matter-dominated universe, differentiating the Friedmann equation gives
$\ddot a/a-(1/2)\dot a^2/a^2=-(1/2)K$, yielding
\begin{equation}
  \ddot\phi+3{\dot a\over a}\dot\phi+\left(k^2c_{\rm s}^2-2K\right)\phi=0\ .
\label{acoust-4}
\end{equation}
When written in terms of the gravitational potential rather than the
density, the wave equation loses its gravitational source term.

The solutions to eq. (\ref{acoust-4}) depend on the time-dependence
of the sound speed as well as on the background cosmology.  To get a rough
idea of the behavior, consider the evolution of the potential in an
Einstein-de Sitter universe filled with an ideal gas.  For a constant
sound speed, the solutions are
\begin{equation}
 \!\phi_+(k,\tau)=\tau^{-2}j_2(kc_{\rm s}\tau)\ ,\ \
  \phi_-(k,\tau)=\tau^{-2}y_2(kc_{\rm s}\tau)\ ,\ \ c_{\rm s}=\hbox{const.}\ ,
\label{poteds-1}
\end{equation}
where $j_2$ and $y_2$ are the spherical Bessel functions of the first
and second kinds of order 2.  Although simple, this is not a realistic
solution even before recombination (in that case, the photons and baryons
behave as a single tightly-coupled relativistic gas, and relativistic
corrections to the fluid equations must be added), except insofar as
it illustrates the generic behavior of the two solutions: (damped)
oscillations for $kc_{\rm s}\tau\gg1$ and power-law behavior for $kc_{\rm
s}\tau\ll1$.

An alternative approximation, valid after recombination, is to assume
that the baryon temperature roughly equals the photon temperature (this
is a reasonable approximation because the small residual ionization
thermally couples the two fluids for a long time even though there is
negligible momentum transfer), $c_{\rm s}^2=c_{\rm 0s}^2a^{-1}$ where
$c_{\rm 0s}$ is a constant.  In this case the solutions are powers of $\tau$:
\begin{equation}
  \phi_\pm(k,\tau)=\tau^n\ ,\ \ n={-5\pm\sqrt{25-4(kc_{\rm 0s}\tau_0)^2}
    \over2}\ ,\quad T_{\rm gas}\propto a^{-1}\ .
\label{poteds-2}
\end{equation}
The solutions oscillate for $kc_{\rm s}\tau_0>5/2$ and they damp for
$kc_{\rm s}\tau_0>0$.

In both of our solutions, and indeed for any reasonable equation of
state in an Einstein-de Sitter universe, long-wavelength ($kc_{\rm s}\tau\ll1$)
growing density modes have corresponding potential $\phi_+=\hbox{constant}$,
while the decaying density modes have $\phi_-\propto\int a^{-3}d\tau$.  The
density perturbation and potential differ by a factor of $\bar\rho
a^2\propto a^{-1}$ from the Poisson equation.  If $K<0$ or $k^2c_{\rm s}^2>0$,
then $\phi_+$ decays with time, although $\delta_+$ still grows.  Note
that the important physical length scale where the {\bf transfer function}
$\phi_+(k,\tau)/\phi(k,0)$ falls significantly below unity is the
{\it acoustic} comoving horizon distance $c_{\rm s}\tau$, not the causal
horizon distance $c\tau$ or the Hubble distance $c/H$.  Setting $c_{\rm
s}$ to the acoustic speed of the coupled photon-baryon fluid at
matter-radiation equality gives the physical scale at which the bend
occurs, $c_{\rm s}\tau_{\rm eq}$, in the power spectrum of the standard
cold dark matter and other models.

\subsection{Linear instability 2: entropy fluctuations and isocurvature mode}

Entropy gradients act as a source term for density perturbation growth.
Using eq. (\ref{p-grad}) and repeating the derivation of the linear
acoustic equation, we obtain (for $c_{\rm s}^2\ll c^2$)
\begin{equation}
  \ddot\delta+{\dot a\over a}\dot\delta-4\pi G\bar\rho a^2\delta-
    c_{\rm s}^2\nabla^2\delta={2\over3}T\nabla^2S\ .
\label{acoust-5}
\end{equation}
For {\it adiabatic} evolution, $\dot S=0$, so what counts is the initial
entropy gradient.  Entropy gradients may be produced in the early universe
by first-order phase transitions resulting in spatial variations in
the photon/baryon ratio or other abundance ratios.  If there were no
entropy gradients present before such a phase transition, then the
entropy variations can only have been produced by nonadiabatic
processes.  (This may explain the ``adiabatic vs. isocurvature"
nomenclature used by some cosmologists.)  In practice, these entropy
fluctuations are taken as initial conditions for subsequent adiabatic
evolution.

Equation (\ref{acoust-5}) is not applicable to the early universe because
it assumes the matter is a one-component nonrelativistic gas.  However,
the behavior of its solutions are qualitatively similar to those for
a relativistic multi-component gas and so its analysis is instructive.

The {\bf isocurvature mode} is given by the particular solution of
density perturbation growth having $\delta=\dot\delta=0$ but $\nabla^2S
\ne0$ at some early initial time $\tau_i$.  The initial conditions may
be regarded as a perturbation in the equation of state in an otherwise
unperturbed Robertson-Walker (constant spatial curvature) spacetime,
accounting for the name ``isocurvature.'' Variations in entropy at
constant density correspond to variations in pressure, which lead through
adiabatic expansion to changes in the density.  Therefore, initial entropy
fluctuations seed density fluctuations.

The solution to eq. (\ref{acoust-5}) is obtained easily in Fourier
space using the source-free (isentropic) solutions $\delta_\pm(k,\tau)$:
\begin{eqnarray}
  \delta_S(k,\tau)=-{2\over3}k^2S(k)\Biggl[&&\delta_+(k,\tau)
    \int_{\tau_i}^{\tau}a'T'\delta_-'\,d\tau'\nonumber\\
    -&&\delta_-(k,\tau)\int_{\tau_i}^{\tau}a'T'\delta_+'\,d\tau'\Biggr]\ ,
\label{isocurv}
\end{eqnarray}
where primes are used to indicate that the variables are evaluated
at $\tau=\tau'$.  We see that both growing and decaying density
perturbations are induced.  After the source ($aT\delta_-$) becomes
small, the density fluctuations evolve the same way as isentropic
fluctuations --- e.g., they oscillate as acoustic waves if $kc_{\rm
s}\tau\gg1$.
To reinforce the point about nomenclature made earlier, I note that
in our approximation, {\it both} isocurvature and ``adiabatic'' (i.e.,
isentropic) modes are {\it adiabatic} in the sense of thermodynamics:
$\dot S=0$ after the initial moment.  For a realistic multi-component
gas the evolution is not truly adiabatic, but that is a complication
we shall not consider further.  In the literature, modes are described
as being adiabatic or isocurvature depending only on whether the
initial density is perturbed with negligible initial entropy perturbation,
or vice versa.

\subsection{Vorticity --- or potential flow?}

With the growing and decaying isentropic perturbations, and the
isocurvature mode, we have accounted for three of the expected five
linear modes.  The remaining two degrees of freedom were lost when
we took the divergence of the Euler equation, thereby annihilating
any transverse (rotational) contribution to $\vec v$.  We consider
them now.

{\bf Theorem}:  Any differentiable vector field $\vec v(\vec x)$
may be written as a sum of longitudinal (curl-free) and transverse
(divergence-free) parts, $\vec v_\parallel$ and $\vec v_\perp$, respectively:
\begin{equation}
  \vec v(\vec x)=\vec v_\parallel(\vec x)+\vec v_\perp(\vec x)\ ,
  \quad \vec\nabla\times\vec v_\parallel=\vec\nabla\cdot\vec v_\perp=0\ .
\label{long-trans}
\end{equation}
The proof follows by construction, by solving $\vec\nabla\cdot\vec
v_\parallel=\theta$ and $\vec\nabla\times\vec v_\perp=\vec\omega$
where $\theta\equiv\vec\nabla\cdot\vec v$ and $\vec\omega\equiv
\vec\nabla\times\vec v$.  In a flat Euclidean space, solutions are
given by
\begin{equation}
  \vec v_\parallel(\vec x)={1\over4\pi}\int\theta(\vec x')
    {(\vec x-\vec x')\over\vert\vec x-\vec x'
    \vert^3}\,d^3x'\ ,\quad\theta(\vec x)\equiv\vec\nabla\cdot\vec v\ ,
\label{v-long}
\end{equation}
\begin{equation}
  \vec v_\perp(\vec x)={1\over4\pi}\int\vec\omega(\vec x')
    \times{(\vec x-\vec x')\over\vert\vec x-\vec x'
    \vert^3}\,d^3x'\ ,\quad\vec\omega(\vec x)\equiv\vec\nabla\times\vec v
      \ .
\label{v-trans}
\end{equation}
Note that this decomposition is not unique; we may always add to
$\vec v_\parallel$ a curl-free solution of $\vec\nabla\cdot\vec
v_\parallel=0$ and to $\vec v_\perp$ a divergence-free solution of
$\vec\nabla\times\vec v_\perp=0$ (e.g., constant vectors).  With
suitable boundary conditions (e.g., $\int \vec v_\parallel\,d^3x=0$
when integrated over all space) this freedom can be eliminated.
The variables $\theta$ and $\vec\omega$ are called the (comoving)
expansion scalar and vorticity vector, respectively.

In our preceding discussion of perturbation evolution we have implicitly
considered only $\vec v_\parallel$.  The remaining two degrees of
freedom correspond to the components of $\vec v_\perp$ (the
transversality condition $\vec\nabla\cdot\vec v_\perp=0$ removes one
degree of freedom from this 3-vector field).  Fortunately, we can get
a simple nonlinear equation for $\vec v_\perp$ --- actually, for its
curl, $\vec\omega$ --- by taking the curl of the Euler
equation:
\begin{eqnarray}
  \dot\vec\omega+{\dot a\over a}\vec\omega&=
    \vec\nabla\times(\vec v\times\vec\omega)+\rho^{-2}
      (\vec\nabla\rho)\times(\vec\nabla p)\nonumber\\
  &=\vec\nabla\times(\vec v\times\vec\omega)+{2\over3}T
      (\vec\nabla\ln\rho)\times(\vec\nabla S)
\label{vort}
\end{eqnarray}
where we have assumed an ideal monatomic gas in writing the second
form.  The term arising from entropy gradients is called the
{\bf baroclinic} term.  It is very important for the dynamics
of the Earth's atmosphere and oceans (Pedolsky 1987).

An important general result follows from eq. (\ref{vort}),
the {\bf Kelvin Circulation Theorem}: If $\vec\omega=0$ everywhere
initially, then $\vec\omega$ remains zero (even in the nonlinear
regime) if the baroclinic term vanishes.  (We are assuming that
other torques such as magnetic ones vanish too.)  The reason for
the importance of this result in cosmology is that many models
assume irrotational, isentropic initial conditions.  With adiabatic
evolution, it follows that $\vec\omega=0$.  Such a flow is also
called potential flow because the velocity field may then be obtained
from a velocity potential: $\vec v=\vec v_\parallel=-\vec\nabla\Phi_v$.

Nonadiabatic processes (heating and cooling) and oblique shock waves
can generate vorticity.  In a collisionless fluid, if the fluid velocity
is defined as the mass-weighted average of all the mass elements at a
point, this averaging behaves like entropy production in regions where
trajectories intersect, and so vorticity can be generated in the mean
(fluid) velocity field.  Vorticity also arises from isocurvature initial
conditions.  Equation (\ref{isocurv}) implies $\delta_S\propto\nabla^2S$
for long wavelengths in the linear regime, giving a baroclinic torque
proportional to $\vec\nabla\delta_S\times\vec\nabla S\propto\vec\nabla
(\nabla^2S)\times\vec\nabla S$, which is nonzero in general (though it
appears only in second-order perturbation theory).

For most structure formation models, vorticity generation is quite small
until shocks form (or trajectories intersect, for collisionless dark
matter).  In this case, one may obtain the velocity potential from the
line integral of the velocity field:
\begin{equation}
  \Phi_v(\vec x)=\Phi_v(0)-\int_0^{\vec x}\vec v\cdot d\vec l\ .
\label{vpot}
\end{equation}
Taking the path to be radial with the observer in the middle allows
one to reconstruct the velocity potential, and therefore the transverse
velocity components, from the radial component.  This idea underlies
the potential flow reconstruction method, POTENT (Bertschinger \&
Dekel 1989).  If the (smoothed) density fluctuations are sufficiently
small for linear theory to be valid, we can estimate the density
fluctuation field from an additional divergence.  If pressure is
unimportant, so $k\ll k_{\rm J}$ and $\delta\propto\delta_+(\tau)$, the
linearized continuity equation gives
\begin{equation}
  \vec\nabla\cdot\vec v=\theta=-\dot\delta=-aH\left(d\ln\delta_+\over
    d\ln a\right)\,\delta\ .
\label{delta-dot}
\end{equation}
For a wide range of cosmological models, $d\ln\delta_+/d\ln a\equiv
f(\Omega)\approx\Omega^{0.6}$ depends primarily on the mass density
parameter and weakly on other cosmological parameters (Peebles 1980,
Lahav et al. 1991).  Thus, combining measurements of $\vec v$ (radial
components from galaxy redshifts and distances) and independent
measurements of $\delta$ (from the galaxy density field plus an
assumption about how dark matter is distributed relative to galaxies)
allows estimation of $\Omega$ (Dekel et al. 1993).  A review of the
POTENT techniques and results is given by Dekel (1994).

\section{Hot dark matter}

The previous lecture studied the evolution of an ideal collisional gas
including gravity and pressure.  A gas of neutrinos, or of collisionless
dark matter particles, behaves differently.  In this lecture we investigate
the evolution of a nonrelativistic collisionless gas whose particles
have significant thermal speeds.  (Relativistic kinetic theory is
discussed by Stewart 1971, Bond \& Szalay 1983, and Ma \& Bertschinger
1994b.) An example is the gas of relic thermal neutrinos that decoupled
at a temperature $k_{\rm B}T\sim1$ MeV in the early universe.   The
present number density of these neutrinos (about 113 $cm^{-3}$ for each
of the three flavors) is such that a single massive type contributes
$m_\nu c^2/(93\,h^2\,\hbox{eV})$ to $\Omega$, where $h=H_0/(100\,
\hbox{km}\,\hbox{s}^{-1}\,\hbox{Mpc}^{-1})$.  Massive neutrinos are
called hot dark matter because their thermal speeds significantly affect
the gravitational growth of perturbations.

Before working out the detailed equations of motion for hot dark matter,
it is useful to consider in general terms the effect of a thermal
distribution.   Suppose we have a cold gas with no thermal motions.
In this case it doesn't matter whether the gas is collisional or
collisionless: gravitational instability amplifies the growing mode
of irrotational density perturbations.  What happens when we add
thermal motions?  We know the answer for a collisional gas: pressure
stabilizes collapse for wavelengths less than the Jeans length, the
distance sound waves travel in one gravitational dynamical time.  For
collisionless particles we also expect suppression.  However, a
collisionless gas cannot support sound waves, because no restoring force is
provided by particle collisions.

A perfect collisional gas is fully described by its mass (or energy) density,
fluid velocity, and temperature as functions of position.  All other properties
follow from the fact that the phase space density distribution is (locally)
the thermal equilibrium distribution, e.g. Maxwell-Boltzmann.  This is not
true for a collisionless gas, whose complete description requires specifying
the full phase space density.

For a collisionless gas, the velocity distribution function may be far
from Maxwellian, so that the spatial stress tensor is not the simple
diagonal form appropriate for an ideal gas.  Instead there may be
significant off-diagonal terms contributing {\bf shear stress} that
acts like viscosity in a weakly collisional fluid: it damps relative
motions.  We expect perturbations in a collisionless gas to be damped
for wavelengths shorter than the distance traveled by particles
with the characteristic thermal speed during one gravitational
collapse time, the collisionless analogue of the Jeans length.  Stated
simply, overdense or underdense perturbations decay because the
particles fly away from them at thermal speeds.  This collisionless damping
process is called free-streaming damping.

The characteristic thermal speed of massive neutrinos after they become
nonrelativistic is
\setcounter{equation}{0}
\begin{equation}
v_{\rm th}={k_{\rm B}T_\nu\over m_\nu c}=50.4(1+z)\,(m_\nu c^2/\hbox{eV})^{-1}
  \,\hbox{km}\,\hbox{s}^{-1}
\label{nu-v}
\end{equation}
where we have used the standard big bang prediction $T_\nu=
(4/11)^{1/3}\,T_\gamma$ (e.g., Kolb \& Turner 1990) with $T_\gamma
\approx2.735$ K today.  Multiplying $v_{\rm th}$ by the gravitational time
$(4\pi G\bar\rho a^2)^{-1/2}$ gives the comoving free-streaming distance,
\begin{equation}
\lambda_{\rm fs}=0.41\,(\Omega h^2)^{-1/2}\,(1+z)^{1/2}\,
  (m_\nu c^2/\hbox{eV})^{-1}\,\hbox{Mpc}\ .
\label{fsd}
\end{equation}
At any time, fluctuations with wavelength less than about $\lambda_{\rm fs}$
are damped; much longer wavelength fluctuations grow with negligible
suppression.

The free-streaming distance does not really grow without bound as $z\to\infty$
because the neutrino thermal speed cannot exceed $c$.  Applying this limit
gives a maximum comoving free-streaming distance of
\begin{equation}
\lambda_{\rm fs,max}=31.8\,(\Omega h^2)^{-1/2}\,
  (m_\nu c^2/\hbox{eV})^{-1/2}\,\hbox{Mpc}\ .
\label{fsdmax}
\end{equation}
Thus, unless they are regenerated by perturbations in other components
(as happens, for example, in a model with hot and cold dark matter),
primeval density fluctuations in massive neutrinos with wavelength
smaller than this rather large scale will be erased by free-streaming
damping.  A more quantitative treatment is presented below using the
actual evolution equations for the neutrino phase space density distribution.

\subsection{Tremaine-Gunn bound}

Before treating the phase space evolution, we discuss another important
consequence of finite neutrino thermal speed: high-speed neutrinos
cannot be tightly packed into galaxy halos.  This fact can be used to
place a lower bound on the neutrino mass if neutrinos make up the dark
matter in galaxy halos (Tremaine \& Gunn 1979).

The initial phase space density for massive neutrinos is a relativistic
Fermi-Dirac distribution (preserved from the time when the neutrinos
decoupled in the early universe):
\begin{equation}
f={2h_{\rm P}^{-3}\over\exp(pc/k_{\rm B}T_0)+1} \equiv f_0(\vec p)\ ,
\label{fermi-dirac}
\end{equation}
where $\vec p$ is the comoving canonical momentum of eq. (\ref{p-conj}),
$h_{\rm P}$ is Planck's constant (with a subscript to distinguish it from
the scaled Hubble constant), and $T_0=aT_\nu$ is the present neutrino
``temperature.'' The decrease of $T_\nu$ with time is compensated for by
the factor $a$ relating proper momentum to comoving momentum.
Ignoring perturbations, the present-day distribution for massive neutrinos
is the relativistic Fermi-Dirac --- not the equilibrium nonrelativistic
distribution --- because the phase space distribution was preserved after
neutrino decoupling.

Tremaine \& Gunn (1979) noted that because of phase mixing (discussed
further below), the maximum coarse-grained phase space density of
massive neutrinos today is less than the maximum of $f_0(\vec p)$,
$h_{\rm P}^{-3}$.  If massive neutrinos dominate the mass in galactic halos,
this must be no less than the phase space density needed for self-gravitating
equilibrium.  This bound can be used to set a lower limit on the neutrino
mass  if one assumes that the neutrinos constitute the halo dark matter.

Although the neutrino mass bound is somewhat model-dependent
because the actual coarse-grained distribution in galactic halos is
unknown, we can get a reasonable estimate by assuming an isothermal
sphere: a Maxwell-Boltzmann distribution with constant velocity
dispersion $\sigma^2$ (at $a=1$ so that there is no distinction between
proper and comoving):
\begin{equation}
  f(\vec r,\vec p)=(2\pi m_\nu^2\sigma^2)^{-3/2}n(r)\,
    \exp\left(-p^2\over2m_\nu^2\sigma^2\right)\ .
\label{m-b}
\end{equation}
In a self-gravitating system there are a family of spherical density profiles
$\rho(r)=m_\nu n(r)$ obeying hydrostatic equilibrium:
\begin{equation}
  {1\over\rho}{d P\over dr}=-{GM(<r)\over r^2}=-{4\pi G\over r^2}
    \int_0^r r^2\rho(r)\,dr\ .
\label{hydstat-eq}
\end{equation}
The simplest case is the singular isothermal sphere with $\rho\propto
r^{-2}$; the reader can easily check that $\rho=\sigma^2/(2\pi Gr^2)$.
Imposing the phase space bound at radius $r$ then gives
\begin{equation}
  m_\nu>(2\pi)^{-5/8}\left(Gh_{\rm P}^3\sigma r^2\right)^{-1/4}\ .
\label{t-g-bound}
\end{equation}
Up to overall numerical factors, this is the Tremaine-Gunn bound.

The singular isothermal sphere is probably a good model where
the rotation curve produced by the dark matter halo is flat, but
certainly breaks down at small radius.   Because the neutrino mass
bound is stronger for smaller $\sigma r^2$, the uncertainty in the
halo core radius (interior to which the mass density saturates) limits
the reliability of the neutrino mass bound.

For the Local Group dwarf galaxies in Draco and Ursa Minor, measurements
of stellar velocity dispersions suggest $\sigma$ is a few to about 10
km s$^{-1}$ (Pryor \& Kormendy 1990).  If these galaxies have isothermal
halos at $r=1$ kpc, the crude bound of eq. (\ref{t-g-bound})
implies $m_\nu$ is greater than a few eV.

\subsection{Vlasov equation}

We now present a rigorous treatment of the evolution of perturbations
in a nonrelativistic collisionless gas, based on the evolution of the phase
space distribution.  The single-particle phase space density $f(\vec x,
\vec p,\tau)$ is defined so that $fd^3xd^3p$ is the number of particles in
an infinitesimal phase space volume element.  We shall use comoving spatial
coordinates $\vec x$ and the associated conjugate momentum $\vec p=am
\dot\vec x$ (eq. \ref{p-conj}).  Note that $d^3xd^3p=m^3d^3rd^3v$ is a
proper quantity so that $f$ is the proper (physical) phase space density.

If the gas is perfectly collisionless, $f$ obeys the Vlasov (or collisionless
Boltzmann) equation of kinetic theory,
\begin{equation}
  {Df\over D\tau}\equiv{\partial f\over\partial\tau}+{d\vec x\over d\tau}
    \cdot{\partial f\over\partial\vec x}+{d\vec p\over d\tau}\cdot
    {\partial f\over\partial\vec p}=0\ .
\label{vlasov-1}
\end{equation}
This equation expresses conservation of particles along the phase space
trajectory $\{\vec x(\tau),\vec p(\tau)\}$.  Using Hamilton's equations
(\ref{hamilton-c}) for nonrelativistic particles, we obtain
\begin{equation}
  {\partial f\over\partial\tau}+{\vec p\over am}\cdot{\partial f\over
    \partial\vec x}-am\vec\nabla\phi\cdot{\partial f\over\partial\vec p}=0\ .
\label{vlasov-2}
\end{equation}

The Vlasov equation is supposed to apply for the coarse-grained phase
space density for a collisionless gas in the absence of two-body correlations
(Ichimaru 1992).  Often, however, the statistical assumptions underlying
the use of the Vlasov equation are vague.  To clarify its application we
digress to present a derivation using the Klimontovich (1967) approach to
kinetic theory.

Consider one realization of a universe filled with particles following
phase space trajectories $\{\vec x_i(\tau),\vec p_i(\tau)\}$ ($i$ labels
the particles).  The {\it exact} single-particle phase space density
(called the Klimontovich density) is written by summing over Dirac
delta functions:
\begin{equation}
  f(\vec x,\vec p,\tau)=\sum_i\delta[\vec x-\vec x_i(\tau)]\,\delta[\vec p-
    \vec p_i(\tau)]\ .
\label{klimontovich}
\end{equation}
No statistical averaging or coarse-graining has been applied; $f$ is the
fine-grained density for one universe.  This phase space density obeys
the Klimontovich (1967) equation, which is of exactly the same form as
eq. (\ref{vlasov-1}).  The proof follows straightforwardly from
substituting eq. (\ref{klimontovich}) into eq. (\ref{vlasov-1}).

The Klimontovich density retains all information about the microstate of
a system because it specifies the trajectories of all particles.  This is far
too much information to be practical.  We must reduce the information
content by performing some averaging or coarse-graining.  This averaging
is taken over a statistical ensemble of microstates corresponding to a
given macrostate --- for example, microstates with the same phase space
density averaged over small phase space volumes containing many
particles on average.  We denote the averages using angle brackets
$\langle\rangle$, without being very precise about the ensemble adopted
for the coarse-graining.

The discreteness effects of individual particles are accounted for by the
$s$-particle distribution functions ($s=1$, 2, $\ldots$) $f_s$, which are
defined using a standard cluster expansion:
\begin{equation}
  \langle f(\vec x,\vec p,\tau)\rangle=\left\langle\sum_i\delta(\vec x-
    \vec x_i)\,\delta(\vec p-\vec p_i)\,\right\rangle\equiv f_1(\vec x,
    \vec p,\tau)\ ,
\label{f1}
\end{equation}
\clearpage
\begin{eqnarray}
\label{f2}
  \langle f(\vec x_1,\vec p_1,\tau)\,f(\vec x_2,\vec p_2,\tau)\rangle=
    \quad\quad\quad\quad\quad\quad\quad\quad\quad\quad\quad\quad\quad\quad
    \nonumber\\
  \left\langle\sum_{i=j}\delta(\vec x_1-\vec x_i)\,\delta(\vec p_1-\vec p_i)
    \,\delta(\vec x_2-\vec x_i)\,\delta(\vec p_2-\vec p_i)\,\right\rangle+
    \nonumber\\
   \left\langle\sum_{i\ne j}\delta(\vec x_1-\vec x_i)\,\delta(\vec p_1-\vec
p_i)
    \,\delta(\vec x_2-\vec x_j)\,\delta(\vec p_2-\vec p_j)\,\right\rangle
    \\
  =\delta(\vec x_1-\vec x_2)\delta(\vec p_1-\vec p_2)f_1(\vec x_1,\vec
p_1,\tau)
  +f_2(\vec x_1,\vec p_1,\vec x_2,\vec p_2,\tau)\ ,\nonumber
\end{eqnarray}
and so on.  We further write $f_2$ as a sum of uncorrelated and correlated
parts,
\begin{equation}
  \!\!\!\!\!\!\!\!\!\!\!\!
  f_2(\vec x_1,\vec p_1,\vec x_2,\vec p_2,\tau)=f_1(\vec x_1,\vec p_1,\tau)
    f_1(\vec x_2,\vec p_2,\tau)+f_{2c}(\vec x_1,\vec p_1,\vec x_2,\vec p_2,
    \tau)\ .
\label{f2c}
\end{equation}
This equation defines $f_{2c}$, known in kinetic theory as the irreducible
two-particle correlation function.  If there are no pair correlations in
phase space, $f_{2c}=0$.

We now ensemble-average the Klimontovich equation, recalling that it
is identical to eq. (\ref{vlasov-2}) provided we use the Klimontovich
density.  If $\phi$ is a specified external potential, neglecting self-gravity,
we see that $f_1$ obeys the Vlasov equation.  However, if $\phi$ is computed
self-consistently from the particles, the $m\vec\nabla\phi\cdot(\partial f/
\partial\vec p)$ term is quadratic in the Klimontovich density, yielding an
additional correlation term from eqs. (\ref{f2}) and (\ref{f2c}) after
coarse-graining.  This term is not present in the Vlasov equation.

The contribution to the gravity field from the particles is (cf. eq.
\ref{egrav-1})
\begin{eqnarray}
  -\vec\nabla\phi(\vec x,\tau)=&&-{Gm\over a}\int d^3x'\,d^3p'\,
    f(\vec x',\vec p',\tau){(\vec x-\vec x')\over\vert\vec x-\vec x'
      \vert^3}\nonumber\\
    &&+G\bar\rho a^2\int d^3x'\,{(\vec x-\vec x')\over\vert\vec x-\vec x'
     \vert^3}\ ,
\label{self-g}
\end{eqnarray}
where the second term, required in comoving coordinates, removes the
contribution from the mean uniform background.

Combining our results now yields the exact kinetic equation for the
one-particle phase space density $f_1$:
\begin{eqnarray}
  &&
  {\partial f_1\over\partial\tau}+{\vec p\over am}\cdot{\partial f_1\over
    \partial\vec x}-am\vec\nabla\phi\cdot{\partial f_1\over\partial\vec p}=
    \nonumber\\
  &&
  Gm^2\int d^3x'd^3p'\,{(\vec x-\vec x')\over\vert\vec x-\vec x'\vert^3}
    \cdot{\partial\over\partial\vec p}f_{2c}(\vec x,\vec p,\vec x',\vec
p',\tau)
\label{bbgky1}
\end{eqnarray}
where $-\vec\nabla\phi$ is given by eq. (\ref{self-g}) using
$f_1$ for $f$, and adding any other contribution from other sources.
Equation (\ref{bbgky1}) is called the first BBGKY hierarchy equation
(Peebles 1980, Ichimaru 1992).  It differs from the Vlasov equation
by a correlation integral term.

If there are no phase space correlations, as would occur if we had a
smooth collisionless fluid, then the one-particle or coarse-grained
distribution obeys the Vlasov equation of kinetic theory.  Correlations
may be introduced by gravitational clustering, which couples $f_{2c}$
to $f_1$.  One may derive an evolution equation for $f_{2c}$ --- the second
BBGKY hierarchy equation --- by averaging $f\partial f/\partial\tau$,
but it involves $f_{3c}$, and so on.  The result is an infinite hierarchy
of coupled kinetic equations, the BBGKY hierarchy.

For some cases, Boltzmann's hypothesis of molecular chaos may hold,
implying $f_{2c}=0$ except at binary collisions, with the right-hand side
of eq. (\ref{bbgky1}) becoming a Boltzmann collision operator.
Fortunately, for the particles of interest here --- neutrinos --- the
gravitational (and non-gravitational, after neutrino decoupling) collision
time is so long that the correlation integral is completely negligible.
Thus, hot dark matter composed of massive neutrinos obeys the Vlasov
equation after decoupling.  From now on we shall drop the subscript
$1$ from $f$.

We now return to our main line of development to discuss phase mixing.
The Vlasov equation implies conservation of phase space density, but a
given initial volume $d^3xd^3p$ evolves in a complicated way (i.e., the
trajectories of particles initially inside this volume may be highly
complicated).  Consider the initial phase space element shown in Figure
2a, extracted from a one-dimensional $N$-body simulation.  Figures
2b and 2c show the phase space distribution at a later time, with each
particle's trajectory evolved according to Hamilton's equations without
(Fig. 2b) and with (Fig. 2c) gravity, respectively.  In both cases the
area $dxdp$ of the phase space element is identical to the initial area
as a consequence of the Vlasov equation.

\begin{figure}
  \vskip 2.1truein \hskip -1.8truein
  \includegraphics{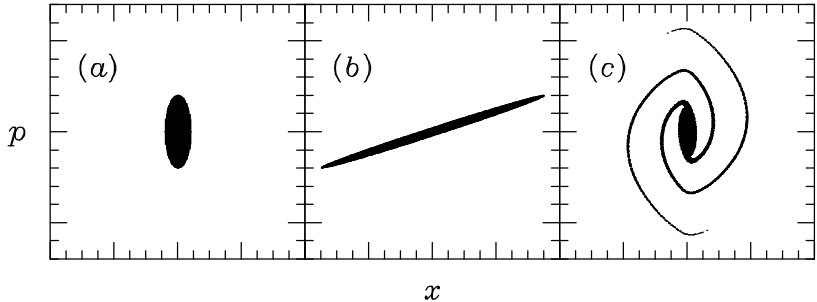}
  \vskip -0.5truein
  \caption{Phase space evolution. (a) Initial conditions.  (b) Evolved
  state without gravity.  (c) Evolved state with gravity.}
\label{fig2}
\end{figure}

Figure 2c illustrates the process known as phase mixing: the phase
space structure becomes highly convoluted as particles make multiple
orbits.  Regions of initially high phase space density can end up
entwined with regions of initially low phase space density.  Although
the density is conserved along each phase space trajectory, if the
distribution is coarse-grained (averaged over finite phase space
volume), the resulting coarse-grained density is not conserved.
The maximum coarse-grained density can only decrease, as we noted
previously in the discussion of the Tremaine-Gunn bound.

The process of phase-mixing is complicated, and the only practical
means of integrating the Vlasov equation for such an evolved collisionless
system is by $N$-body simulation: the phase space is sampled with
discrete particles at some initial time and the particle trajectories are
computed, providing a sample of the evolved phase space.  However,
analytical methods can be used  while the phase space distribution is
only slightly perturbed from the homogeneous equilibrium distribution.
These methods, presented in the next two subsections, will help us to
understand free-streaming damping in detail.

\subsection{Nonrelativistic evolution in an external gravitational field}

In this section we consider hot dark matter made of nonrelativistic
massive neutrinos with $\Omega_\nu\ll\Omega$ so that their self-gravity
is unimportant.  The gravitational potential $\phi(\vec x,\tau)$ (using
comoving coordinates) is assumed to be given from other sources
such as cold dark matter in a mixed hot and cold dark matter model.

We can solve the Vlasov equation (\ref{vlasov-2}) approximately by
replacing $\partial f/\partial\vec p$ with the unperturbed term
$\partial f_0/\partial\vec p$.  This approximation is valid for
$\vert f-f_0\vert\ll f_0$, and should suffice to demonstrate the
collisionless damping of small-amplitude fluctuations.

A quadrature solution of the Vlasov equation can be obtained provided
that we change the time variable from $\tau$ to $s=\int d\tau/a=\int dt/a^2$
and then Fourier transform the spatial variable:
\begin{equation}
  f(\vec x,\vec p,\tau(s))=\int d^3k\,e^{i\vec k\cdot\vec x}\,\hat f(\vec k,
    \vec p,s)\ .
\label{fourier}
\end{equation}
The gravitational potential $\phi$ is transformed similarly.  Integrating
eq. (\ref{vlasov-2}) over $s$, we obtain the solution
\clearpage
\begin{eqnarray}
  &&\hat f(\vec k,\vec p,s)=\hat f(\vec k,\vec p,s_i)\,e^{-i\vec k\cdot\vec u
    (s-s_i)}\nonumber\\
  &&\quad\quad+im\left(\vec k\cdot{\partial f_0\over\partial\vec p}\right)
  \int_{s_i}^sds'\,a^2(s')\hat\phi(\vec k,s')\,e^{-i\vec k\cdot\vec u(s-s')}
    \ ,
\label{fsol}
\end{eqnarray}
where $\vec u=\vec p/m$ and $s_i$ is an initial time.  If the initial phase
space distribution is unperturbed, then $\hat f(\vec k,\vec p,s_i)=f_0(p)
\,\delta(\vec k)$.  Note that the complex exponentials in eq. (\ref
{fsol}) correspond to the propagation of the phase space density along
the characteristics $d\vec x/ds=\vec u$.  This motion is called free-streaming.

To understand the behavior of the free-streaming solution, let us examine
the integral term of eq. (\ref{fsol}), which is proportional to
\begin{equation}
\int_0^{s-s_i}dy\,a^2(y+s_i)\,\hat\phi(\vec k,y+s_i)\,e^{-i\beta y}\ ,
\label{fsol-int}
\end{equation}
where $\beta\equiv \vec k\cdot\vec p/m$ and $y=s'-s_i$.  For sufficiently
slowly moving neutrinos, $\beta$ is small enough so that $\beta y\ll1$.
This condition corresponds to a free-streaming distance along $\vec k$
that is much less than $k^{-1}$.  These neutrinos do not move far from
the crests and troughs of the plane wave perturbation.  Neglecting the
exponential, the time dependence of the solution is the same as for
cold dark matter.

If, however, $\beta y\gg1$, corresponding to neutrinos traveling
across many wavelengths of a perturbation, the rapid oscillations of the
exponential lead to cancellation in the integrand of eq. (\ref{fsol-int})
and suppression of the neutrino phase space density perturbation.
This effect, known as free-streaming damping, occurs because neutrinos
that are initially at the crests or troughs of density waves move so far
that they distribute themselves almost uniformly.  The small gravitational
acceleration induced by the external potential is inadequate to collect
the fast-moving neutrinos in dense regions.

Thus, perturbations can grow only for the neutrinos that move less
than about one wavelength per Hubble time.  Our analysis confirms
the rough picture we sketched in the beginning of this lecture.

We can obtain the net density perturbation (in Fourier space) by
integrating eq. (\ref{fsol}) over momenta:
\begin{eqnarray}
  &&
  {\hat n_\nu(\vec k,s)\over n_0}\equiv{1\over n_0}\int d^3p\,\hat f(\vec k,
    \vec p,s)\nonumber\\
  &&
  \quad\quad=\delta(\vec k)-k^2\int_{s_i}^s ds'\,a^2(s')\,\hat\phi(\vec k,s')
    \,(s-s')\,F\left[k(s-s')\over m\right]\ ,
\label{dnuk-1}
\end{eqnarray}
where $n_0=\int d^3p\,f_0(p)$ is the mean comoving number density
and $F$ is the Fourier transform --- with respect to the momentum!
--- of the unperturbed distribution function:
\begin{equation}
  F(q)={1\over n_0}\int d^3p\,e^{-i\vec p\cdot\vec q}\,f_0(p)\ .
\label{fdist}
\end{equation}
For the relativistic Fermi-Dirac distribution appropriate to hot dark
matter, $F$ has the series representation (Bertschinger \& Watts 1988)
\begin{equation}
  F(q)={4\over 3\,\zeta(3)}\sum_{n=1}^\infty (-1)^{n+1}\,{n\over
    (n^2+q^2p_0^2)^2}\ ,\quad p_0\equiv{k_{\rm B}T_0\over c}\ ,
\label{fseries}
\end{equation}
where $\zeta(3)=1.202\ldots$ is the Riemann zeta function and
$F(0)=1$.

Equation (\ref{dnuk-1}) does not give much insight into free-streaming
damping.  To get a better feel for the physics, as well as a simpler
approximation for treating hot dark matter, we now show how to
convert eq. (\ref{dnuk-1}) into a differential equation for the
evolution of the hot dark matter density perturbation similar to
eq. (\ref{acoust-1}) for a perfect collisional fluid.  This may seem
impossible a priori --- how can the dispersive behavior of a collisionless
gas be represented by fluid-like differential equations? --- but we
shall see that it is possible if we approximate $f_0(p)$ by a form differing
slightly from the Fermi-Dirac distribution.  The results, although not
exact, will give us additional insight into the behavior of collisionless
damping.

The first step is to rewrite eq. (\ref{dnuk-1}) for the Fourier
transform of the density fluctuation $\hat\delta_\nu$:
\begin{equation}
  \hat\delta_\nu(\vec k,s)=-km\int_{s_i}^s ds'\,a^2(s')\,\hat\phi(\vec k,s')
    [qF(q)]\ ,\quad q\equiv{k(s-s')\over m}\ .
\label{dnuk-2}
\end{equation}
Next, we differentiate twice with respect to the time coordinate $s$:
\begin{equation}
  {\partial\hat\delta_\nu\over\partial s}=-k^2\int_{s_i}^s ds'\,a^2(s')\,
    \hat\phi(\vec k,s')\,{d\over dq}[qF(q)]\ ,
\label{dnuk-3}
\end{equation}
\begin{eqnarray}
  {\partial^2\hat\delta_\nu\over\partial s^2}&&=-k^2a^2(s)\hat\phi(\vec k,s)
    \nonumber\\
    &&-{k^3\over m}\int_{s_i}^s ds'\,a^2(s')\,\hat\phi(\vec k,s')\,
    {d^2\over dq^2}[qF(q)]\ .
\label{dnuk-4}
\end{eqnarray}
Note the appearance of a non-integrated source term in the second
derivative, arising because $d(qF)/dq$ does not vanish at $s=s'$ ($q=0$)
while $qF$ does.

Next, we note that if $d^2(qF)/dq^2$ were to equal a linear combination
of $d(qF)/dq$ and $(qf)$, then we could write the integral in equation
(\ref{dnuk-4}) as a linear combination of $\partial\hat\delta_\nu/\partial s$
and $\hat\delta_\nu$.  Unfortunately, this is not the case for $F(q)$ given
by eq. (\ref{fseries}).  However, it is true for the family of
distribution functions whose Fourier transforms are
\begin{equation}
   F_\gamma(q)=\exp(-\gamma qp_0)\ ,
\label{ffgam}
\end{equation}
for any dimensionless constant $\gamma$.  This defines the family of
phase space density distributions
\begin{equation}
  f_\gamma(p)=n_0\int{d^3q\over(2\pi)^3}\,e^{i\vec p\cdot\vec q}\,
    F_\gamma(q)={n_0\over\pi^2(\gamma p_0)^3}\left(1+{p^2\over
    \gamma^2 p_0^2}\right)^{-2}\ .
\label{fgam}
\end{equation}
For this form of unperturbed distribution we have
\begin{equation}
  {d^2\over dq^2}(qF_\gamma)=-2\gamma p_0{d\over dq}(qF_\gamma)
    -(\gamma p_0)^2qF_\gamma\ .
\label{dnuk-5}
\end{equation}
Combining eqs. (\ref{dnuk-2})--(\ref{dnuk-4}) and (\ref{dnuk-5}), we get
\begin{equation}
  {\partial^2\hat\delta_\nu\over\partial s^2}+2{\gamma p_0k\over m}\,
    {\partial\hat\delta_\nu\over\partial s}+{\gamma^2p_0^2k^2\over m^2}\,
    \hat\delta_\nu=-k^2a^2(s)\hat\phi(\vec k,s)\ .
\label{dnu-1}
\end{equation}

To put this result into a form similar to the acoustic wave equation we
derived for a collisional fluid, we define the characteristic proper
thermal speed
\begin{equation}
  c_\nu\equiv\gamma{k_{\rm B}T_\nu\over mc}={\gamma p_0\over ma}\ .
\label{cnu}
\end{equation}
Next, we change the time variable from $s$ back to $\tau$ with
$d\tau/ds=a$.  Finally, we assume that the source term gravitational
potential $\hat\phi$ is given by the Poisson equation for a perturbation
$\delta_{\rm c}$ in a component with mean mass density $\bar\rho_{\rm c}$
(e.g., cold dark matter --- recall that we are neglecting the self-gravity
of the neutrinos).  Dropping the hat on $\hat\delta_\nu$, the result is
\begin{equation}
  \ddot\delta_\nu+\left({\dot a\over a}+2kc_\nu\right)\dot\delta_\nu
    +k^2c_\nu^2\delta_\nu=4\pi Ga^2\bar\rho_{\rm c}\delta_{\rm c}\ .
\label{dnu-2}
\end{equation}
This equation was first derived by Setayeshgar (1990).  It is approximate
(not exact) for the linear evolution of massive neutrinos because we
replaced the Fermi-Dirac distribution by eq. (\ref{fgam}).  It is not
difficult to show that eq. (\ref{fgam}) is the only form of the
distribution function for which eq. (\ref{fsol}) can be reduced
to a differential equation for $\delta_\nu(\vec k,\tau)$.  (Even the
Maxwell-Boltzmann distribution fails --- a collisionless gas with this
distribution initially  does not evolve the same way as a collisional
gas with the Maxwell-Boltzmann distribution function for all times.)
One should also bear in mind that $\delta_\nu$ does not contain all the
information needed to characterize perturbations in a collisionless gas
(Ma \& Bertschinger 1994a).  Complete information resides in $\hat f(\vec k,
\vec p,s)$.

Even if eq. (\ref{dnu-2}) is not exact for massive neutrinos and
does not fully specify the perturbations, it provides an extremely
helpful pedagogic guide to the physics of collisionless damping.  We
see at once that a gravitational source can induce density perturbations
in a collisionless component, but the source competes agains acoustic
($k^2c_\nu^2$) and damping ($\dot a/a+2kc_\nu$) terms.  Roughly
speaking, hot dark matter behaves like a collisional gas with an extra
free-streaming damping term.

Does the $k^2c_\nu^2$ term imply that a collisionless gas can support
acoustic oscillations?  To check this we consider the limit $kc_\nu\tau
\gg1$ so that the Hubble damping and gravitational source terms are
negligible.  We then have
\begin{equation}
  \ddot\delta_\nu+2\omega_\nu\dot\delta_\nu+\omega_\nu^2\delta_\nu
    \approx0\ ,\quad \omega_\nu=kc_\nu\ .
\label{dnu-3}
\end{equation}
Because $\omega_\nu$ changes very slowly with time compared with
the oscillation timescale $\omega^{-1}$, eq. (\ref{dnu-3}) is a
linear differential equation with constant coefficients and is easily
solved to give the two modes
\begin{equation}
  \delta_\nu\propto\tau e^{-\omega_\nu\tau}\quad\hbox{or}\quad
     e^{-\omega_\nu\tau}\ ,\quad\omega_\nu\tau\gg1\ .
\label{dnu-3-sol}
\end{equation}
Neither solution oscillates!  The first one begins to grow but is rapidly
damped on a timescale $\omega_\nu^{-1}$, after the typical neutrino
has had time to cross one wavelength.

Because the damping time $(kc_\nu)^{-1}$ is proportional to the wavelength,
short-wavelength perturbations are damped most strongly.  At any given
time $\tau$, perturbations of comoving wavelength less than about
$c_\nu\tau$ are attenuated.  This is precisely the free-streaming
distance we introduced in the beginning of this lecture, equation
(\ref{fsd}).

Our results enable us to understand why the hot dark matter transfer
function is similar to that of cold dark matter for long wavelengths but
cuts off sharply for short wavelengths (Bond \& Szalay 1983).  During the
radiation-dominated era, $a(\tau)\propto\tau$.  While the massive
neutrinos were relativistic, $c_\nu\approx c$ was constant.  The
comoving free-streaming distance increased, $c_\nu\tau\propto a$,
with hot dark matter perturbations being erased on scales up to the
Hubble distance.  After the neutrinos became nonrelativistic, however,
$c_\nu$ is given by eq. (\ref{cnu}), $c_\nu\propto a^{-1}$.
Thus, the free-streaming distance saturates at the Hubble distance
when the neutrinos become nonrelativistic.  During the matter-dominated
era, $a(\tau)\propto\tau^2$ (while $\Omega\approx1$) so that the
free-streaming distance decreases: $c_\nu\tau\propto a^{-1/2}$.
However, free-streaming has already erased the hot dark matter
perturbations on scales up to the maximum free-streaming distance,
eq. (\ref{fsdmax}).  Only if the perturbations are re-seeded,
e.g. by cold dark matter or topological defects, will small-scale power
be restored to the hot dark matter.

\subsection{Nonrelativistic evolution including self-gravity}

Now that we have developed the basic techniques for solving the
linearized nonrelativistic Vlasov equation, adding self-gravity of the
collisionless particles is easy.  We simply add a contribution to $\phi$
arising from $\delta_\nu$.  In eq. (\ref{fsol}),  if we have a
mixture of hot and cold dark matter, $\hat\phi\to(\hat\phi_{\rm c}+
\hat\phi_\nu$); additional contributions may be added as appropriate.
Equation (\ref{dnuk-2}) becomes
\begin{eqnarray}
  \hat\delta_\nu(\vec k,s)={m\over k}&&\int_{s_i}^s ds'\,a^2(s')\,
    [qF(q)]\,4\pi Ga^2(s')\nonumber\\
    &&\times\left[\bar\rho_{\rm c}(s')\hat\delta_{\rm c}
    (\vec k,s')+\bar\rho_\nu(s')\hat\delta_\nu(\vec k,s')\right]\ .
\label{dnuk-6}
\end{eqnarray}
This equation was first derived (in a slightly different form) by
Gilbert (1966) and is known as the Gilbert equation.  Note that in
the self-gravitating case $\delta_\nu$ appears both inside and outside
an integral.  Equation (\ref{dnuk-6}) is a Volterra integral equation of
the second kind.  Bertschinger \& Watts (1988) present a numerical
quadrature solution method.

Using the same trick as in the previous subsection, we can convert the
Gilbert equation to a differential equation for $\delta_\nu$, if the
unperturbed phase space density distribution is approximated by the
form $f_\gamma(p)$ of eq. (\ref{fgam}).  The result is
\begin{equation}
  \ddot\delta_\nu+\left({\dot a\over a}+2kc_\nu\right)\dot\delta_\nu
    +k^2c_\nu^2\delta_\nu=4\pi Ga^2\left[\bar\rho_{\rm c}\delta_{\rm c}
    +\bar\rho_\nu\delta_\nu\right]\ .
\label{dnu-4}
\end{equation}

With a suitable choice for the parameter $\gamma$, the solution of
eq. (\ref{dnu-4}) provides a good match (to within a few percent,
in general) to the solution of the Gilbert equation using the correct
Fermi-Dirac distribution for massive neutrinos (Setayeshgar 1990).
Therefore, it may be used for obtaining quick estimates of the density
perturbations of nonrelativistic hot dark matter.

\section{Relativistic cosmological perturbation theory}

\def\htens{{\sf h}}
\def\stens{{\sf\Sigma}}

\subsection{Introduction}
\label{int}

This section is an expanded version of my fifth lecture at Les Houches.
One lecture gave barely enough time to introduce the essential ideas of
relativistic perturbation theory: classification of metric perturbations,
the linearized Einstein equations, and gauge modes.  Understanding the
physics of these topics, as well as the relativistic generalizations
of my previous lectures, requires a much deeper immersion.  Unable to
find a pedagogical treatment in the existing literature that matches
these needs to my satisfaction, I have developed the subject more fully
in these written lecture notes.  They are not a complete guide to
relativistic perturbation theory but rather a starting point from which
the reader may delve into the increasingly rich literature of applications.
This section is self-contained and may be read independently of the previous
sections, although the reader may find it interesting to contrast the
nonrelativistic presentations of sections 1 and 2 with the relativistic
treatment given below.

\subsubsection{Synopsis}

According to the Newtonian perspective of gravity and cosmology,
spacetime is flat and absolute, gravity is action at a distance, and
particle dynamics is given by Newton's second law $\vec F=m\vec a$ or,
equivalently, by Hamilton's principle of least action.  The Einsteinian
perspective is quite different: spacetime is a curved manifold which
evolves causally through the Einstein field equations in response to
sources, and particle dynamics is given in absence of nongravitational
forces by geodesic motion.  In this section I attempt not only to present
the essentials of relativistic gravitational dynamics, but also to show
how it reduces to and extends Newtonian cosmology in the appropriate limit.

One of the main purposes of these notes is to provide a clear explanation
of the scalar, vector, and tensor modes of gravitational perturbations.
(We shall follow the customary usage in this subject by referring to
different spatial symmetry components as ``modes'' even when they are
not expanded in any basis eigenfunctions.  Thus, the ``scalar mode''
is described, in part, by a field $\phi(x^\mu)$ that is a scalar under
spatial coordinate transformations but is not restricted to being a
single Fourier component or other harmonic basis function.)
Newtonian gravity corresponds to the former (the scalar mode), while
the latter (vector and tensor modes) represent the relativistic effects
of gravitomagnetism and gravitational radiation, which have no counterpart
in Newtonian gravity although they are similar to electromagnetic phenomena.
If the motion of sources is expanded in powers of $v/c$, the vector and
tensor gravitational fields are $O(v/c)$ and $O(v/c)^2$ times the
Newtonian field, respectively.  On terrestrial scales the vector and
tensor modes are extremely weak --- they have not been detected in the
laboratory, although satellite experiments are planned to search for
the former through the Lense-Thirring ``gravitomagnetic moment''
precession, and large interferometric detectors are being built
to measure gravitational radiation --- but they could have important
consequences for the evolution of large-scale matter and radiation
fluctuations, including the production of anisotropy in the microwave
background radiation.

The Newtonian limit corresponds to weak gravitational fields (black holes
are to be avoided) and slow motions ($v^2\ll c^2$, for both sources and
test particles).  For nearly all cosmological applications it is sufficient
to consider only weak fields --- small perturbations of the spacetime metric
around a homogeneous and isotropic background spacetime.  At the same time
it is usually safe to assume that the gravitational sources are
nonrelativistic, although the test particles (e.g., photons) need not be.
Because the weak-field, slow source motion limit does not necessarily imply
small density fluctuations, we can (and will) investigate nonlinear particle
and fluid dynamics even while treating the metric perturbations and source
velocities as being small.

In sections \ref{class}--\ref{gmodes} we shall develop the machinery for
cosmological perturbation theory using the methods developed by Lifshitz,
Peebles, Bardeen, Kodama \& Sasaki, and others.  We discuss the consequences
of gauge invariance --- the invariance of physical quantities to small
changes in the spacetime coordinates --- and summarize the standard results
in the synchronous gauge of Lifshitz (1946).\footnote{Apparently it is not
widely known that Lifshitz' paper is published in English and is available
in many libraries.  This classic paper was remarkably complete, including
a full treatment of the scalar, vector, and tensor decomposition in open
and closed universes and a concise solution to the gauge mode problem; it
presented solutions for perfect fluids in matter- and radiation-dominated
universes; and it contrasted isentropic (adiabatic) and entropy
fluctuations.} In section \ref{poiss} we introduce a new gauge that clarifies
how general relativity extends Newtonian gravity in the weak-field limit and
in section \ref{phys} we attempt to clarify the physical content of general
relativity theory in this limit.  In section \ref{hamil} we shall see how
simply and clearly the Hamiltonian formulation of particle dynamics follows
from general relativity.  Finally, in section \ref{ellis} we introduce an
alternative fully nonlinear formulation of general relativity due to Ehlers,
Ellis and others, and we demonstrate its connection with the Lagrangian
fluid dynamics that was discussed in my fourth lecture.

We shall not discuss the relativistic Boltzmann equation nor the
classification of isentropic and isocurvature initial conditions.
In the nonrelativistic limit, these topics have already been covered
in my preceding lectures.  Neither shall we discuss the physics of
microwave background anisotropy or the evolution of perturbations in
specific models.  Our aim here is to derive and comprehend the
gravitational field equations, not their solution.  Although this goal
is restricted, we shall see that the physical content is sufficiently
rich.  After working through these notes the reader may wish to consult
one of the many books or articles discussing the detailed evolution for
a variety of models (e.g., Lifshitz \& Khalatnikov 1963; Peebles \& Yu
1970; Weinberg 1972; Peebles 1980; Press \& Vishniac 1980; Wilson \& Silk
1981; Wilson 1983; Bond \& Szalay 1983; Zel'dovich \& Novikov 1983;
Kodama \& Sasaki 1984, 1986; Efstathiou \& Bond 1986; Bond \&
Efstathiou 1987; Ratra 1988; Holtzman 1989; Efstathiou 1990; Mukhanov,
Feldman \& Brandenberger 1992; Liddle \& Lyth 1993; Peebles 1993;
Ma \& Bertschinger 1994b).

Understanding these notes will not require much experience with general
relativity, although some background is helpful.  The reader can test the
waters by examining the following summary of essential general relativity
and differential geometry.  While some mathematical formalism is needed to
get started, the focus thereafter will remain as much as possible on physics.

\subsubsection{Summary of essential relativity}

We adopt the following conventions and notations, similar to those of
Misner, Thorne \& Wheeler (1973).  Units are chosen so that $c=1$.
The metric signature is $(-,+,+,+)$.  The unperturbed background spacetime
is Robertson-Walker with scale factor $a(\tau)$ expressed in terms of
conformal time.  A dot (or $\partial_\tau$) indicates a conformal time
derivative.  The comoving expansion rate is written $\eta(\tau)\equiv
\dot a/a=aH$.  The scale factor obeys the Friedmann equation,
\setcounter{equation}{0}
\begin{equation}
  \eta^2={8\pi\over3}\,Ga^2\bar\rho-K\ .
\label{friedman}
\end{equation}
The Robertson-Walker line element is written in the general form using
conformal time $\tau$ and comoving coordinates $x^i$:
\begin{equation}
  ds^2=g_{\mu\nu}dx^\mu dx^\nu=
  a^2(\tau)\left[-d\tau^2+\gamma_{ij}(x^k)dx^idx^j\right]\ .
\label{rwmetric}
\end{equation}
Latin indices ($i$, $j$, $k$, etc.) indicate spatial components while Greek
indices ($\mu$, $\nu$, $\lambda$, etc.) indicate all four spacetime
components; we assume a coordinate basis for tensors.  Summation is
implied by repeated upper and lower indices.  The inverse 4-metric
$g^{\mu\nu}$ (such that $g^{\mu\nu}g_{\nu\kappa}=\delta^\mu_{\ \,\kappa}$)
is used to raise spacetime indices while the inverse 3-metric $\gamma^{ij}$
($\gamma^{ij}\gamma_{jk}=\delta^i_{\ k}$) is used to raise indices of
3-vectors and tensors.  Three-tensors are defined in the spatial
hypersurfaces of constant $\tau$ with metric $\gamma_{ij}$ and they shall
be clearly distinguished from the spatial components of 4-tensors.  We
shall see as we go along how this ``3+1 splitting'' of spacetime works
when there are metric perturbations.

Many different spatial coordinate systems may be used to cover a
uniform-curvature 3-space.  For example, there exist quasi-Cartesian
coordinates $(x,y,z)$ in terms of which the 3-metric components are
\begin{equation}
  \gamma_{ij}=\delta_{ij}\left[1+{K\over4}\left(x^2+y^2+z^2\right)
  \right]^{-2}\ .
\label{gamma}
\end{equation}
We shall use 3-tensor notation to avoid restricting ourselves to any
particular spatial coordinate system.  Three-scalars, vectors, and
tensors are invariant under transformations of the spatial coordinate
system  in the background spacetime (e.g., rotations).  A 3-vector may
be written $\vec A=A^i\vec e_i$ where $\vec e_i$ is a basis 3-vector
obeying the dot product rule $\vec e_i\cdot\vec e_j=\gamma_{ij}$.  A
second-rank 3-tensor may be written (using dyadic notation and the
tensor product) $\htens=h^{ij} \vec e_i\otimes\vec e_j$.  We write the
spatial gradient 3-vector operator $\vec\nabla=\vec e^i\partial_i$
($\partial_i\equiv\partial/\partial x^i$) where $\vec e^i\cdot\vec
e_j=\delta^i_{\ j}$.  The experts will recognize $\vec e^i$ as a
basis one-form but we can treat it as a 3-vector $\vec e^i=\gamma^
{ij}\vec e_j$ because of the isomorphism between vectors and one-forms.
Because the basis 3-vectors in general have nonvanishing gradients, we
define the covariant derivative (3-gradient) operator $\nabla_i$ with
$\nabla_i\gamma_{jk}=0$.  If the space is flat ($K=0)$ and we use
Cartesian coordinates, then $\gamma_{ij}=\delta_{ij}$, $\nabla_i=
\partial_i$, and the 3-tensor index notation reduces to elementary
Cartesian notation.  If $K\ne0$, the 3-tensor equations will continue
to look like those in flat space (that is why we use a 3+1 splitting
of spacetime!) except that occasionally terms proportional to $K$ will
appear in our equations.

Our application is not restricted to a flat Robertson-Walker background
but allows for nonzero spatial curvature.  This complicates matters
for two reasons.  First, we cannot assume Cartesian coordinates.  As a
result, for example, the Laplacian of a scalar and the divergence and
curl of a 3-vector involve the determinant of the spatial metric,
$\gamma\equiv\det\{\gamma_{ij}\}$:
\begin{eqnarray}
  \nabla^2\phi\equiv&\gamma^{-1/2}\partial_i\left(\gamma^{1/2}\gamma^{ij}
    \partial_j\phi\right)\ ,\quad
  \vec\nabla\cdot\vec v\equiv\gamma^{-1/2}\partial_i\left(
    \gamma^{1/2}v^i\right)\ ,\nonumber\\
  &\vec\nabla\times\vec v\equiv\epsilon^{ijk}(\partial_iv_j)\vec e_k\ ,
\label{divcurl}
\end{eqnarray}
where $\epsilon^{ijk}=\gamma^{-1/2}\,[ijk]$ is the three-dimensional
Levi-Civita tensor, with $[ijk]=+1$ if $\{ijk\}$ is an even permutation
of $\{123\}$, $[ijk]=-1$ for an odd permutation, and 0 if any two indices
are equal.  The factor $\gamma^{-1/2}$ ensures that $\epsilon^{ijk}$
transforms like a tensor; as an exercise one can show that $\epsilon_{ijk}
=\gamma^{1/2}\,[ijk]$.

The second complication for $K\ne0$ is that gradients do not commute when
applied to 3-vectors and 3-tensors (though they do commute for 3-scalars).
The basic results are
\begin{eqnarray}
  \left[\nabla_j,\nabla_k\right]A^i&&=\ ^{(3)}\!R^i_{\ njk}A^n\ ,\nonumber\\
  \left[\nabla_k,\nabla_l\right]h^{ij}&&=\ ^{(3)}\!R^i_{\ nkl}h^{nj}+
    \ ^{(3)}\!R^j_{\ nkl}h^{in}\ ,
\label{commute}
\end{eqnarray}
where $\left[\nabla_j,\nabla_k\right]\equiv(\nabla_j\nabla_k-\nabla_k
\nabla_j)$.  The commutator involves the spatial Riemann tensor, which
for a uniform-curvature space with 3-metric $\gamma_{ij}$ is simply
\begin{equation}
  ^{(3)}\!R^i_{\ jkl}=K\left(\delta^i_{\ k}\gamma_{jl}-\delta^i_{\ l}
    \gamma_{jk}\right)\ .
\label{3-riemann}
\end{equation}

Finally, we shall need the evolution equations for the full spacetime metric
$g_{\mu\nu}$.  These are given by the Einstein equations,
\begin{equation}
  G^\mu_{\ \,\nu}=8\pi G\,T^\mu_{\ \,\nu}\ ,
\label{einstein}
\end{equation}
where $T^\mu_{\ \,\nu}$ is the stress-energy tensor and $G^\mu_{\ \,\nu}$ is
the Einstein tensor, related to the spacetime Ricci tensor $R_{\mu\nu}$ by
\begin{equation}
  G_{\mu\nu}\equiv R_{\mu\nu}-{R\over2}g_{\mu\nu}\ ,\quad
    R\equiv R^\mu_{\ \,\mu}\ ,\quad
   R_{\mu\nu}\equiv R^\kappa_{\ \mu\kappa\nu}\ .
\label{ricci}
\end{equation}
The spacetime Riemann tensor is defined according to the convention
\begin{equation}
  R^\mu_{\ \,\nu\kappa\lambda}\equiv
     \partial_\kappa\Gamma^\mu_{\ \nu\lambda}
    -\partial_\lambda\Gamma^\mu_{\ \nu\kappa}
    +\Gamma^\mu_{\ \alpha\kappa}\Gamma^\alpha_{\ \nu\lambda}
    -\Gamma^\mu_{\ \alpha\lambda}\Gamma^\alpha_{\ \nu\kappa}\ ,
\label{riemann}
\end{equation}
where the affine connection coefficients are
\begin{equation}
  \Gamma^\mu_{\ \nu\lambda}\equiv{1\over2}g^{\mu\kappa}\left(
    \partial_\nu g_{\kappa\lambda}+\partial_\lambda g_{\kappa\nu}-
    \partial_\kappa g_{\nu\lambda}\right)\ .
\label{affine}
\end{equation}
We see that the Einstein tensor involves second derivatives of the metric
tensor components, so that eq. (\ref{einstein}) provides second-order
partial differential equations for $g_{\mu\nu}$.

The reader who is not completely comfortable with the material summarized
above may wish to consult an introductory general relativity textbook,
e.g. Schutz (1985).

\subsection{Classification of metric perturbations}
\label{class}

Now we consider small perturbations of the spacetime metric away from the
Robertson-Walker form:
\begin{eqnarray}
  &&\!\!\!\!\!\!\!\!
    ds^2=a^2(\tau)\left\{-(1+2\psi)d\tau^2+2w_id\tau dx^i+\left[(1-2\phi)
    \gamma_{ij}+2h_{ij}\right]dx^idx^j\right\},\nonumber\\
  &&\quad\quad\quad\quad\quad\quad\quad\quad\quad\quad\quad\quad\quad
     \quad\quad\quad\quad\quad\quad\quad\gamma^{ij}h_{ij}=0\ .
\label{pmetric}
\end{eqnarray}
We have introduced two 3-scalar fields $\psi(\vec x,\tau)$ and $\phi(\vec
x,\tau)$, one 3-vector field $\vec w(\vec x,\tau)=w_i\vec e^i$, and
one symmetric, traceless second-rank 3-tensor field $\htens(\vec x,\tau)=
h_{ij}\vec e^i\otimes\vec e^j$.  No generality is lost by making
$h_{ij}$ traceless since any trace part can be put into $\phi$.  The
factors of 2 and signs have been chosen to simplify later expressions.

Equation (\ref{pmetric}) is completely general: $g_{\mu\nu}$ has 10
independent components and we have introduced 10 independent fields
($1+1+3+5$ for $\psi+\phi+\vec w+\htens$).  In fact, only 6 of these
fields can represent physical degrees of freedom because we are free
to transform our 4 coordinates $(\tau,x^i)$ without changing any
physical quantities.  Infinitesimal coordinate transformations, called
gauge transformations, result in changes of the fields $(\psi,\phi,\vec w,
\htens)$ because the spacetime scalar $ds^2=g_{\mu\nu}dx^\mu dx^\nu$ must
be invariant under general coordinate transformations.  We shall explore
the consequences of this invariance later.  Coordinate invariance
complicates general relativity compared with other gauge theories (e.g.,
electromagnetism) in which the spacetime coordinates are fixed while other
variables change under the appropriate gauge transformations.

Unless stated explicitly to the contrary, in the following we shall
treat the perturbation variables $(\psi,\phi,w_i,h_{ij})$ exclusively as
3-tensors (of rank 0, 1, or 2 according to the number of indices) with
components raised and lowered using $\gamma^{ij}$ and $\gamma_{ij}$.
In doing this we {\it choose} to use $\gamma_{ij}$ as the 3-metric in
the perturbed hypersurface of constant $\tau$ despite the fact that the
spatial part of the 4-metric (divided by $a^2$) is given by $(1-2\phi)
\gamma_{ij}+2h_{ij}$.  This treatment is satisfactory because we will
assume that the metric perturbations are small and we will neglect all
terms quadratic in them.  However, we will use $g^{\mu\nu}$ to raise
4-vector components: $G^\mu_{\ \,\nu}=g^{\mu\kappa}G_{\kappa\nu}$.
Do take care to distinguish Latin from Greek!

We have introduced 3-scalar, 3-vector, and 3-tensor perturbations.
(From now on we will drop the prefix 3- since it should be clear from
the context whether 3- or 4- is implied.)  Are these the famous scalar,
vector, and tensor metric perturbations?  Not quite!  Recall the
decomposition of a vector into longitudinal and transverse parts:
\begin{equation}
  \vec w=\vec w_\parallel+\vec w_\perp\ ,\quad
  \vec\nabla\times\vec w_\parallel=\vec\nabla\cdot\vec w_\perp=0\ .
\label{vecdec}
\end{equation}
Since $\vec w_\parallel=-\vec\nabla w$ for some scalar $w$, how can it be
called a vector perturbation?  By definition, only the {\it transverse}
component $\vec w_\perp$ represents a vector perturbation.

There is a similar decomposition theorem for tensor fields:
Any differentiable traceless symmetric 3-tensor field $h_{ij}(\vec x)$
may be decomposed into a sum of parts, called longitudinal, solenoidal,
and transverse:
\begin{equation}
  \htens(\vec x)=\htens_{\parallel}+\htens_{\perp}+\htens_{\rm T}\ .
\label{tendec1}
\end{equation}
The various parts are defined in terms of a scalar field $h(\vec x)$
and transverse (or solenoidal) vector field $\vec h(\vec x)$ such that
\begin{equation}
  h_{ij,\,\parallel}=D_{ij}h\ ,\quad
  h_{ij,\,\perp}=\nabla_{(i}h_{j)}\ ,\quad
  \nabla_i h^i_{\ j,\,\rm T}=0\ ,
\label{tendec2}
\end{equation}
where we have denoted symmetrization with parentheses and have employed
the traceless symmetric double gradient operator:
\begin{equation}
  \nabla_{(i}h_{j)}\equiv{1\over2}\left(\nabla_ih_j+\nabla_jh_i \right)\ ,\quad
  D_{ij}\equiv\nabla_i\nabla_j-{1\over3}\,\gamma_{ij}\nabla^2\ .
\label{tendec3}
\end{equation}
Note that the divergences of $\htens_\parallel$ and $\htens_\perp$ are
longitudinal and transverse vectors, respectively (it doesn't matter which
index is contracted on the divergence since $\htens$ is symmetric):
\begin{equation}
  \vec\nabla\cdot\htens_\parallel={2\over3}\,\vec\nabla\left(\nabla^2+
    3K\right)h\ ,\quad
  \vec\nabla\cdot\htens_\perp={1\over2}\left(\nabla^2+2K\right)\vec h\ ,
\label{divh}
\end{equation}
where $\nabla^2\vec h\equiv(\nabla^2 h^i)\vec e_i$.  (We do not call
$\htens_\perp$ the transverse part, as we would by extension from $\vec
w_\perp$, because ``transverse'' is conventionally used to refer to the
tensor part.) The longitudinal tensor $\htens_\parallel$ is also called
the scalar part of $\htens$, the solenoidal part $\htens_\perp$ is also
called the vector part, and the transverse-traceless part $\htens_{\rm T}$
is also called the tensor part.  This classification of the spatial metric
perturbations $h_{ij}$ was first performed by Lifshitz (1946).

The purpose of this decomposition is to separate $h_{ij}$ into parts that
can be obtained from scalars, vectors, and tensors.  Is the decomposition
unique?  Not quite.  It is clear, first of all, that $h$ and $h_i$ are
defined only up to a constant.  But there may be additional freedom
(Stewart 1990).

First, the vector $\vec h$ is defined only up to solutions of Killing's
equation $\nabla_i h_j+\nabla_j h_i=0$, called Killing vectors (Misner et
al. 1973).  The reader can easily verify that one such solution (using the
quasi-Cartesian coordinates of eq. \ref{gamma}) is $(h_x,h_y,h_z)=(y,-x,0)$.
In an open space ($K\le0$) this solution would be excluded because it is
unbounded --- our perturbations should not diverge! ---  but in a closed
space $(K>0$) the coordinates have a bounded range.  This Killing vector,
and its obvious cousins, correspond to global rotations of the spatial
coordinates and not to physical perturbations.

Next, there may also be non-uniqueness associated with the tensor (and
scalar) component:
\begin{equation}
  h_{ij,\,\rm T}\to h_{ij,\,\rm T}+\zeta_{ij}\ ,\quad
  \zeta_{ij}\equiv\left[\nabla_i\nabla_j-\gamma_{ij}\left(\nabla^2+2K\right)
    \right]\zeta\ ,
\label{ttmode}
\end{equation}
where $\zeta$ is some scalar field.  From eqs. (\ref{commute}) and
(\ref{3-riemann}) one can show $\nabla^2(\nabla_i\zeta)=\nabla_i(\nabla^2
+2K)\zeta$ so that $\nabla_i\zeta^i_{\ j}=0$ as required for the tensor
component.  However, we also require $h_{ij,\,\rm T}$ to be traceless,
implying $(\nabla^2+3K)\zeta=0$.  Thus, the tensor mode is defined only
up to eq. (\ref{ttmode}) with bounded solutions of $(\nabla^2+3K)
\zeta=0$.  In fact, this condition also implies $\zeta_{ij}=D_{ij}\zeta$,
so we may equally well attribute $\zeta_{ij}$ to the scalar mode $h_{ij,\,
\parallel}$.  Thus, we are free to add any multiple of $\zeta$ to $h$
(the scalar mode) provided we subtract $D_{ij}\zeta$ from the tensor mode.
In an open space ($K\le0$) there are no nontrivial bounded solutions to
$(\nabla^2+3K)\zeta=0$ but in a closed space ($K>0$) there are four linearly
independent solutions (Stewart 1990).  Once again, these solutions correspond
to redefinitions of the coordinates with no physical significance.  Kodama
\& Sasaki (1984, Appendix B) gave a proof of the tensor decomposition theorem,
but they missed the additional vector and scalar/tensor mode solutions
present in a closed space.  In practice, it is easy to exclude these modes,
and so we shall ignore them hereafter.

Thus, we conclude that the most general perturbations of the
Robertson-Walker metric may be decomposed at each point in space into four
scalar parts each having 1 degree of freedom ($\psi,\phi,\vec w_\parallel,
\htens_\parallel$), two vector parts each having 2 degrees of freedom
($\vec w_\perp,\htens_\perp$), and one tensor part having 2 degrees of
freedom ($\htens_{\rm T}$, which lost 3 degrees of freedom to the
transversality condition).  The total number of degrees of freedom is 10.

Why do we bother with this mathematical classification?  First and foremost,
the different metric components represent distinct physical phenomena.
(By way of comparison, in previous lectures we have already seen that
$\vec v_\parallel$ and $\vec v_\perp$ play very different roles in fluid
motion.)  Ordinary Newtonian gravity obviously is a scalar phenomenon
(the Newtonian potential is a 3-scalar), while gravitomagnetism and
gravitational radiation  --- both of which are absent from Newton's laws,
and will be discussed below --- are vector and tensor phenomena,
respectively.  Moreover, this spatial decomposition can also be applied
to the Einstein and stress-energy tensors, allowing us to see clearly
(at least in some coordinate systems) the physical sources for each type
of gravity.  Finally, the classification will help us to eliminate
unphysical gauge degrees of freedom.  There are at least four of them,
corresponding to two of the scalar fields and one transverse vector field.

We will not write the weak-field Einstein equations for the general metric
of eq. (\ref{pmetric}).  Instead, we will consider only two particular
gauge choices, each of which allows for all physical degrees of freedom
(and more, in the case of synchronous gauge).  First, however, we must
examine the stress-energy tensor.

\subsection{Stress-energy tensor}
\label{entens}

The Einstein field eqs. (\ref{einstein}) show that the stress-energy
tensor provides the source for the metric variables.  For a perfect fluid
the stress-energy tensor takes the well-known form
\begin{equation}
  T^{\mu\nu}=(\rho+p)u^\mu u^\nu+pg^{\mu\nu}\ ,
\label{perfluid}
\end{equation}
where $\rho$ and $p$ are the proper energy density and pressure in the
fluid rest frame and $u^\mu=dx^\mu/d\lambda$ (where $d\lambda^2\equiv-ds^2$)
is the fluid 4-velocity.  In any locally flat coordinate system, $T^{00}$
represents the energy density, $T^{0i}$ the energy flux density (which
equals the momentum density $T^{i0}$), and $T^{ij}$ represents the spatial
stress tensor.  In locally flat coordinates in the fluid frame, $T^{00}=
\rho$, $T^{0i}=0$, and $T^{ij}=p\delta^{ij}$ for a perfect fluid.

For an imperfect fluid such as a sum of several uncoupled components (e.g.,
photons, neutrinos, baryons, and cold dark matter), the stress-energy
tensor must include extra terms corresponding in a weakly collisional
gas to shear and bulk viscosity, thermal conduction, and other physical
processes.  We may write the general form as
\begin{equation}
  T^{\mu\nu}=(\rho+p)u^\mu u^\nu+pg^{\mu\nu}+\Sigma^{\mu\nu}\ .
\label{tmunu}
\end{equation}
Without loss of generality we can require $\Sigma^{\mu\nu}$ to be
traceless and flow-orthogonal: $\Sigma^\mu_{\ \,\mu}=0$, $\Sigma^\mu_
{\ \,\nu}u^\nu=0$.  In locally flat coordinates in the fluid rest frame
only the spatial components $\Sigma^{ij}$ are nonzero (but their trace
vanishes) and the spatial stress is $T^{ij}=p\delta^{ij}+\Sigma^{ij}$.
With these restrictions on $\Sigma^{\mu\nu}$ (in particular, the absence
of a $\Sigma^{0i}$ term in the fluid rest frame) we implicitly define
$u^\mu$ so that $\rho u^\mu$ is the {\it energy current} 4-vector (as
opposed, for example, to the particle mass times the {\it number} current
4-vector for the baryons or other conserved particles).  As a result of
these conditions, $\rho u^\mu$ includes any heat conduction, $p$ includes
any bulk viscosity (the isotropic stress generated when an imperfect
fluid is rapidly compressed or expanded), and $\Sigma^{\mu\nu}$ (called
the shear stress) includes shear viscosity.  Some workers add to eq.
(\ref{tmunu}) terms proportional to the 4-velocity, $q^\mu u^\nu+u^\mu
q^\nu$, where $q^\mu$ is the energy current in the particle frame (taking
$u^\mu$ to be proportional to the particle number current).  Either
choice is fully general, although our choice is the simplest.

We shall need to evaluate the stress-energy components in the comoving
coordinate frame implied by eq. (\ref{pmetric}).  This requires
specifying the form of the 4-velocity $u^\mu$.  Therefore we must
digress to discuss the 4-velocity components in a perturbed spacetime.

Consider first the case where the fluid is at rest in the comoving frame,
i.e., $u^i=0$.  (This condition {\it defines} the comoving frame.)
Normalization ($g_{\mu\nu}u^\mu u^\nu=-1$) then requires $u^0=a^{-1}
(1-\psi)$ to first order in $\psi$.  Lowering the components using the
full 4-metric gives $u_0=-a(1+\psi)$ and $u_i=aw_i$ in the weak-field
approximation.

The appearance of $\psi$ and $w_i$ in the components $u_\mu$ for a fluid
at rest in the comoving frame may appear odd.  They arise because, in our
coordinates, clocks run at different rates in different places if $\nabla_i
\psi\ne0$ (the coordinate time interval $d\tau$ corresponds to a proper
time interval $a(\tau)(1+\psi)d\tau$) and they also have a position-dependent
offset if $w_i\ne0$ (an observer at $x^i=\hbox{constant}$ sees the clocks
at $x^i+dx^i$ running fast by an amount $w_idx^i$).  At first these may
seem like strange coordinate artifacts one should avoid (this may be a
motivation for the synchronous gauge in which $\psi=w_i=0$!) but they
have straightforward physical interpretations: $\psi$ represents the
gravitational redshift and $w_i$ represents the dragging of inertial
frames.  We shall see later that they also can be interpreted as giving
rise to ``forces,'' allowing us to apply Newtonian intuition in general
relativity.  Do not forget that in general relativity we are forced to
accept coordinates whose relation to proper times and distances is
complicated by spacetime curvature.  Therefore, it is advantageous when
we can reinterpret these effects in Newtonian terms.

We define the coordinate 3-velocity
\begin{equation}
  \vec v\equiv{d\vec x\over d\tau}={dx^i\over dx^0}\,\vec e_i=
    {u^i\over u^0}\,\vec e_i\ ,
\label{3vel}
\end{equation}
whose components are to be raised and lowered using $\gamma^{ij}$ and
$\gamma_{ij}$: $v_i=\gamma_{ij}v^j=\gamma_{ij}u^j/u^0$, $v^2\equiv
\gamma_{ij}v^iv^j$, $\vec w\cdot\vec v\equiv w_iv^i$, $\vec v\cdot\htens
\cdot\vec v\equiv h_{ij}v^iv^j$, etc.  The 4-vector component $u^0$
follows from applying the normalization condition $u_\mu u^\mu=-1$:
\begin{equation}
  u^0={1\over a\sqrt{1-v^2}}\,\left[1-{\psi-\vec w\cdot\vec v+\phi v^2
    -\vec v\cdot\htens\cdot\vec v\over 1-v^2}\right]\ .
\label{u0}
\end{equation}
In the absence of metric perturbations this looks like the standard result
in special relativity aside from the factor $a^{-1}$ that appears because
we use comoving coordinates.  With metric perturbations we can no longer
interpret $\vec v$ exactly as the {\it proper} 3-velocity because $a dx^i$
is not proper distance and $ad\tau$ is not proper time.  However, the
corrections are only first order in the metric perturbations.

We will assume that the mean fluid velocity is nonrelativistic so that we
can neglect all terms that are quadratic in $\vec v$.  (This does not
exclude the radiation era, since we allow individual particles to be
relativistic and require only the bulk velocity to be nonrelativistic.)
We will also neglect terms involving products of $\vec v$ and the metric
perturbations.  With these approximations, the 4-velocity components become
\begin{equation}
  u^0=a^{-1}(1-\psi)\ ,\ \ u^i=a^{-1}v^i\ ,\ \
  u_0=-a(1+\psi)\ ,\ \ u_i=a(v_i+w_i)\ .
\label{4vel}
\end{equation}
The apparent lack of symmetry in the spatial components arises because
$u_i=g_{i0}u^0+g_{ij}u^j$ and $g_{i0}=a^2w_i\ne0$ in general.

{}From eq. (\ref{4vel}) we can see how $w_i$ is interpreted as a
frame-dragging effect.  For $w_i\ne0$ the worldline of a comoving observer
(defined by the condition $v_i=0$) is not normal to the hypersurfaces
$\tau=\hbox{constant}$: $u_\mu\xi^\mu=aw_i\xi^i\ne0$ for a 3-vector
$\xi^i$.  In a locally inertial frame, on the other hand, the worldline
of a freely-falling observer obviously would be normal to the spatial
directions. (This is true in special relativity and also in general
relativity as a consequence of the equivalence principle.)  By making a
local Galilean transformation, $dx^i\to dx^i+w^id\tau$, we can remove
$w_i$ from the metric at a point.  This transformation corresponds to
choosing a locally inertial frame, called the {\it normal} frame, moving
with 3-velocity $-\vec w$ relative to the comoving frame.  In the normal
frame the fluid 3-velocity is $\vec v+\vec w$.

If $w_i=w_i(\tau)$ is independent of $\vec x$, one can remove $w_i$
everywhere from the metric by a global Galilean transformation.  (Try
it and see!) However, we may be interested in situations where $w_i=w_i
(\vec x,\tau)$ so that different transformations are required in different
places.  In this case there is no {\it global} inertial frame.  Spatially
varying $w_i$ corresponds to shearing and/or rotation of the comoving
frame relative to the normal frame.  This is called the ``dragging of
inertial frames.'' Although we can choose coordinates in which $w_i=0$
everywhere, we shall see that there are advantages in not hiding the
dragging of inertial frames.  In general, the comoving frame is noninertial:
an observer can remain at fixed $x^i$ only if accelerated by
nongravitational forces.  The synchronous gauge is an exception in that
$w_i=0$ everywhere and the comoving frame is locally inertial.  We shall
see later that these features of synchronous gauge obscure rather than
eliminate the physical dragging of inertial frames.

Now that we have all the ingredients we can finally write the stress-energy
tensor components in our perturbed comoving coordinate system in terms of
physical quantities:
\begin{eqnarray}
  T^0_{\ \,0}=-\rho\ ,\ \ T^i_{\ \,0}&&=-(\rho+p)v^i\ ,\nonumber\\
  T^0_{\ \,i}=(\rho+p)(v_i+w_i)&&\ ,\ \
  T^i_{\ \,j}=p\delta^i_{\ j}+\Sigma^i_{\ j}\ .
\label{setens}
\end{eqnarray}
We use mixed components in order to avoid extraneous factors of $a(1+\psi)$
and $a(1-\phi)$.  Note that the traceless shear stress $\Sigma^i_{\ j}$
may be decomposed as in eqs. (\ref{tendec1}) and (\ref{tendec2}) into
scalar, vector, and tensor parts.  Similarly, the energy flux density
$(\rho+p)v^i$ may be decomposed into scalar and vector parts.  (The pressure
appears here, just as in special relativity, to account for the $pdV$ work
done in compressing a fluid.  For a nonrelativistic fluid $p\ll\rho$, but
we shall not make this restriction.) We may already anticipate that these
sources are responsible in the Einstein equations for scalar, vector, and
tensor metric perturbations.

In writing the components of the stress-energy tensor we have not assumed
$\vert\delta\rho\vert\ll\bar\rho$.  The only approximations we make in the
stress-energy tensor are to neglect (relative to unity) $v^2$ and all terms
involving products of the metric perturbations with $\vec v$ and $\Sigma^i_
{\ j}$.  Of course, owing to the weak-field approximation, we are also
neglecting any terms that are quadratic in the metric perturbations themselves.

Before moving on to discuss the Einstein equations we should rewrite the
conservation of energy-momentum, $\nabla_\mu T^\mu_{\ \,\nu}=0$, in
terms of our metric perturbation and fluid variables.  (We use $\nabla_\mu$
to denote the full spacetime covariant derivative relative to the 4-metric
$g_{\mu\nu}$.  It should not be confused with the spatial gradient $\nabla_i$
defined relative to the 3-metric $\gamma_{ij}$.) Using the approximations
mentioned in the preceding paragraph, one finds
\begin{equation}
  \partial_\tau\rho+3(\eta-\dot\phi)(\rho+p)+\vec\nabla\cdot\left[(\rho+p)
    \vec v\,\right]=0
\label{econs}
\end{equation}
and
\begin{eqnarray}
  \partial_\tau\left[(\rho+p)(\vec v+\vec w)\right]
    &&+4\eta(\rho+p)(\vec v+\vec w)\nonumber\\
    &&+\vec\nabla p+\vec\nabla\cdot\stens+(\rho+p)\vec\nabla\psi=0\ .
\label{pcons}
\end{eqnarray}
(Deriving these gives useful practice in tensor algebra.)
It is easy to interpret the various terms in these equations.  The terms
proportional to the expansion rate $\eta$ arise because we are using
comoving coordinates and conformal time and have not factored out $a^{-3}$
from $\rho$ or $p$.  The pressure $p$ is present with $\rho$ because we
let $\rho$ be the energy density (not the rest-mass density), which is
affected by the work pressure does in compressing the fluid.  Excluding
these terms, the energy-conservation eq. (\ref{econs}) looks exactly
like the Newtonian continuity equation aside from the change in the
expansion rate from $\eta$ to $\eta-\dot\phi$.  This modification is
easily understood by noting from eq. (\ref{pmetric}) that the effective
isotropic expansion factor is modified by spatial curvature perturbations
to become $a(1-\phi)$.  The momentum-conservation eq. (\ref{pcons}) similarly
looks like the Newtonian version with a gravitational potential $\psi$,
aside from the special-relativistic effects of pressure and the addition
of $\vec w$ to all the velocities to place them in the normal (inertial) frame.

\subsection{Synchronous gauge}
\label{synch}

Synchronous gauge, introduced by Lifshitz (1946) in his pioneering
calculations of cosmological perturbation theory, is defined by the
conditions $\psi=w_i=0$, which eliminate two scalar fields ($\psi$ and
the longitudinal part of $\vec w$) and one transverse vector field
($\vec w_\perp)$.  It is not difficult to show that synchronous
coordinates can be found for any weakly-perturbed spacetime.  However,
the synchronous gauge conditions do not eliminate all gauge freedom.
This has in the past led to considerable confusion (for discussion see
Press \& Vishniac 1980 and Bardeen 1980).

Synchronous gauge has the property that there exists a set of comoving
observers who fall freely without changing their spatial coordinates.
(This is nontrivial when one notes that in order to remain at a fixed
terrestrial latitude, longitude, and altitude above the surface of the
earth it is necessary to accelerate everywhere except in geosynchronous
orbits.)  These observers are called ``fundamental'' comoving observers.
The existence of fundamental observers follows from the geodesic equation
\begin{equation}
  {du^\mu\over d\lambda}+\Gamma^\mu_{\ \alpha\beta}u^\alpha u^\beta=0
\label{geodesic}
\end{equation}
for the trajectory $x^\mu(\lambda)$, where $d\lambda=(-ds^2)^{1/2}$ for
a timelike geodesic and $u^\mu=dx^\mu/d\lambda$.  With $\psi=w_i=0$,
eq. (\ref{affine}) gives $\Gamma^i_{\ 00}=0$, implying that $u^i=0$
is a geodesic.

Each fundamental observer carries a clock reading conformal
time $\tau=\int dt/a(t)$ and a fixed spatial coordinate label $x^i$.
The clocks and labels of the fundamental observers are taken to {\it
define} the coordinate values at all spacetime points (assuming that
these hypothetical observers densely fill space).  The residual gauge
freedom in synchronous gauge arises from the freedom to adjust the initial
settings of the clocks and the initial coordinate labels of the
fundamental observers.

Because the spatial coordinates $x^i$ of each fundamental observer are
held fixed with time, the $x^i$ in synchronous gauge are Lagrangian
coordinates.  This implies that the coordinate lines become highly deformed
when the density perturbations become large.  When the trajectories of two
fundamental observers intersect the coordinates become singular: two
different sets of $x^\mu$ label the same spacetime event.  This flaw of
synchronous gauge is not apparent if $\vert\delta\rho/\bar\rho\vert\ll1$
and the initial coordinate labels are nearly unperturbed, so this gauge
may be used successfully (with some care required to avoid contamination
of physical variables by the residual gauge freedom) in linear perturbation
theory.

To be consistent with the conventional notation used for synchronous gauge
(Lifshitz 1946; Lifshitz \& Khalatnikov 1963; Weinberg 1972; Peebles 1993),
in this section only we shall absorb $\phi$ into $h_{ij}$ and double
$h_{ij}$:
\begin{equation}
  ds^2=a^2(\tau)\left[-d\tau^2+(\gamma_{ij}+h_{ij})dx^idx^j\right]\ ,\quad
    h\equiv h^i_{\ i}\ne0\ .
\label{smetric}
\end{equation}
Using this line element and the definitions of the Ricci and Einstein
tensors, it is straightforward (if rather tedious) to derive the components
of the perturbed Einstein tensor:
\begin{equation}
  -a^2G^0_{\ \,0}=3(\eta^2+K)+\eta\dot h-{1\over2}\left(\nabla^2+2K\right)h+
    {1\over2}\nabla_i\nabla_jh^{ij}\ ,
\label{sG00}
\end{equation}
\begin{equation}
  a^2G^0_{\ \,i}={1\over2}\left(\nabla_i\dot h-\nabla_j\dot h^j_{\ i}
    \right)\ , \quad G^i_{\ \,0}=-\gamma^{ij}G^0_{\ \,j}\ ,
\label{sG0i}
\end{equation}
\begin{eqnarray}
  -a^2G^i_{\ \,j}=&&\left(2\dot\eta+\eta^2+K\right)\delta^i_{\ j}+
    \left({1\over2}\partial_\tau^2+\eta\partial_\tau-{1\over2}\nabla^2
    \right)\left(h\delta^i_{\ j}-h^i_{\ j}\right)\nonumber\\
  &&-Kh^i_{\ j}+{1\over2}\gamma^{ik}\left(\nabla_k\nabla_jh-\nabla_k
    \nabla_lh^l_{\ j}-\nabla_j\nabla_lh^l_{\ k}\right)\nonumber\\
  &&+{1\over2}\left(\nabla_k\nabla_l h^{kl}\right)\delta^i_{\ j}\ .
\label{sGij}
\end{eqnarray}
One can easily verify that the unperturbed parts of the Einstein
equations $G^0_{\ \,0}=8\pi GT^0_{\ \,0}=-8\pi G\bar\rho$ and
$G^i_{\ \,j}=8\pi GT^i_{\ \,j}=8\pi G\bar p\delta^i_{\ j}$ give the
Friedmann and energy-conservation equations for the background
Robertson-Walker spacetime.

Our next goal is to separate the perturbed Einstein equations into
scalar, vector, and tensor parts.  First we must decompose the metric
perturbation field $h_{ij}$ as in eqs. (\ref{tendec1})--(\ref{tendec3}),
with a term added (and the notation changed slightly) to account for the
trace of $h_{ij}$:
\begin{equation}
  h_{ij}={1\over3}\,h\gamma_{ij}+D_{ij}\left(\nabla^{-2}\xi\right)+
    \nabla_{(i}h_{j)}+h_{ij,\,\rm T}\ ,
\label{hij}
\end{equation}
where $D_{ij}$ was defined in eq. (\ref{tendec3}).  We require
$\nabla_i h^i=\nabla_i h^i_{\ j,\,\rm T}=0$ to ensure that the last
two parts of $h_{ij}$ are purely solenoidal (vector mode) and
transverse-traceless (tensor mode) contributions.  The scalar mode variables
are $h$ and $\nabla^{-2}\xi$, whose Laplacian is $\xi$.  We shall not
worry about how to invert the Laplacian on a curved space but simply assume
that it can be done if necessary.

The perturbed Einstein equations now separate into 7 different parts
according to the spatial symmetry:

\begin{eqnarray}
\label{sG00s}
  G^0_{\ \,0}:\ \ &&
    {1\over3}\left(\nabla^2+3K\right)(\xi-h)+\eta\dot h=
      8\pi Ga^2(\rho-\bar\rho)\ ,\\
\label{sG0is}
  G^0_{\ \,i,\,\parallel}:\ \ &&
    {1\over3}\,\nabla_i(\dot h-\dot\xi)-K\nabla_i\left(\nabla^{-2}\dot
      \xi\right)=8\pi Ga^2\left[(\rho+p)v_i\right]_\parallel\ ,\\
\label{sG0iv}
  G^0_{\ \,i,\,\perp}:\ \ &&
    -{1\over4}\left(\nabla^2+2K\right)\dot h_i=8\pi Ga^2\left[(\rho+p)v_i
      \right]_\perp\ ,\\
\label{sGii}
  G^i_{\ \,i}:\ \ &&
    -(\partial_\tau^2+2\eta\partial_\tau)h+{1\over3}\left(\nabla^2+3K
      \right)(h-\xi)\nonumber\\
      &&\quad\quad\quad\quad=24\pi Ga^2(p-\bar p)\ ,\\
\label{sGijs}
  G^i_{\ \,j\ne i,\,\parallel}:\ \ &&
    \left({1\over2}\partial_\tau^2+\eta\partial_\tau\right)D_{ij}
      \left(\nabla^{-2}\xi\right)+{1\over6}D_{ij}(\xi-h)\nonumber\\
      &&\quad\quad\quad\quad=8\pi Ga^2\Sigma_{ij,\,\parallel}\ ,\\
\label{sGijv}
  G^i_{\ \,j,\,\perp}:\ \ &&
    \left({1\over2}\partial_\tau^2+\eta\partial_\tau\right)
      \nabla_{(i}h_{j)}=8\pi Ga^2\Sigma_{ij,\,\perp}\ ,\\
\label{sGijt}
  G^i_{\ \,j,\,\rm T}:\ \ &&
    \left({1\over2}\partial_\tau^2+\eta\partial_\tau-{1\over2}\nabla^2+
      K\right)h_{ij,\,\rm T}=8\pi Ga^2\Sigma_{ij,\,\rm T}\ .
\end{eqnarray}
The derivation of these equations is straightforward but tedious.  They
have decomposed naturally into separate equations for the scalar, vector,
and tensor parts of the metric perturbation, with the sources for each
given by the appropriate part of the energy-momentum tensor.  However,
there are more equations than unknowns!  There are four scalar equations for
$\xi$ and $h$, two vector equations for $h_i$, and one tensor equation
for $h_{ij,\,\rm T}$.  How can this be?

Before answering this question, let us note another interesting feature
of the equations above, which will provide a clue.  The equations arising
from $G^0_{\ \,\mu}$ involve only a single time derivative of the scalar
and vector mode variables, while those arising from $G^i_{\ \,\mu}$ have
two time derivatives, as we might have expected for equations of motion
for the gravitational fields.  This means that we could discard eqs.
(\ref{sG00s})--(\ref{sG0iv}) and be left with exactly as many second-order
in time equations as unknown fields.  Alternatively, we could discard
eqs. (\ref{sGii})--(\ref{sGijv}) and be left with exactly enough
first-order in time equations for the scalar and vector modes.  Only the
tensor mode evolution is uniquely specified by a second-order wave equation.

The reason for this redundancy is that the twice-contracted Bianchi
identities of differential geometry, $\nabla_\mu G^\mu_{\ \,\nu}=0$,
force the Einstein eqs. (\ref{einstein}) to imply $\nabla_\mu
T^\mu_{\ \,\nu}=0$.  The Einstein equations themselves contain redundancy,
as we can check explicitly here.  By combining the time derivative of eq.
(\ref{sG00s}) and the divergence of eqs. (\ref{sG0is}) and (\ref{sG0iv})
one obtains the perturbed part of eq. (\ref{econs}) (note, however,
that $\phi\to-h/6$).  Similarly, eq. (\ref{pcons}) follows from the
time derivative of eqs. (\ref{sG0is}) and (\ref{sG0iv}) combined with
the gradient of eqs. (\ref{sGii})--(\ref{sGijv}).  {\it Because we
require the equations of motion for the matter and radiation to locally
conserve the net energy-momentum, three of the perturbed Einstein eqs.\,}
(\ref{sG00s})--(\ref{sGijt}) {\it are redundant.}

In the literature, $G^0_{\ \,0}=8\pi GT^0_{\ \,0}$ is often called the
``ADM energy constraint'' and $G^0_{\ \,i}=8\pi GT^0_{\ \,i}$ is called
the ``ADM momentum constraint'' equation.  The 3+1 space-time decomposition
of the Einstein equations into constraint and evolution equations was
developed in detail by Arnowitt, Deser \& Misner (1962, ADM) and applied
to cosmology by Durrer \& Straumann (1988) and Bardeen (1989).  The ADM
constraint equations may be regarded as providing initial-value constraints
on ($h,\xi,\dot h,\dot\xi,\dot h_i$) and the matter variables.
If these constraints are satisfied initially (this is required for a
consistent metric), and if eqs. (\ref{sGii})--(\ref{sGijv}) are
used to evolve ($h,\xi,\dot h,\dot\xi,\dot h_i$) while the matter
variables are evolved so as to locally conserve the net energy-momentum,
then the ADM constraints will be fulfilled at all later times.  (This
follows from the results stated in the preceding paragraph.) In effect,
the Einstein equations have built into themselves the {\it requirement}
of energy-momentum conservation for the matter.  If one were to integrate
eqs. (\ref{sGii})--(\ref{sGijv}) correctly but to violate
energy-momentum conservation, then eqs. (\ref{sG00s})--(\ref{sG0iv})
would be violated.

In practice, we may find it preferable to regard the ADM constraints
alone --- and not eqs. (\ref{sGii})--(\ref{sGijv}) --- as giving
the actual field equations for the scalar and vector metric perturbations.
They have fewer time derivatives and hence are easier to integrate.
Equations (\ref{sGii})--(\ref{sGijv}) are not necessary at all (although
they may be useful for numerical checks) because they can always be
obtained by differentiating eqs. (\ref{sG00s})--(\ref{sG0iv}) and
using energy-momentum conservation.

This situation becomes clearer if we compare it with Newtonian gravity.
The field equation $\nabla^2\phi=4\pi Ga^2\delta\rho$ is analogous to
eq. (\ref{sG00s}).  (We shall see this equivalence much more clearly
in the Poisson gauge below.)  Let us take the time derivative: $\nabla^2
\dot\phi=4\pi G\partial_\tau(a^2\delta\rho)$.  If we now replace
$\partial_\tau(\delta\rho)$ using the continuity equation, we obtain
a time evolution equation for $\nabla^2\phi$ analogous to the divergence
of eq.  (\ref{sG0is}).  The solutions to this evolution equation obey
the Poisson equation if and only if the initial $\phi$ obeys the Poisson
equation.  Why should one bother to integrate $\nabla^2\dot\phi$ in time
when the solution can always be obtained instantaneously from the Poisson
equation? Viewed in this way, we would say that the extra time derivatives
in the $G^i_{\ \,\mu}$ equations have nothing to do with gravity {\it per se}.
The {\it real} field equations for the scalar and vector modes come from
the ADM constraint equations.

If the scalar and vector metric perturbations evolve according to
first-order in time equations, their solutions are not manifestly
causal (e.g., retarded solutions of the wave equation).  We shall
discuss this point in detail in section \ref{phys}.  However, for now
we may note that the tensor mode obeys the wave eq. (\ref{sGijt}).
The solutions are the well-known gravity waves which, as we shall
see, play a key role in enforcing causality.  The source for these waves
is given by the transverse-traceless stress (generated, for example,
by two masses orbiting around each other).  The $\eta\partial_\tau$
term arises because we use comoving coordinates and the $K$ term arises
as a correction to the Laplacian in a curved space; otherwise the vacuum
solutions are clearly waves propagating at the speed of light.  Abbott
\& Harari (1986) show that eq. (\ref{sGijt}) is the Klein-Gordon
equation for a massless spin-two particle.

\subsection{Gauge modes}
\label{gmodes}

As we noted above, the synchronous gauge conditions do not completely
fix the spacetime coordinates because of the freedom to redefine the
perturbed constant-time hypersurfaces and to reassign the spatial
coordinates within these hypersurfaces.  This freedom is not obvious
in the linearized Einstein equations for the scalar and vector modes,
but it is present in the form of additional solutions that must be
fixed by appropriate choice of initial conditions and that represent
nothing more than relabeling of the coordinates in an unperturbed
Robertson-Walker spacetime.

To see this effect more clearly, we consider a general infinitesimal
coordinate transformation from $(\tau,x^i)$ to $(\hat\tau,\hat x^i)$,
known as a {\bf gauge transformation}:
\begin{eqnarray}
  \hat\tau=\tau+\alpha(\vec x,\tau)\ ,\quad
  &&\hat x^i=x^i+\gamma^{ij}\nabla_j\beta(\vec x,\tau)+\epsilon^i
    (\vec x,\tau)\ ,\nonumber\\
  &&\hbox{with}\ \ \vec\nabla\cdot\vec\epsilon=0\ .
\label{xcoord}
\end{eqnarray}
For convenience we have split the spatial transformation into longitudinal
and transverse parts.  Note that the transformed time and space coordinates
depend in general on all four of the old coordinates.

Coordinate freedom leads to ambiguity in the meaning of density
perturbations.  Consider, for example, the simple case of an unperturbed
Robertson-Walker universe in which the density depends only on $\tau$
(if one uses the ``correct'' $\tau$ coordinate).  In the transformed
system it depends also on $\hat x^i$: $\bar\rho(\hat\tau)=\bar\rho(\tau)+
(\partial_\tau\bar\rho)\alpha(\vec x,\tau)$.  In other words, even in an
unperturbed universe we can be fooled into thinking there are
spatially-varying density perturbations.

This example may seem contrived, but the ambiguity is not trivial to avoid:
When spacetime itself is perturbed, and time is not absolute, what is the
best choice of time? The same question arises for the spatial coordinates.

To clarify this situation we must examine gauge transformations further.
First note that when we transform the coordinates we must also transform
the metric perturbation variables so that the line element $ds^2$ (a
spacetime scalar) is invariant.  It is straightforward to do this using
eqs. (\ref{pmetric}) and (\ref{xcoord}).  The result is
\begin{eqnarray}
  &&\hat\psi=\psi-\dot\alpha-\eta\alpha\ ,\ \
  \hat\phi=\phi+{1\over3}\,\nabla^2\beta+\eta\alpha\ ,\nonumber\\
  &&\hat w_i=w_i+\nabla_i(\alpha-\dot\beta)-\dot\epsilon_i\ ,\ \
  \hat h_{ij}=h_{ij}-D_{ij}\beta-\nabla_{(i}\epsilon_{j)}\ ,
\label{gaugex}
\end{eqnarray}
where $D_{ij}$ is the traceless double gradient operator defined in eq.
(\ref{tendec3}).  The transformed fields (with carets) are to be evaluated
at the same coordinate values $(\tau,x^i)$ as the original fields.

Suppose now that our original coordinates satisfy the synchronous gauge
conditions $\psi=w_i=0$.  [To recover the notation of eq. (\ref{smetric})
used specially for synchronous gauge we now double $h_{ij}$ and put the
trace of $h_{ij}$ into $h=-6\phi$.]  From eqs. (\ref{gaugex}) and
(\ref{smetric}) it follows that there is a whole {\it family} of
synchronous gauges with metric variables related to the original ones by
\begin{equation}
  \hat h=h-2\nabla^2\beta-6\eta\dot\beta\ ,\ \
  \hat\xi=\xi-2\nabla^2\beta\ ,\ \
  \hat h_i=h_i-2\epsilon_i\ ,
\label{gauges1}
\end{equation}
where
\begin{equation}
  \beta=\beta_0(\vec x)\int{d\tau\over a(\tau)}\ ,\quad
  \epsilon_i=\epsilon_i(\vec x)\ .
\label{gauges2}
\end{equation}
Thus, the {\it synchronous gauge has residual freedom in the form of one
scalar ($\beta_0)$ and one transverse vector ($\epsilon_i$) function
of the spatial coordinates.}

The presence of these extraneous solutions (called gauge modes) has
created a great deal of confusion in the past, which might have been
avoided had more cosmologists read the paper of Lifshitz (1946).
In 1980, Bardeen wrote an influential paper showing how one may take
linear combinations of the metric and matter perturbation variables
that are free of gauge modes.  For example, Bardeen defined two scalar
perturbations $\Phi_A$ and $\Phi_H$ related to our synchronous gauge
variables $h$ and $\xi$ (Bardeen actually used the variables
$H_L\equiv h/6$ and $H_T\equiv-\xi/2$) as follows:
\begin{equation}
  \Phi_A=-{1\over2}\,\nabla^{-2}(\ddot\xi+\eta\dot\xi)\ ,\quad
  \Phi_H={1\over6}\,(h-\xi)-{1\over2}\,\eta\nabla^{-2}\dot\xi\ .
\label{bardeen}
\end{equation}
It is easy to check that these variables are invariant under the
synchronous gauge transformation given by eqs.
(\ref{gauges1})--(\ref{gauges2}).

Bardeen's work led to a flurry of papers concerning gauge-invariant
variables in cosmology.  A standard reference is the classic paper by
Kodama \& Sasaki (1984).  Elegant treatments based on general 3+1
splitting of spacetime were given later by Durrer \& Straumann (1988)
and Bardeen (1989).  The simpler form of the gauge-invariant variables
often makes it easier to find analytical solutions (e.g., Rebhan 1992).
However, it is not {\it necessary} to use gauge-invariant variables
during a calculation, and many cosmologists have continued successfully
to use synchronous gauge.  In the end, when the results are converted
to measurable quantities --- spacetime scalars --- the gauge modes
automatically get canceled.  In a numerical solution, however, one must
be careful that the gauge modes do not swamp the physical ones, otherwise
roundoff can produce significant numerical errors.

Gauge invariant variables actually appear somewhat strange if we consider
the analogous situation in electromagnetism.  The electric and magnetic
fields in flat spacetime may be obtained from potentials $\phi$ and
$\vec A$ (note we are implicitly using a 3+1 split of spacetime),
\begin{equation}
  \vec E=-\vec\nabla\phi-\partial_\tau\vec A\ ,\quad
  \vec B=\vec\nabla\times\vec A\ .
\label{ebfields}
\end{equation}
With this choice, the source-free Maxwell equations are automatically
satisfied; the other two (the Coulomb and Amp\`ere laws) become
\begin{equation}
  \nabla^2\phi+\partial_\tau(\vec\nabla\cdot\vec A)=-4\pi\rho\ ,\quad
  \left(\partial_\tau^2-\nabla^2\right)\vec A+\vec\nabla\dot\phi=
  4\pi\vec J\ ,
\label{maxwellp}
\end{equation}
where $\rho$ is the charge density and $\vec J$ is the current density.
These equations are invariant under the gauge transformation $\hat\phi=
\phi-\partial_\tau\alpha$, $\hat A_i=A_i+\nabla_i\alpha$.

If we didn't know about electric and magnetic fields, but were alarmed
by the gauge-dependence of the potentials, we could try to find linear
combinations of $\phi$ and $\vec A$ that are gauge-invariant.  However,
there are two well-known and more direct ways to eliminate gauge modes.
The first is ``gauge fixing'' --- i.e., placing constraints on the
potentials so as to eliminate gauge degrees of freedom.  One popular
choice, for example, is the Coulomb gauge $\vec\nabla\cdot\vec A=0$, so
that $\vec A=\vec A_\perp$ is transverse.  The transversality condition
means that the gauge transformation variable $\alpha$ cannot depend on
position (though it can depend on time); thus, most of the gauge freedom
is eliminated.  The second possibility is to work with the physical
fields themselves instead of the potentials: $\vec E$ and $\vec B$ are
automatically gauge-invariant.  This procedure requires that we analyze
the equation of motion for charges to determine which combinations of
$\phi$ and $\vec A$ are physically most significant.

In the next section we shall adopt the first procedure (gauge-fixing)
using the gravitational analogue of the Coulomb gauge.  Later we shall
introduce Ellis' covariant approach based on gravitational fields themselves.

\subsection{Poisson gauge}
\label{poiss}

Recall that our general perturbed Robertson-Walker metric (\ref{pmetric})
contains four extraneous degrees of freedom associated with coordinate
invariance.  In the synchronous gauge these degrees of freedom are
eliminated from $g_{00}$ (one scalar) and $g_{0i}$ (one scalar and one
transverse vector) by requiring $\psi=w_i=0$.  There are other ways to
eliminate the same number of fields.  As we shall see, a good choice is
to constrain $g_{0i}$ (eliminating one scalar) and $g_{ij}$ (eliminating
one scalar and one transverse vector) by imposing the following gauge
conditions on eq. (\ref{pmetric}):
\begin{equation}
  \vec\nabla\cdot\vec w=0\ ,\ \ \vec\nabla\cdot\htens=0\ .
\label{pgauge}
\end{equation}
I call this choice the {\bf Poisson gauge} by analogy with the Coulomb
gauge of electromagnetism ($\vec\nabla\cdot\vec A=0$).\footnote{The
same gauge has been proposed recently by Bombelli, Couch \& Torrence
(1994), who call it ``cosmological gauge.''  However, I prefer the
name Poisson gauge because cosmology --- i.e., nonzero $\dot a$ ---
is irrelevant for the definition and physical interpretation of this
gauge.  Although I have seen no earlier discussion of Poisson gauge
in the literature, its time slicing corresponds with the minimal shear
hypersurface condition of Bardeen (1980).}
More conditions are required here than in electromagnetism because gravity
is a {\it tensor} rather than a {\it vector} gauge theory.  Note that in
the Poisson gauge there are {\it two scalar} potentials ($\psi$ and
$\phi$), {\it one transverse vector} potential ($\vec w\,$), and {\it one
transverse-traceless} tensor potential $\htens$.

A restricted version of the Poisson gauge, with $w_i=h_{ij}=0$, is known
in the literature as the longitudinal or conformal Newtonian gauge
(Mukhanov, Feldman \& Brandenberger 1992).  These conditions can be
applied only if the stress-energy tensor contains no vector or tensor parts
and there are no free gravitational waves, so that only the scalar metric
perturbations are present.  While this condition may apply, in principle,
in the linear regime ($\vert\delta\rho/\bar\rho\vert\ll1$), nonlinear
density fluctuations generally induce vector and tensor modes even if
none were present initially.  Setting $\vec w=\htens=0$ is analogous to
zeroing the electromagnetic vector potential, implying $\vec B=0$.  In
general, this is not a valid {\it gauge condition} --- it is rather the
elimination of physical phenomena.  The longitudinal/conformal Newtonian
gauge really should be called a ``restricted gauge.''  The Poisson gauge,
by contrast, allows all physical degrees of freedom present in the metric.

To prove the last statement, and to find out how much residual gauge
freedom is allowed, we must find a coordinate transformation from an
{\it arbitrary} gauge to the Poisson gauge.  Using eq. (\ref{gaugex})
with hats indicating Poisson gauge variables, we see that a suitable
transformation exists with
\begin{equation}
  \alpha=w+\dot h\ ,\quad \beta=h\ ,\quad \epsilon_i=h_i\ ,
\label{gaugep}
\end{equation}
where $w$ comes from the longitudinal part of $\vec w$ ($\vec w_\parallel=
-\vec\nabla w$), while $h$ and $h_i$ come from the longitudinal and
solenoidal parts of $\htens$ in eq. (\ref{tendec2}).  Because these
conditions are {\it algebraic} in $\alpha$, $\beta$, and $\vec\epsilon$
(they are not differentiated, in contrast with the transformation to
synchronous gauge of eq. \ref{gauges1}), we have found an almost unique
transformation from an arbitrary gauge to the Poisson gauge.  One can
still add arbitrary functions of time alone (with no dependence on $x^i$)
to $\alpha$ and $\epsilon_i$.  (Adding a function of time alone to $\beta$
has no effect at all because the transformation, eq. \ref{xcoord},
involves only the gradient of $\beta$.)

Spatially homogeneous changes in $\alpha$ represent changes in the units
of time and length, while spatially homogeneous changes in $\epsilon$
represent shifts in the origin of the spatial coordinate system.  These
trivial residual gauge freedoms --- akin to electromagnetic gauge
transformations generated by a function of time, the only gauge freedom
remaining in Coulomb gauge --- are physically transparent and should cause
no conceptual or practical difficulty.

It is interesting to see the coordinate transformation from a synchronous
gauge to the Poisson gauge.  As an exercise the reader can show that this
is given by
\begin{equation}
  \psi=-{1\over2}\,\nabla^{-2}(\ddot\xi+\eta\dot\xi)\ ,\ \
  \phi={1\over6}\,(\xi-h)+{1\over2}\,\eta\nabla^{-2}\dot\xi\ ,\ \
  w_i=-{1\over2}\,\partial_\tau h_i\ .
\label{synpois}
\end{equation}
Comparing with eq. (\ref{bardeen}), we see that the two Poisson-gauge
scalar potentials are $\psi=\Phi_A$ and $\phi=-\Phi_H$.  (Kodama \& Sasaki
1984 call these variables $\Psi=\psi$ and $\Phi=-\phi$.) The vector potential
$w_i$ in Poisson gauge is related simply to the solenoidal potential
$h_i$ of the synchronous gauge (eq. \ref{hij}).

Thus, the metric perturbations in the Poisson gauge correspond exactly
with several of the gauge-invariant variables introduced by Bardeen.
By imposing the explicit gauge conditions (\ref{pgauge}), we have
simplified the mathematical analysis of these variables.

Now that we have seen that the Poisson gauge solves the gauge-fixing problem,
let us give the components of the perturbed Einstein equations.  They
are no more complicated than those of the synchronous gauge:
\begin{eqnarray}
\label{pG00s}
  G^0_{\ \,0}:\ \ &&
    \left(\nabla^2+3K\right)\phi-3\eta\left(\dot\phi+\eta\psi\right)=
      4\pi Ga^2(\rho-\bar\rho)\ ,\\
\label{pG0is}
  G^0_{\ \,i,\,\parallel}:\ \ &&
    -\nabla_i(\dot\phi+\eta\psi)=
      4\pi Ga^2\left[(\rho+p)(v_i+w_i)\right]_\parallel\ ,\\
\label{pG0iv}
  G^0_{\ \,i,\,\perp}:\ \ &&
    \left(\nabla^2+2K\right)w_i=
      16\pi Ga^2\left[(\rho+p)(v_i+w_i)\right]_\perp\ ,\\
\label{pGii}
  G^i_{\ \,i}:\ \ &&
    \ddot\phi-K\phi+\eta(\dot\psi+2\dot\phi)+(2\dot\eta+\eta^2)\psi
      -{1\over3}\,\nabla^2(\phi-\psi)\nonumber\\
      &&\quad\quad\quad\quad=4\pi Ga^2(p-\bar p)\ ,\\
\label{pGijs}
  G^i_{\ \,j\ne i,\,\parallel}:\ \ &&
    D_{ij}\left(\phi-\psi\right)=8\pi Ga^2\Sigma_{ij,\,\parallel}\ ,\\
\label{pGijv}
  G^i_{\ \,j,\,\perp}:\ \ &&
    -\left(\partial_\tau+2\eta\right)\nabla_{(i}w_{j)}=
      8\pi Ga^2\Sigma_{ij,\,\perp}\ ,\\
\label{pGijt}
  G^i_{\ \,j,\,\rm T}:\ \ &&
    \left(\partial_\tau^2+2\eta\partial_\tau-\nabla^2+2K\right)h_{ij}=
      8\pi Ga^2\Sigma_{ij,\,\rm T}\ .
\end{eqnarray}

As in the synchronous gauge, the scalar and vector modes satisfy
initial-value (ADM) constraints (eqs. \ref{pG00s}--\ref{pG0iv}) in
addition to evolution equations.  However, it is remarkable that in the
Poisson gauge we can obtain the scalar and vector potentials {\it directly}
from the instantaneous stress-energy distribution with no time integration
required.  This is clear for $\phi-\psi$ and $\vec w$, both of which obey
elliptic equations with no time derivatives (eqs. \ref{pGijs} and
\ref{pG0iv}, respectively).  By combining the ADM energy and longitudinal
momentum constraint equations we can also get an instantaneous equation
for $\phi$:
\begin{eqnarray}
  \left(\nabla^2+3K\right)\phi=4\pi Ga^2\left[\delta\rho+3\eta\Phi_f
    \right]\ ,\quad
  -\vec\nabla\Phi_f\equiv\left[(\rho+p)(\vec v+
    \vec w)\right]_\parallel\ .
    \nonumber\\
\label{poisson}
\end{eqnarray}
Bardeen (1980) defined the matter perturbation variable $\epsilon_m\equiv
(\delta\rho+3\eta\Phi_f)/\bar\rho$ and noted that it is the natural
measure of the energy density fluctuation in the normal (inertial)
frame at rest with the matter such that $\vec v+\vec w=0$ (recall the
discussion in section \ref{entens}).  However, for our analysis
we will remain in the comoving frame of the Poisson gauge, in which case
$\delta\rho/\bar\rho$ and not $\epsilon_m$ is the density fluctuation.

We can show that for nonrelativistic matter the field equations we have
obtained reduce to the Newtonian forms.  First, it is clear that in the
non-cosmological limit ($\eta=K=0$), eq. (\ref{poisson}) reduces to
the Poisson equation.  For $\eta\ne0$ the longitudinal momentum density
$\Phi_f$ is also a source for $\phi$, but it is unimportant for perturbations
with $\vert\delta\rho/\bar\rho\vert\gg v_Hv/c^2$ where $v_H$ is the Hubble
velocity across the perturbation.  Next, consider the implications of
the fact that the shear stress for any physical system is {\it at most}
$O(\rho c_{\rm s}^2)$ where $c_{\rm s}$ is the characteristic thermal speed
of the gas particles.  (For a collisional gas the shear stress is much less
than this.)  Equation (\ref{pGijs}) then implies that the relative difference
between $\psi$ and $\phi$ is no more than $O(c_{\rm s}/c)^2$.  Third, eq.
(\ref{pG0iv}) implies that the vector potential $\vec w\sim (v_H/c)^2\vec v$.
Thus, the deviations from the Newtonian results are all $O(v/c)^2$.
{\it Poisson gauge gives the relativistic cosmological generalization of
Newtonian gravity}.

There are still more remarkable features of the Poisson gauge.  First,
the Poisson gauge metric perturbation variables are almost always small
in the nonrelativistic limit ($\vert\phi\vert\ll c^2$, $v^2\ll c^2$),
in contrast with the synchronous gauge variables $h_{ij}$, which become
large when $\vert\delta\rho/\bar\rho\vert>1$.  (However, Bardeen 1980
shows that the relative numerical merits of these two gauges can reverse
for isocurvature perturbations of size larger than the Hubble distance.)
Second, if $(\psi,\phi,\vec w,\htens)$ are very small, they --- but not
necessarily their derivatives! --- may be neglected to a good approximation,
in which case the Poisson gauge coordinates reduce precisely to the Eulerian
coordinates used in Newtonian cosmology.  Finally, it is amazing that the
scalar and vector potentials depend solely on the {\it instantaneous}
distribution of stress-energy  --- in fact, only the energy and momentum
densities and the shear stress are required.  {\it Only the tensor mode}
--- gravitational radiation --- {\it follows unambiguously from a time
evolution equation}.  In fact, it obeys precisely the same equation as in
the synchronous gauge (with a factor of 2 difference owing to our different
definitions) because tensor perturbations are gauge-invariant ---
coordinate transformations involving 3-scalars and a 3-vector cannot change
a 3-tensor (leaving aside the special case of eq. \ref{ttmode} for a closed
space).

\subsection{Physical content of the Einstein equations}
\label{phys}

In the last section we showed that the Poisson gauge variables ($\psi,\phi,
\vec w\,$) are given by the {\it instantaneous} distributions of energy
density, momentum density, and shear stress (longitudinal momentum flux
density).  Is this action at a distance in general relativity?

We showed in eq. (\ref{gaugep}) that the Poisson gauge can be
transformed to any other gauge.  In the cosmological Lorentz gauge
(see Misner et al. 1973 for the noncosmological version) all metric
perturbation components obey wave equations.  Therefore, the solutions
in Poisson gauge {\it must be causal} despite appearances to the contrary.

There is a precedent for this type of behavior: the Coulomb gauge of
electromagnetism.  With $\vec\nabla\cdot\vec A=0$, eqs. (\ref{maxwellp})
become
\begin{equation}
  \nabla^2\phi=-4\pi\rho\ ,\quad \vec\nabla\dot\phi=4\pi\vec J_\parallel\ ,
    \quad \left(\partial_\tau^2-\nabla^2\right)\vec A=4\pi\vec J_\perp\ .
\label{maxwellc}
\end{equation}
We have separated the current density into longitudinal and transverse
parts.  The similarity of the first two (scalar) equations to eqs.
(\ref{pG00s}) and (\ref{pG0is}) is striking.  The similarity would be even
more striking if we were to use comoving coordinates rather than treating
$\vec x$ and $\tau$ here as flat spacetime coordinates.  As an exercise one
can show that with comoving coordinates, $\rho$ and $\vec J$ will be
multiplied by $a^2$ and that $\dot\phi$ becomes $\dot\phi+\eta\phi$.  The
last step follows when one distinguishes time derivatives at fixed $\vec
x$ from those at fixed $a\vec x$.

Are we to conclude that electromagnetism {\it also} violates causality,
because the electric potential $\phi$ depends only on the instantaneous
distribution of charge?  No! To understand this let us examine the Coulomb
and Amp\`ere laws in flat spacetime for the {\it fields} rather than the
potentials:
\begin{equation}
  \vec\nabla\cdot\vec E=\vec\nabla\cdot\vec E_\parallel=4\pi\rho\ ,\
  -\partial_\tau\vec E_\parallel=4\pi\vec J_\parallel\ ,\
  \vec\nabla\times\vec B-\partial_\tau\vec E_\perp=4\pi\vec J_\perp\,.
\label{maxwell}
\end{equation}
The Amp\`ere law has been split into longitudinal and transverse parts.
We see that the {\it longitudinal} electric field indeed {\it is} given
instantaneously by the charge density.  Because the photon is a massless
vector particle, only the {\it transverse} part of the electric and magnetic
fields is radiative, and its source is given by the transverse current
density:
\begin{equation}
  \left(\partial_\tau^2-\nabla^2\right)\vec B=4\pi\vec\nabla\times
    \vec J_\perp\ ,\quad
  \left(\partial_\tau^2-\nabla^2\right)\vec E_\perp=-4\pi\partial_\tau
    \vec J_\perp\ .
\label{ebwave}
\end{equation}

But how does this restore causality?  To see how, let us consider the
following example.  Suppose that there is only one electric charge in
the universe and initially it is at rest in the lab frame.  If the
charge moves --- even much more slowly than the speed of light ---
$\vec E_\parallel$ --- the solution to the Coulomb equation --- is
changed everywhere instantaneously.  It must be therefore that
$\vec E_\perp$ {\it also} changes instantaneously in such a way as to
exactly {\it cancel} the acausal behavior of $\vec E_\parallel$.

This indeed happens, as follows.  First, note that the motion of the
charge generates a current density $\vec J=\vec J_\parallel+\vec J_\perp$.
The longitudinal and transverse parts separately extend over all space
(and are in this sense acausal) while their sum vanishes away from the
charge (as do $\vec\nabla\cdot\vec J_\parallel$ and $\vec\nabla\times
\vec J_\perp$).  The magnetic and transverse electric fields obey eqs.
(\ref{ebwave}).  Because $\vec J_\perp$ is distributed over all space
but $\vec\nabla\times\vec J_\perp$ is not, retarded-wave solutions for
$\vec B$ are localized and causal while those for $\vec E_\perp$ are not.
However, when $\vec E_\parallel$ is added to $\vec E_\perp$, one finds
that the net electric field is causal (Brill \& Goodman 1967).  It is a
useful exercise to show this in detail.

Now that we understand how causality is maintained, what is the use of the
longitudinal part of the Amp\`ere law, $-\partial_\tau\vec E_\parallel=4\pi
\vec J_\parallel$?  The answer is, to ensure charge conservation, which is
implied by combining the time derivative of the Coulomb law with the
divergence of the Amp\`ere law:
\begin{equation}
  \partial_\tau\rho+\vec\nabla\cdot\vec J=\partial_\tau\rho+
    \vec\nabla\cdot\vec J_\parallel=0\ .
\label{charge}
\end{equation}
{\it Charge conservation is built into the Coulomb and Amp\`ere laws.}
This remarkable behavior occurs because electromagnetism is a {\it gauge}
theory.  Gauge invariance effectively provides a redundant scalar field
equation whose physical role is to enforce charge conservation.  From
Noether's theorem (e.g., Goldstein 1980), a continuous symmetry (in this
case, electromagnetic gauge invariance) leads to a conserved current.

General relativity is also a gauge theory.  Coordinate invariance --- a
continuous symmetry --- leads to conservation of energy and momentum.
As a result there are redundant scalar and vector equations [eqs.
(\ref{pG0is}), (\ref{pGii}), and (\ref{pGijv})] {\it whose role is to enforce
the conservation laws} [eqs. (\ref{econs}) and (\ref{pcons})].  We are free
to use the action-at-a-distance field equations for the scalar and vector
potentials in Poisson gauge because, when they are converted to fields and
combined with the gravitational radiation field, the resulting behavior is
entirely causal.

The analogy with electromagnetism becomes clearer if we replace the
gravitational potentials by fields.  We define the ``gravitoelectric''
and ``gravitomagnetic'' fields (Thorne, Price \& Macdonald 1986;
Jantzen, Carini \& Bini 1992)
\begin{equation}
  \vec g=-\vec\nabla\psi-\partial_\tau\vec w\ ,\quad
  \vec H=\vec\nabla\times\vec w\ ,
\label{ghfields}
\end{equation}
{\it using the Poisson gauge variables} $\psi$ and $\vec w$.  In section
\ref{hamil} we shall see how these fields lead to ``forces'' on particles
similar to the Lorentz forces of electromagnetism.  For now, however, we
are interested in the fields themselves.

Note that $\vec g$ and $\vec H$ are invariant under the transformation
$\psi\to\psi-\dot\alpha$, $\vec w\to\vec w+\vec\nabla\alpha$.  In the
noncosmological limit ($\eta=0$) this is a gauge transformation
corresponding to transformation of the time coordinate (cf. eqs.
\ref{xcoord} and \ref{gaugex}).  However, gauge transformations in general
relativity are complicated by the fact that they change the coordinates
and fields as well as the potentials.  For example, the $\eta\alpha$ terms
in eq. (\ref{gaugex}) arise because the transformed metric is evaluated
at the old coordinates.  Thus, $\vec g$ should acquire a term $\eta\vec
\nabla\alpha$ under a true gauge (coordinate) transformation, which is
incompatible with eq. (\ref{ghfields}).  The actual transformation
($\psi\to\psi-\dot\alpha,\,\vec w\to\vec w+\vec\nabla\alpha$) is {\it not}
a coordinate transformation.  General relativity differs from
electromagnetism in that gauge transformations change not just the
potentials but also the coordinates used to evaluate the potentials;
remember that the potentials {\it define} the perturbed coordinates!
Only in a simple coordinate system, such as Poisson gauge --- the
gravitational analogue of Coulomb gauge --- is it possible to see a
simple relation between fields and potentials similar to that of
electromagnetism.

In the limit of comoving distance scales small compared with the curvature
distance $\vert K\vert^{-1/2}$ and the Hubble distance $\eta^{-1}$, and
nonrelativistic shear stresses, the gravitoelectric and gravitomagnetic
fields obey a gravitational analogue of the Maxwell equations:
\begin{eqnarray}
  &&\vec\nabla\cdot\vec g=-4\pi Ga^2\delta\rho\ ,\quad
    \vec\nabla\times\vec g+\partial_\tau\vec H=0\ ,\nonumber\\
  &&\vec\nabla\cdot\vec H=0\ ,\quad
    \vec\nabla\times\vec H=-16\pi Ga^2\vec f_\perp\ ,
\label{gmaxwell}
\end{eqnarray}
where $\vec f=(\rho+p)(\vec v+\vec w)$ is the momentum density in the
normal (inertial) frame.  (You may derive these equations using eqs.
\ref{pG00s}, \ref{pG0is}, \ref{pGijs}, and \ref{ghfields}.) These equations
differ from their electromagnetic counterparts in three essential ways:
(1) the sources have opposite sign (gravity is attractive), (2) the
transverse momentum density has a coefficient 4 times larger than the
transverse electric current (gravity is a tensor and not a vector theory),
and (3) there is no ``displacement current'' $-\partial_\tau\vec g$ in
the transverse Amp\`ere law for $\vec\nabla\times\vec H$.  Recalling that
Maxwell added the electric displacement current precisely to conserve charge
and thereby obtained radiative (electromagnetic wave) solutions, we understand
the difference here: the vector component of gravity is nonradiative.
Unlike the photon, the graviton is a spin-2 particle (or would be if we
could quantize general relativity!), so radiative solutions appear only
for the (transverse-traceless) tensor potential $h_{ij}$.  In fact, the
vector potential is nonradiative precisely because it is needed to ensure
momentum conservation; mass conservation is already taken care of by the
scalar potential.  Recall the role of the ADM constraint equations discussed
in section \ref{synch}. Gravity has more conservation laws to maintain than
electromagnetism and consequently needs more fields to constrain.

Obtaining this physical insight into general relativity is {\it much}
easier in the Poisson gauge than in the synchronous gauge.  This fact alone
is a good reason for preferring the former.  When combined with the other
advantages (simpler equations, no time evolution required for the scalar
and vector potentials, reduction to the Newtonian limit, no nontrivial
gauge modes, and lack of unphysical coordinate singularities), the
superiority of the Poisson gauge should be clear.

Although the physical picture we have developed for gravity in analogy
with electromagnetism is beautiful, it is inexact.  Not only have we
linearized the metric, we have also neglected cosmological effects in
eqs. (\ref{gmaxwell}).  We shall see in section \ref{ellis} how to
obtain exact nonlinear equations for (the gradients of) the gravitational
fields.

\subsection{Hamiltonian dynamics of particles}
\label{hamil}

In this section we extend to general relativity the Hamiltonian formulation
of particle dynamics that is familiar in Newtonian mechanics.  In the process
we shall obtain further insight into the physical meaning of the gravitational
fields discussed in the previous section.  A preliminary version of this
material appears in (Bertschinger 1993).  A related presentation in the
context of gravitational fields near black holes is given by Thorne et
al. (1986).

As in the nonrelativistic case, we choose a Hamiltonian that is related
to the energy of a particle.  Consequently, our approach is not manifestly
covariant; the energy depends on how spacetime is sliced into hypersurfaces
of constant conformal time $\tau$ because the energy is the time component
of a 4-vector.  Nevertheless, our approach is fully compatible with general
relativity; we must only select a specific gauge.  For simplicity we shall
adopt the Poisson gauge, eq. (\ref{pmetric}) with gauge conditions eq.
(\ref{pgauge}).  We assume that the metric perturbations are given by
a solution of the field eqs. (\ref{pG00s})--(\ref{pGijt}).  Our
Hamiltonian will include only the degrees of freedom associated with
one particle; one can generalize this to include many particles (even
treated as a continuum) and the metric variables (Arnowitt et al. 1962;
Misner et al. 1973; Salopek \& Stewart 1992) but this involves more
machinery than necessary for our purposes.

The goal of the Hamiltonian approach is to obtain equations of motion
for trajectories in the single-particle phase space consisting of the
spatial coordinates $x^i$ and their conjugate momenta.  The first question
is, what are the appropriate conjugate momenta?  This question practically
answers itself when we express the action scalar in terms of our
coordinates:
\begin{equation}
  S=\int P_\mu dx^\mu=\int\left(P_\tau+P_i{dx^i\over d\tau}\right)\,d\tau
    =\int(-H+P_i\dot x^i)\,d\tau\ .
\label{action}
\end{equation}
Note that we have automatically expressed the action in terms of the
covariant (lower-index) components of the 4-momentum (also known as the
components of the momentum one-form).  We can read off the Hamiltonian
and conjugate momenta using the fact that $S=\int L d\tau$ where $L(x^i,
\dot x^j,\tau)$ is the Lagrangian, which is related to the Hamiltonian
$H(x^i,P_j,\tau)$ by the Legendre transformation $L=P_i\dot x^i-H$.
The Hamiltonian therefore is $H=-P_\tau$  --- despite appearances, we
shall see that this is {\it not} in general the {\it proper} energy ---
and the conjugate momenta equal the covariant spatial components of the
4-momentum.  Indeed, we may simply {\it define} the conjugate momenta
and Hamiltonian in this way.  (Care should be taken not to confuse the
Hamiltonian $H$ with the Hubble parameter $H$ and the gravitomagnetic
field $\vec H$!)

With these definitions, $H$ and $P_i$ correspond to the usual quantities
encountered in elementary nonrelativistic mechanics, but we need not rely
on this fact.  For any choice of spacetime geometry and coordinates we may
determine the corresponding Hamiltonian and conjugate momenta from the
4-momentum components: for a particle of mass $m$, $H=-mg_{0\mu}dx^\mu/
d\lambda$, $P_i=mg_{i\mu}dx^\mu/d\lambda$ where $d\lambda$ measures proper
time along the particle trajectory.  As an exercise, one may show that
with cylindrical coordinates $(r,\theta,z)$ for a nonrelativistic particle
of mass $m$ in Minkowski spacetime, $P_r=m\dot r$ is the radial momentum,
$P_\theta=mr^2\dot\theta$ is the angular momentum about $\vec e_z$,
$P_z=m\dot z$ is the linear momentum along $\vec e_z$, and $H=E\approx
m+(P_r^2+P_\theta^2/r^2+P_z^2)/2m$ is the proper energy (including the
rest mass energy).  We shall determine the functional form $H(x^i,P_j,\tau)$
for our perturbed Robertson-Walker spacetime below.

First, however, let us show that our approach leads to the usual canonical
Hamilton's equations of motion, rigorously justifying our choices $H=-P_\tau$
and $P_i$ being the momentum conjugate to $x^i$.  To do this we simply
vary the phase space trajectory $\{x^i(\tau),\,P_j(\tau)\}$ to $\{x^i+
\delta x^i,\,P_j+\delta P_j\}$, treating $\delta x^i(\tau)$ and $\delta
P_j(\tau)$ as independent variations and computing the variation of the
action of eq. (\ref{action}):
\begin{eqnarray}
  \delta S&&=\int\left(-{\partial H\over\partial x^i}\delta x^i-
    {\partial H\over\partial P_i}\delta P_i+{dx^i\over d\tau}\delta P_i
    +P_i{d\over d\tau}\delta x^i\,\right)\,d\tau\nonumber\\
  &&=\int\left[\left({dx^i\over d\tau}-{\partial H\over\partial P_i}
    \right)\delta P_i(\tau)-\left({dP_i\over d\tau}+{\partial H\over
    \partial x^i}\right)\delta x^i(\tau)\,\right]\,d\tau\ ,
\label{varyS}
\end{eqnarray}
where we have assumed $P_i\delta x^i=0$ at the endpoints of integration.
Requiring the action to be stationary under all variations, $\delta S=0$,
we obtain the standard form of Hamilton's equations:
\begin{equation}
  {dx^i\over d\tau}={\partial H\over\partial P_i}\ ,\quad
  {dP_i\over d\tau}=-{\partial H\over\partial x^i}\ .
\label{hamileqs}
\end{equation}
Thus, Hamilton's equations give phase space trajectories in general
relativity just as they do in nonrelativistic mechanics.

Our next step is to determine the Hamiltonian for the problem at hand.
We shall assume that the particle falls freely in the perturbed
Robertson-Walker spacetime described in the Poisson gauge.  For
comparison with the nonrelativistic results, it is useful to relate
the 4-momentum components to the proper energy and 3-momentum measured
by a comoving observer (i.e., one at fixed $x^i$), $E$ and $p_i$:
\begin{equation}
  P_\tau=-a(1+\psi)E\ , \quad P_i=a\left[(1-\phi)(p_i+Ew_i)+h_{ij}p^j
    \right]\ .
\label{4-mom}
\end{equation}
The first equation follows from $E=-u^\mu P_\mu$ where $u^\mu$ is the
4-velocity of a comoving observer from eq. (\ref{u0}) with $\vec v=0$,
while the second equation follows from projecting $P_\mu$ into the
hypersurface normal to $u^\mu$ and normalizing to give the proper
3-momentum.  The weak-field approximation has been made (i.e., terms
quadratic in the metric perturbations are neglected), but the particle
motion is allowed to be relativistic.  The factors $a(1+\psi)$ and
$a(1-\phi)$ are obviously needed from eq. (\ref{pmetric}) to convert
proper quantities into coordinate momenta, the $Ew_i$ term arises because
our space and time coordinates are not orthogonal if there is a vector
mode, and the $h_{ij}p^j$ term arises because our spatial
coordinates are not orthogonal if there is a tensor mode.  The reader
may verify that the 4-momentum satisfies the normalization condition
$g^{\mu\nu}P_\mu P_\nu=-E^2+p^2=-m^2$, and that this condition would be
violated in general without the vector and tensor terms in $P_i$.

Using these results it is easy to show that, to first order in the metric
perturbations, the Hamiltonian is
\begin{equation}
  H(x^i,P_j,\tau)=\left[\left\vert(1+\phi)\vec P-\epsilon\,\vec w
  -\htens\cdot\vec P\right\vert^2+a^2m^2\,\right]^{1/2}+\epsilon\,\psi\ ,
\label{hamiltonian}
\end{equation}
where
\begin{equation}
  \epsilon=\epsilon(\vec P,\tau)\equiv\left(P^2+a^2m^2\,\right)^{1/2}
\label{epsp}
\end{equation}
and the squares and dot products of 3-vectors such as $p_i$, $P_i$, and
$h^{ij}P_j$ are computed using the 3-metric, e.g., $P^2=\gamma^{ij}P_iP_j$.
Using the Hamiltonian of eq. (\ref{hamiltonian}), eqs. (\ref{hamileqs})
may be shown to be fully equivalent to the geodesic equations for a
freely falling particle moving in the metric of eq. (\ref{pmetric}),
and they could also be obtained starting from a Lagrangian approach.
The advantage of the Hamiltonian approach is that it treats positions
and conjugate momenta equally as is needed for a phase space description.

Equation (\ref{hamiltonian}) appears strange at first glance.  To understand
it better, let us recall the standard form for the Hamiltonian of a particle
with charge $e$ in electromagnetic fields (with $\phi$ being the electrostatic
potential):
\begin{equation}
  H_e(x^i,P_j,t)=\left[\left\vert\vec P-e\vec A\right\vert^2+m^2\,
    \right]^{1/2}+e\phi\ .
\label{EMham}
\end{equation}
Note that the proper momentum is $\vec p=\vec P-e\vec A$ where $\vec P$
is the conjugate momentum.  Comparing eqs. (\ref{hamiltonian}) and
(\ref{EMham}), we see that they are very similar aside from the tensor
term $\htens\cdot\vec P$ present in the gravitational case.  The few
remaining differences are easily understood.  To compensate for spatial
curvature --- effectively a local change of the units of length --- in the
gravitational case $\vec P$ is multiplied by $(1+\phi)$.  The electric
charge $e$ is replaced by the gravitational charge $\epsilon$ (energy!);
to zeroth order in the perturbations $\epsilon=H=aE$.  The use of
comoving coordinates is responsible for the factors of $a(\tau)$.
The gravitational (gravitomagnetic) vector potential is $\vec w$ ---
as we anticipated in eq. (\ref{ghfields}).  Finally, the electrostatic
potential energy $e\phi$ is replaced in the gravitational case by
$\epsilon\,\psi$.  The strong analogy between the vector mode and
magnetism accounts for the adjective ``gravitomagnetic.''

A different interpretation of the gravitomagnetic contribution to the
Hamiltonian will clarify the relation of gravitomagnetism and the dragging
of inertial frames.  In section \ref{entens} we noted that $\vec w$ is
the velocity of the comoving frame relative to a locally inertial frame
(the normal frame).  For $w^2\ll1$, $\vec p'\equiv\vec p+E\vec w$ is
therefore the proper momentum in the normal frame.  According to eq.
(\ref{4-mom}), then, neglecting the scalar and tensor modes, $\vec P$
is the comoving momentum (i.e., multiplied by $a$) in the {\it normal}
frame, $\vec P=a\vec p'$, while $\vec P-\epsilon\,\vec w$ (the combination
present in the Hamiltonian) is the comoving momentum in the {\it comoving}
frame.  It is logical that the Hamiltonian should depend on the latter
quantity; after all, we are using non-orthogonal comoving spacetime
coordinates.  However, it is equally reasonable that the conjugate
momentum should be measured in the frame normal to the hypersurface
$\tau=\hbox{constant}$.  Thus, it is simply the offset between these
two frames --- if one likes, the dragging of inertial frames --- that
is responsible for the $-\epsilon\,\vec w$ term in eq. (\ref{hamiltonian}).
Gravitomagnetism --- and similarly magnetism, if one interprets
$(e/m)\vec A$ as a velocity --- can be viewed as a kinematical effect!


The tensor mode, corresponding to gravitational radiation, gives an
extra term in the Hamiltonian --- really in the relation between the
proper and conjugate momenta --- that is not present in the case of
electromagnetism.  Geometrically, $\htens$ corresponds simply to a
local volume-preserving deformation of the spatial coordinate lines,
and in this way it simply extends the effect of the spatial curvature
term $\phi\vec P$ in eq. (\ref{hamiltonian}) ($\phi$ represents an
orientation-preserving dilatation of the coordinate lines).  However,
what is more important is the dynamical effect of these terms, neither
of which is familiar in either Newtonian gravity or electromagnetism.

To study the dynamics of particle motion we use Hamilton's eqs.
(\ref{hamileqs}) with the Hamiltonian of eq. (\ref{hamiltonian}).
In terms of the proper momentum $\vec p$ measured by a comoving observer,
Hamilton's equations in the Poisson gauge become
\begin{eqnarray}
  &&{d\vec x\over d\tau}=\left(1+\psi+\phi-\htens\cdot\right)
    {\vec p\over E'}\ ,\quad
  E'\equiv\left[\left\vert\vec p+E\vec w\right\vert^2+m^2\,\right]^{1/2}\ ,
    \nonumber\\
  &&{d\over d\tau}\left[a\left(1-\phi+\htens\cdot\right)\vec p\,\right]
  =\epsilon\left[\vec g+\vec v\times\vec H-v^2\,\vec\nabla\phi+
    v^iv^j\vec\nabla h_{ij})\right]
    -\dot\epsilon\,\vec w\ ,
    \nonumber\\
\label{eoms}
\end{eqnarray}
where we have defined $E'$ to be the proper energy in the normal frame,
$\vec v$ is the peculiar velocity (in the weak-field limit it doesn't
matter whether it is the coordinate or proper peculiar velocity nor
whether it is measured in the comoving or normal frame) and $\vec g$
and $\vec H$ are the gravitoelectric and gravitomagnetic fields given by
eqs. (\ref{ghfields}).  The dot following $\htens$ indicates the
three-dimensional dot product, with $\htens\cdot\vec p$ being a 3-vector.

Equations (\ref{eoms}) appear rather complicated at first but each
term can be understood without much difficulty.  First, note that the
factor $(1+\psi+\phi-\htens\cdot)$ in the first equation is present solely
to convert from a proper velocity to a coordinate velocity $d\vec x/d\tau$
according to the metric eq. (\ref{pmetric}).  Using the transformation
from the normal (primed) to comoving frame, $\vec p=\vec p'-E\vec w
\approx\vec p'-E'\vec w$, the equation for $d\vec x/d\tau$ implies that
the proper velocity in the comoving frame must equal $\vec p/E'=\vec
p'/E'-\vec w$.  This is identically true because $\vec p'/E'$ is the
proper velocity in the normal frame, whose velocity relative to the
comoving frame is $-\vec w$.

Similarly, the factor $a(1-\phi+\htens\cdot)$ in the momentum equation
simply converts the proper momentum $\vec p$ to the comoving momentum in
the comoving frame, $\vec P-\epsilon\,\vec w$ (cf. eq. \ref{4-mom}).
The first two terms on the right-hand side have exactly the same form
as the Lorentz force law of electrodynamics, with the electric charge
$e$ replaced by the comoving energy $\epsilon$ and the electric and
magnetic fields $\vec E$ and $\vec B$ replaced by their gravitational
counterparts $\vec g$ and $\vec H$.  Thus, general relativity in the
weak-field limit gives ``forces'' on freely-falling bodies (when
expressed in the Poisson gauge) that are very similar to those of
electromagnetism!

The remaining terms in the momentum equation have no counterpart in
electrodynamics or Newtonian gravity.  There are two gravitational
force terms quadratic in the velocity arising from spatial curvature.
The first one is present for a scalar mode and is responsible for the
fact that photons are deflected twice as much as nonrelativistic
particles in a gravitostatic field ($\phi=\psi$ in the Newtonian limit).
The second term represents, in effect, scattering of moving particles
by gravitational radiation.  A gravity wave traveling in the $z$-direction
will accelerate a particle in this direction if the particle has nonzero
velocity in the $x$-$y$ plane (the direction of polarization of the
transverse gravity wave).  If the particle is at rest in our coordinate
system, it remains at rest when a gravity wave passes by.  However,
because the gravity wave corresponds to a deformation of the spatial
coordinate lines, the {\it proper} distance between two particles at
rest in the coordinate system does change (Misner et al. 1973).

Finally, the last term in the momentum equation, $-\dot\epsilon\,\vec w$,
represents a sort of cosmic drag that causes velocities of massive
particles to tend toward zero in the normal (inertial) frame (by
driving $\vec p$ toward $-E\vec w$).  The timescale for this term
(the time over which $\epsilon$ changes appreciably) is the Hubble
time, so it should not be regarded as the frame dragging normally
spoken of when loosely describing the vector mode.  In fact, in the
normal frame this term is absent, but then the gravitomagnetic term
changes from $\epsilon\vec v\times\vec H$ to $\epsilon\vec\nabla(\vec
w\cdot\vec v)$.  The relative velocity of the comoving and normal
(inertial) frames $\vec w$ is responsible for the frame-dragging
and other effects; let us consider a particularly interesting one.

In general, $\vec w$ varies with position so that at different places
the inertial frames rotate relative to the comoving frame with
angular velocity $-\frac{1}{2}\vec\nabla\times\vec w=-\frac{1}{2}\vec
H$; this is easily shown from a first-order Taylor series expansion of
$\vec w$ with the constraint $\vec\nabla\cdot\vec w=0$.  As a result,
a spin $\vec S$ will precess relative to the comoving frame at a rate
$d\vec S/d\tau=-\frac{1}{2}\vec H\times\vec S$ (the Lense-Thirring
effect).  Using the magnetic analogy, one would predict a gravitomagnetic
precession rate $\gamma\vec S\times\vec H$ in the comoving frame, where
$\gamma$ is the gyrogravitomagnetic ratio.  (The analogous magnetic
precession rate is $\vec\mu\times\vec B$, where $\vec\mu=\gamma\vec S$.)
Note that this result leads to the conclusion that there is a universal
gyrogravitomagnetic ratio $\gamma=\frac{1}{2}$!

Thus, one may interpret the vector mode perturbation variable $\vec w$
either as a source for (rather mysterious) frame-dragging effects, or
as a vector potential for the gravitomagnetic field $\vec H$.  In the
former case one can eliminate $\vec w$ altogether by choosing orthogonal
space and time coordinates such as given by the synchronous gauge.
However, I prefer the latter interpretation because of the close analogy
it brings to electrodynamics, allowing us to transfer our flat spacetime
intuition to general relativity.  The price to pay is that one must be
careful to distinguish the comoving and normal frames.

We have discussed the gravitomagnetic and gravitational wave contributions
to the equations of motion in order to illustrate the similarities and
differences between gravity and electrodynamics.  (They are clearest in
the Poisson gauge; the interested reader may wish to rederive the results
of this section in synchronous or some other gauge.)  Why aren't we familiar
with these forces in the Newtonian limit?  The answer is because the
sources of $\vec H$ and $\htens$ are smaller than the source of the
``gravitostatic'' field $-\vec\nabla\psi$ by $O(v/c)$ and $O(v/c)^2$,
respectively (cf. eqs. \ref{gmaxwell} and \ref{pGijt}).  From eqs.
(\ref{eoms}), the forces they induce are smaller by additional factors
of $O(v/c)$ and $O(v/c)^2$.  Thus, for nonrelativistic sources and
particles, the dynamical effects of gravitomagnetism and gravitational
radiation are negligible.  While ordinary magnetic effects are suppressed
by the same powers of $v/c$, the existence of opposite electric charges
leads in most cases to a nearly complete cancellation of the electric
charge density but not the current density.  No such cancellation occurs
with gravity because energy density is always positive.

Since typical gravitational fields in the universe have $\psi\approx\phi
\sim10^{-5}$ and $h_{ij}$ is much smaller than this, the curvature
factors $(1+\psi+\phi-\htens)$ and $(1-\phi+\htens)$ may be replaced by
unity to high precision in eqs. (\ref{eoms}) (and they are absent anyway
in locally flat comoving coordinates).  In the weak-field and slow-motion
limit, then, eqs. (\ref{eoms}) reduce to the standard Newtonian equations
of motion in comoving coordinates.

\subsection{Lagrangian field equations}
\label{ellis}

General relativity makes no fundamental distinction between time and
space, although we do.  To obtain field equations that are similar to
those of Newtonian gravity and electrodynamics, we have until now employed
a ``3+1 split'' of the Einstein and energy conservation equations.
Ellis (1971, 1973), following earlier work of Ehlers (1961, 1971),
Kundt \& Tr\"umper (1961), and Hawking (1966), has developed an
alternative approach based on a ``1+3 split'' of the Bianchi and Ricci
identities.  The cosmological applications have been developed
extensively by Ellis and others in recent years (Ellis \& Bruni 1989;
Hwang \& Vishniac 1990; Lyth \& Stewart 1990; Bruni, Dunsby \& Ellis 1992;
and references therein).  Ellis' approach has some important advantages,
as we shall see.

The 3+1 split corresponds to the ``slicing'' of spacetime into a series
of spatial hypersurfaces, each labeled by a coordinate time $\tau$.
(The different splitting procedures are most easily visualized with
one spatial dimension suppressed using a 2+1 spacetime diagram, with
time corresponding to the vertical axis.  The spatial hypersurfaces are
then horizontal slices through spacetime.) Spacetime is described by
Eulerian observers sitting in these hypersurfaces with constant spatial
coordinates.

The 1+3 split, called ``threading,'' is complementary to slicing (Jantzen
et al.  1992).  In this case the fundamental geometrical objects used for
charting spacetime are a series of timelike worldlines $x^\mu(\lambda;
\vec q)$, where $\lambda$ is an affine parameter measuring proper time
along the worldline and $\vec q$ gives a unique label (e.g., a spatial
Lagrangian position vector) to each different worldline (or ``thread'').
In this case spacetime is described by Lagrangian observers moving along
these worldlines.

The threading description is more general than the slicing one.  If we
take the threads to correspond to the worldlines of comoving observers
in the slicing framework (lines of fixed $\vec x$), then the two descriptions
are the same.  In the 1+3 description, however, different threads may cross
with no harmful consequences while in the 3+1 description a spatial
hypersurface must not be allowed to cross itself or other slices.  Thus,
the threading description may be used to follow the evolution of cold dust
beyond the time when matter trajectories intersect, when the perfect-fluid
Euler equations break down.  The advantage of a Lagrangian description is
well known for collisionless matter --- the Lagrangian approach exclusively
is used for nonlinear gravitational simulations --- and the same advantages
accrue even when describing the spacetime geometry itself.

In the 1+3 approach each worldline threading spacetime has a timelike unit
tangent vector (4-velocity) $u^\mu=dx^\mu/d\lambda=u^\mu(\lambda;\vec q)$
such that $u^\mu u_\mu=-1$.  Spacetime tensors are then decomposed into
parts parallel and normal to the worldline passing through a given point.
This decomposition is accomplished in a covariant form using the tangent
vector $u^\mu$ and the orthogonal projection tensor
\begin{equation}
  P_{\mu\nu}(u)=g_{\mu\nu}+u_\mu u_\nu\ ,
\label{project}
\end{equation}
such that $P_{\mu\nu}u^\nu=0$ and $P^{\mu\kappa}P_{\kappa\nu}=
P^\mu_{\ \,\nu}$.  $P_{\mu\nu}$ is effectively the spatial metric for
observers moving with 4-velocity $u^\mu$ (Ellis 1973).  We may use it and
$u^\mu$ to split any 4-vector $A^\mu$ into timelike and spacelike parts,
labeled by the tangent vector of the appropriate thread:
\begin{equation}
  A(u)\equiv -u_\mu A^\mu\ ,\quad
  A^\mu(u)\equiv P^\mu_{\ \,\nu}A^\nu\ .
\label{1+3v}
\end{equation}
Even though $A^\mu(u)$ looks like (and is, in fact) a 4-vector, we can
regard it as a 3-vector in the rest frame of an observer moving along
the worldline $x^\mu(\lambda;\vec q)$ because $u_\mu A^\mu(u)=0$.
[Note that $A^\mu$ denotes the original 4-vector while $A^\mu(u)$
denotes its projection normal to $u^\mu$.  We shall include the argument
$(u)$ for the projection whenever needed to remove ambiguity.]  We require
that at each point in spacetime there is at least one thread with
corresponding tangent $u^\mu(\lambda;\vec q)$.  If there are several
threads then there are several different decompositions of $A(u)$ and
$A^\mu(u)$ at $x^\mu$, each labeled by $\vec q$ (implicitly, if not
explicitly) through $u^\mu(\lambda;\vec q)$.  This causes no problems
as long as we refer to a single distinct thread, which we do by retaining
$u$ in the argument list.

\clearpage
The decomposition of a second-rank tensor $T^{\mu\nu}$ is similar:
\begin{eqnarray}
  T(u)=u_\mu u_\nu T^{\mu\nu}\ ,\ \
  &&T_\mu(u)=g_{\mu\nu}T^\nu(u)=-P_{\mu\nu}u_\alpha T^{\alpha\nu}\ ,\nonumber\\
  &&T^\mu_{\ \,\nu}(u)=P^\mu_{\ \,\alpha}P_{\nu\beta}T^{\alpha\beta}\ .
\label{1+3t}
\end{eqnarray}
As an exercise one may apply this decomposition to the stress-energy
tensor of eq. (\ref{tmunu}) using the comoving observers to define
the threading.  For $v^2\ll1$, one obtains nonzero elements $T=\rho$,
$T_i=a (\rho+p)v_i$ (with no $w_i$), and $T^i_{\ j}=p\delta^i_{\ j}+
\Sigma^i_{\ j}$.  Be careful to distinguish the 4-velocity of the threads
(with $v^i=0$) from those of the matter (eq. \ref{4vel}).

Now that we have described the 1+3 spacetime splitting procedure, we
are ready to apply it to gravity following Hawking (1966) and Ellis
(1971, 1973).  What equations should we use?  One might think to split
the Einstein equations using 1+3 threading, but this does not add anything
fundamentally new to what we have already done.  The correct approach
suggests itself when we think in Lagrangian terms following a
freely-falling observer, whose worldline defines one of the threads.
Such an observer feels no gravitational force at all but does notice
that adjacent freely-falling observers do not necessarily move in
straight lines with constant speed.  In Newtonian terms this is explained
by ``tidal forces'' while in general relativity it is called geodesic
deviation.  We shall not present a derivation of geodesic deviation here
(one may find it in any general relativity textbook) but simply note
that it follows from the non-commutativity of covariant spacetime derivatives
of the 4-velocity.  The relevant equation is the 4-dimensional version
of the first of eqs.  (\ref{commute}), called the Ricci identity:
\begin{equation}
  \left[\nabla_\kappa,\nabla_\lambda\right]u^\mu=
    R^\mu_{\ \,\nu\kappa\lambda}u^\nu\ .
\label{ricci-id}
\end{equation}
This identity holds for any differentiable vector field $u^\mu$.
{\it In the Lagrangian field approach we seek evolution equations for
the Riemann tensor itself rather than the metric tensor components.}

One advantage of working with the Riemann tensor is the fact that part
of it --- the Ricci tensor --- is given {\it algebraically} by the local
stress-energy through eqs. (\ref{einstein}) and (\ref{ricci}).
However, one cannot (in 4 dimensions) reconstruct the entire Riemann
tensor from the Ricci tensor alone.  One could obtain it by differentiating
the metric found by solving the Einstein equations (cf. eqs. \ref{riemann},
\ref{affine}).  As we shall see, there is another method that does not
require integrating the Einstein equations.

This alternative method is based on an evolution equation for that part
of the Riemann tensor that cannot be obtained from the Ricci tensor,
the Weyl tensor $C_{\mu\nu\kappa\lambda}$:
\clearpage
\begin{eqnarray}
  C_{\mu\nu\kappa\lambda}\equiv R_{\mu\nu\kappa\lambda}&&-{1\over2}(
    g_{\mu\kappa}R_{\nu\lambda}+g_{\nu\lambda}R_{\mu\kappa}-
    g_{\mu\lambda}R_{\nu\kappa}-g_{\nu\kappa}R_{\mu\lambda})\nonumber\\
  &&+{R\over6}\,(g_{\mu\kappa}g_{\nu\lambda}-g_{\mu\lambda}g_{\nu\kappa})\ .
\label{weyl}
\end{eqnarray}
This tensor obeys all the symmetries of the Riemann tensor ---
$C_{\mu\nu\kappa\lambda}=C_{[\mu\nu][\kappa\lambda]}=C_{\kappa\lambda
\mu\nu}$ and $C_{\mu[\nu\kappa\lambda]}=0$ (where square brackets denote
antisymmetrization) --- and in addition is traceless: $C^\kappa_{\ \,
\mu\kappa\nu}=0$.  Thus, the trace part of the Riemann tensor is given by
the Ricci tensor $R_{\mu\nu}$ (through the Ricci terms on the right-hand
side of eq. \ref{weyl}) while the traceless part is given by the Weyl
tensor.  Physically, the Ricci tensor gives the contribution to the
spacetime curvature from local sources (through the Einstein eqs.
\ref{einstein} combined with \ref{ricci}) while the Weyl tensor gives the
contribution due to nonlocal sources.  It is clear that Newtonian tidal
forces will be represented in the Weyl tensor.  It may be shown that in
4 dimensions the Ricci and Weyl tensors each have 10 independent components.

How do we get an evolution equation for the Weyl tensor?  The Einstein
equations will not do because the Weyl tensor makes no appearance at all
in the Einstein tensor.  The correct method, due to Kundt \& Tr\"umper
(1961), makes use of the Bianchi identities,
\begin{equation}
  \nabla_\sigma R_{\mu\nu\kappa\lambda}+\nabla_\mu R_{\nu\sigma\kappa\lambda}
    +\nabla_\nu R_{\sigma\mu\kappa\lambda}=0\ .
\label{bianchi}
\end{equation}
These identities follow directly from the definition of the Riemann tensor
(see any general relativity or differential geometry textbook).  For our
purposes the key point is that they provide differential equations for
the Riemann tensor.  Contracting eq. (\ref{bianchi}) on $\kappa$ and
$\sigma$ and using eqs. (\ref{weyl}) and (\ref{ricci}), we get
\begin{equation}
  \nabla^\kappa C_{\mu\nu\kappa\lambda}=\nabla_{[\mu}G_{\nu]\lambda}
    +{1\over3}\,g_{\lambda[\mu}\nabla_{\nu]}G^\kappa_{\ \,\kappa}\ .
\label{weyl-ev1}
\end{equation}
Note that if we contract now on $\lambda$ and $\mu$, using the symmetry
of $G_{\mu\nu}$ and $g_{\mu\nu}$ we get $\nabla_\mu G^\mu_{\ \,\nu}=0$,
as noted before.  However, here we regard eq. (\ref{weyl-ev1}) as
an equation of motion for the Weyl tensor.  Using the Einstein eqs.
(\ref{einstein}), we see that the source is given in terms of the
energy-momentum tensor, so
\begin{equation}
  \nabla^\kappa C_{\mu\nu\kappa\lambda}=8\pi G\left(\nabla_{[\mu}T_{\nu]
    \lambda}+{1\over3}\,g_{\lambda[\mu}\nabla_{\nu]}T^\kappa_{\ \,\kappa}
    \right)\ .
\label{weyl-evol}
\end{equation}

The next step is to split the Weyl tensor into two second-rank tensors
using a 1+3 threading of spacetime (Hawking 1966, Ellis 1971),
\begin{equation}
  E_{\mu\nu}(u)\equiv u^\kappa u^\lambda C_{\mu\kappa\nu\lambda}\ ,\quad
  H_{\mu\nu}(u)\equiv {1\over2}\,\epsilon_{\alpha\beta\kappa(\mu}\,
    u^\kappa u^\lambda C^{\alpha\beta}_{\ \ \ \nu)\lambda}\ .
\label{weyl-split}
\end{equation}
We have used the fully antisymmetric tensor $\epsilon_{\mu\nu\kappa\lambda}
=(-g)^{1/2}\,[\mu\nu\kappa\lambda]$, where $g$ is the determinant of
$g_{\mu\nu}$ and $[\mu\nu\kappa\lambda]$ is the completely antisymmetric
Levi-Civita symbol defined by three conditions: (1) $[0123]=+1$, (2)
$[\mu\nu\kappa\lambda]$ changes sign if any two indices are exchanged,
and (3) $[\mu\nu\kappa\lambda]=0$ if any two indices are equal.  (Note
that Ellis uses the tensor $\eta_{\mu\nu\kappa\lambda}=-\epsilon_{\mu\nu
\kappa\lambda}$.  We have compensated for the sign change in defining
$H_{\mu\nu}$.  Beware that $\epsilon^{\mu\nu\kappa\lambda}=-(-g)^{-1/2}
\,[\mu\nu\kappa\lambda]$.) The two new tensors $E_{\mu\nu}$ and
$H_{\mu\nu}$ are both symmetric ($H_{\mu\nu}$ must be explicitly
symmetrized), traceless, and flow-orthogonal, i.e., $E_{\mu\nu}u^\nu=
H_{\mu\nu}u^\nu=0$ and $P^\nu_{\ \,\kappa}E_{\mu\nu}=E_{\mu\kappa}$,
$P^\nu_{\ \,\kappa}H_{\mu\nu}=H_{\mu\kappa}$.  Therefore $E_{\mu\nu}$
and $H_{\mu\nu}$ each has 5 independent components, half as many as the
Weyl tensor.  Indeed, the Weyl tensor is fully determined by them for
non-null threads:
\begin{eqnarray}
  C_{\mu\nu\kappa\lambda}=&&(g_{\mu\nu\alpha\beta}\,g_{\kappa\lambda
    \gamma\delta}-\epsilon_{\mu\nu\alpha\beta}\,\epsilon_{\kappa\lambda
    \gamma\delta})\,u^\alpha u^\gamma E^{\beta\delta}(u)
      \nonumber\\
  +&&(\epsilon_{\mu\nu\alpha\beta}\,g_{\kappa\lambda\gamma\delta}+
    g_{\mu\nu\alpha\beta}\,\epsilon_{\kappa\lambda\gamma\delta})\,
    u^\alpha u^\gamma H^{\beta\delta}(u)\ ,
\label{weyl-join}
\end{eqnarray}
where $g_{\mu\nu\alpha\beta}\equiv g_{\mu\alpha}g_{\nu\beta}-g_{\mu\beta}
g_{\nu\alpha}=-\frac{1}{2}\epsilon_{\mu\nu}^{\ \ \ \kappa\lambda}
\epsilon_{\kappa\lambda\alpha\beta}=g_{[\mu\nu][\alpha\beta]}=
g_{\alpha\beta\mu\nu}$, with $g_{\mu[\nu\alpha\beta]}=0$.  Eq.
(\ref{weyl-join}) is the inverse of eqs. (\ref{weyl-split}) provided
$g_{\mu\nu}u^\mu u^\nu=\pm1$.  Ellis (1971) has a sign error in the
first term of his version of eq. (\ref{weyl-join}) at the end of his
section 4.2.3.

The tensors $E_{\mu\nu}(u)$ and $H_{\mu\nu}(u)$ are called the electric
and magnetic parts of the Weyl tensor, respectively.  Together with the
Ricci tensor they fully determine the spacetime curvature for a given
threading (i.e., a system of threads with tangent vectors) $u^\mu(\lambda;
\vec q)$.  It is worth noting that, if there are several threads at a
given spacetime point, $E_{\mu\nu}(u)$ and $H_{\mu\nu}(u)$ have different
values for each thread, and so they may be considered Lagrangian functions:
$E_{\mu\nu}(\lambda;\vec q)$ and $H_{\mu\nu}(\lambda;\vec q)$.  The
Weyl tensor components are, however, unique, with the same value for all
threads passing through the same spacetime point.  This condition is
satisfied automatically if the same 4-velocity $u^\mu$ is used in both
eqs. (\ref{weyl-split}) and (\ref{weyl-join}).

Our goal is to rewrite eq. (\ref{weyl-evol}) in terms of $E_{\mu\nu}$
and $H_{\mu\nu}$.  Because the results involve the covariant derivative
of the 4-velocity field $\nabla_\mu u_\nu$, we first decompose this
quantity into acceleration, expansion, shear, and vorticity:
\begin{eqnarray}
  &&\nabla_\mu u_\nu=-u_\mu{Du_\nu\over d\lambda}+
    P^\alpha_{\ \,\mu}P^\beta_{\ \,\nu}\nabla_\alpha u_\beta=
    -u_\mu a_\nu+{1\over3}\,\Theta P_{\mu\nu}+\sigma_{\mu\nu}+
    \omega_{\mu\nu}\,;\nonumber\\
  &&\Theta=\nabla_\mu u^\mu\ ,\ \
    \sigma_{\mu\nu}=\sigma_{(\mu\nu)}\ ,\ \
    \omega_{\mu\nu}=\omega_{[\mu\nu]}=
      \epsilon_{\mu\nu\alpha\beta}u^\alpha\omega^\beta\ .
\label{gradv}
\end{eqnarray}
We have introduced the covariant derivative in the direction $u^\nu$,
$D/d\lambda\equiv u^\nu\nabla_\nu$.  Since this is just the proper time
derivative along the worldline, $a_\nu=Du_\nu/d\lambda$ is the 4-acceleration.
The flow-orthogonal part of the velocity gradient, $P^\alpha_{\ \,\mu}
P^\beta_{\ \,\nu}\nabla_\alpha u_\beta$, has been decomposed into the
expansion scalar $\Theta$, the traceless shear tensor $\sigma_{\mu\nu}$,
and the vorticity tensor $\omega_{\mu\nu}$ or its flow-orthogonal dual,
$\omega^\mu$.  Note that the expansion scalar includes a contribution
due to cosmic expansion in addition to the peculiar velocity: neglecting
metric perturbations, $\Theta=a^{-1}(\eta+\vec\nabla\cdot\vec v)$.
Note also that in the fluid rest frame, $\omega^i\vec e_i=\frac{1}{2}
\vec\nabla\times\vec v$ is half the usual three-dimensional vorticity.
(Ellis defines $\omega_{\mu\nu}$ and $\omega^\mu$ with the opposite sign
to us.)

We shall apply this gradient expansion to the tangent field of the 1+3
spacetime threading.  This requires that $u^\mu$ be differentiable,
which will be true (almost everywhere) if it corresponds to the 4-velocity
field of a flow.  In a frame comoving with the fluid, $\Theta$,
$\sigma_{ij}$ and $\omega_{ij}$ are then the usual fluid expansion, shear,
and vorticity, respectively.

By projecting $\nabla^\kappa C_{\mu\nu\kappa\lambda}$ with various
combinations of $u^\alpha$ and $P^{\alpha\beta}(u)$ (these are dependent
on the spacetime threading), one can derive the following identities:
\begin{equation}
  u^\nu u^\lambda\nabla^\kappa C^\mu_{\ \,\nu\kappa\lambda}=
    P^\mu_{\ \,\alpha}P^\nu_{\ \,\beta}\nabla_\nu E^{\alpha\beta}
    +\epsilon^{\mu\nu\alpha\beta}\,u_\nu\sigma_{\alpha\gamma}H^\gamma_{\
    \,\beta}-3H^\mu_{\ \,\nu}\omega^\nu\ ,
\label{grad-weyl1}
\end{equation}
\begin{eqnarray}
  \frac{1}{2}P^\mu_{\ \,\alpha}u_\beta u^\lambda
    \epsilon^{\alpha\beta\gamma\delta}
    \nabla^\kappa C_{\gamma\delta\kappa\lambda}=
    -P^\mu_{\ \,\alpha}P^\nu_{\ \,\beta}\nabla_\nu H^{\alpha\beta}
    &&+\epsilon^{\mu\nu\alpha\beta}\,u_\nu\sigma_{\alpha\gamma}
    E^\gamma_{\ \,\beta}\nonumber\\
  &&-3E^\mu_{\ \,\nu}\omega^\nu\ ,
\label{grad-weyl2}
\end{eqnarray}
\begin{eqnarray}
  P^{\mu\lambda}P^{\nu\alpha}u^\beta\nabla^\kappa
    C_{\alpha\beta\kappa\lambda}&&=P^\mu_{\ \,\alpha}P^\nu_{\ \,\beta}
    {DE^{\alpha\beta}\over d\lambda}+P^{\alpha\nu}\epsilon^{\mu\beta
    \gamma\delta}u_\beta\nabla_\gamma H_{\alpha\delta}\nonumber\\
  &&+2u_\alpha a_\beta H_\gamma^{\ \,(\mu}\epsilon^{\nu)\alpha\beta\gamma}
    +\Theta E^{\mu\nu}+P^{\mu\nu}(\sigma^{\alpha\beta}E_{\alpha\beta})
    \nonumber\\
  &&-2E^{\alpha\nu}(\sigma^\mu_{\ \,\alpha}-\omega^\mu_{\ \,\alpha})
    -E^{\alpha\mu}(\sigma^\nu_{\ \,\alpha}+\omega^\nu_{\ \,\alpha})\ ,
\label{grad-weyl3}
\end{eqnarray}
\begin{eqnarray}
  &&\frac{1}{2} P^\mu_{\ \,\alpha}P^{\nu\lambda}u_\beta\epsilon^{\alpha
    \beta\gamma\delta}\nabla^\kappa C_{\gamma\delta\kappa\lambda}=
    -P^\mu_{\ \,\alpha}P^\nu_{\ \,\beta}{DH^{\alpha\beta}\over d\lambda}
    +P^{\alpha\mu}\epsilon^{\nu\beta\gamma\delta}u_\beta\nabla_\gamma
    E_{\alpha\delta}\nonumber\\
  &&\quad\quad\quad\quad\quad\quad\quad\quad\ \ \,
    +2u_\alpha a_\beta E_\gamma^{\ \,(\mu}\epsilon^{\nu)\alpha\beta\gamma}
    -\Theta H^{\mu\nu}-P^{\mu\nu}(\sigma^{\alpha\beta}H_{\alpha\beta})
    \nonumber\\
  &&\quad\quad\quad\quad\quad\quad\quad\quad\ \ \,
    +2H^{\alpha\mu}(\sigma^\nu_{\ \,\alpha}-\omega^\nu_{\ \,\alpha})
    +H^{\alpha\nu}(\sigma^\mu_{\ \,\alpha}+\omega^\mu_{\ \,\alpha})\ .
\label{grad-weyl4}
\end{eqnarray}
These identities follow from eqs. (\ref{weyl-join}) and (\ref{gradv}).
All quantities on the right-hand sides are to be evaluated for a given
thread $u^\mu(\lambda;\vec q)$.

Finally we are ready to obtain equations of motion for the electric and
magnetic parts of the Weyl tensor from eq. (\ref{weyl-evol}).  In fact,
infinitely many sets of equations are possible because are free to use any
spacetime threading!  For example, we may choose {\it Eulerian} threading
with $\vec q=\vec x$, in which case in the Poisson gauge we have
$u^0=a^{-1}(1-\psi)$ and $u^i=0$, so that $D/d\lambda=a^{-1}(1-\psi)
\partial_\tau$ is the Eulerian proper time derivative.  In this case the
1+3 split coincides with our previous 3+1 split.  The Eulerian description
is not covariant, for it depends on our choice of gauge.  Because the Weyl
tensor formalism is more complicated than our previous treatment based on
the Einstein equations, there is no clear advantage to its use with Eulerian
threading.

If, however, we use the fluid velocity itself --- the $u^\mu$ appearing
in eq. (\ref{tmunu}), which is well-defined even for an imperfect or
collisionless fluid --- to define the threading, then the Weyl tensor
approach becomes more attractive.  This choice corresponds to {\it
Lagrangian} threading: the threads are the worldlines of fluid elements,
so that $D/d\lambda$ now is the proper time derivative measured in the
fluid rest frame.  There are two important advantages to this choice.
First, it is covariant: the fluid worldlines define a unique spacetime
threading with no gauge ambiguities (Ellis \& Bruni 1989), while any
coordinates may be used to express the tensor components $E_{\mu\nu}$
and $H_{\mu\nu}$.  Second, the right-hand side of eq. (\ref{weyl-evol})
--- the source for the Weyl tensor --- is expressed in terms of the same
4-velocity used in the threading, greatly simplifying the projections
appearing in eqs. (\ref{grad-weyl1})--(\ref{grad-weyl4}).

Ellis (1971) and Hwang \& Vishniac (1990) give the Lagrangian gravitational
field equations for a general stress-energy tensor.  For a perfect fluid
(with $\Sigma^{\mu\nu}=0$ in eq. \ref{tmunu}) the results are
\begin{eqnarray}
\label{div-E}
  \hbox{(div-$E$)}:\ \ &&
    P^\mu_{\ \,\alpha}P^\nu_{\ \,\beta}\nabla_\nu E^{\alpha\beta}
      +\epsilon^{\mu\nu\alpha\beta}u_\nu\sigma_{\alpha\gamma}H^\gamma_{\
      \,\beta}-3H^\mu_{\ \,\nu}\omega^\nu\nonumber\\
    &&\quad\quad\quad\quad\quad\quad\quad\quad\quad\quad
      ={8\pi\over3}\,GP^{\mu\nu}\nabla_\nu\rho\ ,\\
\label{H-dot}
  (\dot H):\ \ &&
    P^\mu_{\ \,\alpha}P^\nu_{\ \,\beta}{DH^{\alpha\beta}\over d\lambda}
      -P^{\alpha(\mu}\epsilon^{\nu)\beta\gamma\delta}u_\beta\nabla_\gamma
      E_{\alpha\delta}\nonumber\\
    &&-2u_\alpha a_\beta E_\gamma^{\ \,(\mu}\epsilon^{\nu)\alpha\beta
      \gamma}+\Theta H^{\mu\nu}+P^{\mu\nu}(\sigma^{\alpha
      \beta}H_{\alpha\beta})-3H^{\alpha(\mu}\sigma^{\nu)}_{\ \ \alpha}
      \nonumber\\
    &&\quad\quad\quad\quad\quad\quad\quad\quad\quad\quad
      +H^{\alpha(\mu}\omega^{\nu)}_{\ \ \alpha}=0\ ,\\
\label{div-H}
  \hbox{(div-$H$)}:\ \ &&
    P^\mu_{\ \,\alpha}P^\nu_{\ \,\beta}\nabla_\nu H^{\alpha\beta}
      -\epsilon^{\mu\nu\alpha\beta}u_\nu\sigma_{\alpha\gamma}E^\gamma_{\
      \,\beta}+3E^\mu_{\ \,\nu}\omega^\nu\nonumber\\
    &&\quad\quad\quad\quad\quad\quad\quad\quad\quad\quad
      =-8\pi G(\rho+p)\omega^\mu\ ,\\
\label{E-dot}
  (\dot E):\ \ &&
    P^\mu_{\ \,\alpha}P^\nu_{\ \,\beta}{DE^{\alpha\beta}\over d\lambda}
      +P^{\alpha(\mu}\epsilon^{\nu)\beta\gamma\delta}u_\beta\nabla_\gamma
      H_{\alpha\delta}\nonumber\\
    &&+2u_\alpha a_\beta H_\gamma^{\ \,(\mu}\epsilon^{\nu)\alpha\beta
      \gamma}+\Theta E^{\mu\nu}+P^{\mu\nu}(\sigma^{\alpha
      \beta}E_{\alpha\beta})-3E^{\alpha(\mu}\sigma^{\nu)}_{\ \ \alpha}
      \nonumber\\
    &&\quad\quad\quad\quad\quad
      +E^{\alpha(\mu}\omega^{\nu)}_{\ \ \alpha}=
      -4\pi G(\rho+p)\sigma^{\mu\nu}\ .
\end{eqnarray}
These have been obtained by substituting eqs. (\ref{tmunu}) and
(\ref{grad-weyl1})--(\ref{grad-weyl4}) into eq. (\ref{weyl-evol}),
and using $\nabla_\nu T^{\mu\nu}=0$ to simplify the right-hand sides
of the div-$E$ and $\dot E$ equations.  The results agree with eqs.
(4.21) of Ellis (1971).  For an imperfect fluid it is necessary to add
terms to the right-hand sides involving the shear stress $\Sigma^{\mu\nu}$.
For a pressureless fluid (e.g., cold dust before the intersection of
trajectories) the 4-acceleration $a_\beta$ vanishes.

In his beautifully lucid pedagogical articles presenting the Lagrangian
fluid approach, Ellis (1971, 1973) has noted the similarity of eqs.
(\ref{div-E})--(\ref{E-dot}) to the Maxwell equations, particularly if
the covariant form of the latter are split using 1+3 threading.  Compare
them with eqs. (\ref{gmaxwell}) for the vector (not tensor) gravitational
fields in the Poisson gauge.  Although the latter equations are more
reminiscent of the Maxwell equations in flat spacetime, they are only
approximate (they are based on a linearized metric and neglect several
generally small terms), they are tied to a particular coordinate system
(Poisson gauge), and they do not incorporate gravitational radiation.
By contrast, eqs. (\ref{div-E})--(\ref{E-dot}) are exact, they are valid
in any coordinate system (all quantities appearing in them are spacetime
tensors), and they include all gravitational effects.  The exact equations
involve second-rank tensors rather than vectors because, in the terminology
of particle physics, gravity is a spin-2 rather than a spin-1 gauge theory.

The quasi-Maxwellian equations (\ref{div-E})--(\ref{E-dot}) show that
the evolution of the Weyl tensor depends on the fluid velocity gradient.
This quantity could be computed by evolving the equations of motion for
the matter (e.g., eqs. \ref{econs} and \ref{pcons}) to get the velocity
field $u^\mu(x)$ and then taking its derivatives.  However, there is
a more natural way in the context of the Lagrangian approach: integrate
evolution equations for the velocity gradient itself.  In fact, such
equations follow simply from projecting the Ricci identity (\ref{ricci-id})
for the fluid velocity $u^\mu$ with $u^\kappa P^{\alpha\lambda}P_{\beta\mu}$
and separating the result as in eqs. (\ref{gradv}).  It is straightforward
to derive the following equations (Ellis 1971, 1973):
\begin{equation}
  {D\Theta\over d\lambda}-\nabla_\mu a^\mu+\frac{1}{3}\,\Theta^2
    +\sigma^{\mu\nu}\sigma_{\mu\nu}-2\omega^2=-4\pi G(\rho+3p)\ ,
\label{raych}
\end{equation}
\begin{equation}
  P^\mu_{\ \,\nu}{D\omega^\nu\over d\lambda}+\frac{1}{2}\,\epsilon^
    {\mu\nu\alpha\beta}u_\nu\nabla_\alpha a_\beta+\frac{2}{3}\,
    \Theta\omega^\mu-\sigma^\mu_{\ \,\nu}\omega^\nu=0\ ,
\label{vort-evol}
\end{equation}
\begin{eqnarray}
  &&P^\mu_{\ \,\alpha}P^\nu_{\ \,\beta}{D\sigma^{\alpha\beta}\over d\lambda}
    -\nabla^{(\mu}a^{\nu)}+\frac{2}{3}\,\Theta\sigma^{\mu\nu}+\sigma^{\mu
    \alpha}\sigma^\nu_{\ \,\alpha}+\omega^\mu\omega^\nu\nonumber\\
  &&-\frac{1}{3}\,P^{\mu\nu}\left(\sigma^{\alpha\beta}\sigma_{\alpha\beta}
    +\omega^2-\nabla_\alpha a^\alpha\right)=-E^{\mu\nu}\ ,
\label{shear-evol}
\end{eqnarray}
where $\omega^2\equiv\omega^\mu\omega_\mu$.  Equation (\ref{raych}) is
known as the Raychaudhuri equation.  It shows that the expansion is
decelerated by the shear and by the local density and pressure (if
$\rho+3p>0$), but is accelerated by the vorticity.  Vorticity, on the
other hand, is unaffected by gravity; eq. (\ref{vort-evol}) implies
that vorticity can be described by field lines that (if $a^\mu$ vanishes
or if the fluid has vanishing shear stress) are frozen into the fluid
(Ellis 1973).  Finally, shear, being the traceless symmetric part of the
velocity gradient tensor, has as its source the electric part of the
Weyl tensor.  These equations are essentially identical to their Newtonian
counterparts (Ellis 1971; Bertschinger \& Jain 1994).  Note that the
magnetic part of the Weyl tensor does not directly influence the matter
evolution.

Closing the Lagrangian field equations also requires specifying the
evolution of density and pressure (and shear stress, if present).
These follow from energy conservation, $\nabla_{\nu}T^{\mu\nu}=0$,
combined with an equation of state.  For a perfect fluid, using
eq. (\ref{tmunu}) with $\Sigma^{\mu\nu}=0$ and projecting the divergence
of the stress-energy tensor with $u_\mu$ gives
\begin{equation}
  {D\rho\over d\lambda}+(\rho+p)\Theta=0\ .
\label{econs1}
\end{equation}
Equations (\ref{div-E})--(\ref{econs1}) now provide a set of Lagrangian
equations of motion for the matter and spacetime curvature variables
following a mass element.  These Lagrangian equations of motion offer
a powerful approach to general relativity --- and to relativistic cosmology
and perturbation theory --- that is quite different from the usual methods
based on integration of the Einstein equations in a particular gauge (or
with gauge-invariant variables).

To relate the relativistic Lagrangian approach to dynamics to the
standard Newtonian one, we now evaluate the electric and magnetic parts
of the Weyl tensor in the weak-field, slow-motion limit.  They involve
second derivatives of the metric and not simply the first derivatives
present in eqs. (\ref{ghfields}).  In the Poisson gauge, to lowest
order in the metric perturbations and the velocity, from eqs.
(\ref{weyl-split}) one obtains (Bertschinger \& Hamilton 1994)
\begin{eqnarray}
  &&E_{ij}=\frac{1}{2}\,D_{ij}(\psi+\phi)+\frac{1}{2}\,\nabla_{(i}
    \dot w_{j)}-\frac{1}{2}\,(\ddot h_{ij}+\nabla^2h_{ij}-2Kh_{ij})\ ,
    \nonumber\\
  &&H_{ij}=-\frac{1}{2}\,\nabla_{(i}H_{j)}+\epsilon_{kl(i}\nabla^k\dot
    h_{j)}^{\ \ l}\ ,
\label{ehij}
\end{eqnarray}
where $H_j$ is the gravitomagnetic field defined in eq. (\ref{ghfields}).
The time-time and space-time components of $E_{\mu\nu}$ and $H_{\mu\nu}$
vanish in the fluid frame because these tensors are flow-orthogonal.

Do these results imply that in the Newtonian limit $H_{ij}=0$ and
$E_{ij}=D_{ij}\phi$ is simply the gravitational tidal field?  If we
say that the Newtonian limit implies $\psi=\phi$ and $w_i=h_{ij}=0$ (no
relativistic shear stress, no gravitomagnetism, and no gravitational
radiation), then the answer would appear to be yes.  This possibility,
considered by Matarrese, Pantano, \& Saez (1993) and Bertschinger \&
Jain (1994), has an important implication: for cold dust, the Lagrangian
evolution of the tidal tensor obtained from eq. (\ref{E-dot}) would then
be purely local (Barnes \& Rowlingson 1989).  That is, the evolution
of the tide (the electric part of the Weyl tensor) along the thread
$u^\mu(\lambda; \vec q)$ would depend only on the density, velocity
gradient, and tide defined at each point along the trajectory with no
further spatial gradients (since they arise only from the magnetic terms
in eq. \ref{E-dot}).  The evolution of the density and of the velocity
gradient tensor are clearly local (eqs. \ref{raych}--\ref{econs1}, with
$a^\mu=0$) aside from the tidal tensor, but we have just seen that its
evolution depends only on other local quantities.  In other words,
if $H_{ij}=0$, the matter and spacetime curvature variables would evolve
independently along different fluid worldlines.  Bruni, Matarrese, and
Pantano (1994) call this a ``silent universe.''

Local evolution does occur if the metric perturbations are one-dimensional
(e.g., the Bondi-Tolman solution in spherical symmetry, or the Zel'dovich
solution in plane symmetry; see Matarrese et al. 1993 and Croudace et
al. 1994), but it would be surprising were this to happen for
arbitrary matter distributions in the Newtonian limit.

Bertschinger \& Hamilton (1994) and Kofman \& Pogosyan (1995) have
shown that, in fact, the general evolution of the tidal tensor in the
Newtonian limit is nonlocal.  The reason is that, while one may neglect
the metric perturbation $w_i$ in the Newtonian limit, its gradient
should not be neglected.  Doing so violates the transverse momentum
constraint equation (\ref{pG0iv}), unless the transverse momentum density
(the source term for $\vec w$ in the Poisson gauge) vanishes.  This
condition does not hold for general motion in the Newtonian limit.

A convincing proof of nonlocality is given by the derivation of eq.
(\ref{E-dot}) in locally flat coordinates in the fluid frame by
Bertschinger \& Hamilton (1994) using only the Newtonian continuity
and Poisson equations plus the second pair of eqs. (\ref{gmaxwell})
and a modified form of eq. (\ref{ehij}):
\begin{equation}
  H_{ij}=-\frac{1}{2}\,\nabla_{(i}H_{j)}-2v_k\epsilon^{kl}_{\ \ (i}
    E_{j)l}+O(v/c)^2\ .
\label{hij-new}
\end{equation}
This is taken as the definition of $H_{ij}$ in the Newtonian limit
(where we also have $E_{ij}=D_{ij}\phi$).  Note that in the Newtonian
limit we neglect gravitational radiation, but we must include terms
that are first-order in the velocity.  Even though we define the
magnetic part of the Weyl tensor using the fluid 4-velocity, we are
evaluating its components in a particular gauge --- Poisson gauge ---
in which the 3-velocity does not necessarily vanish.  The extra term
in eq. (\ref{hij-new}) arises from evaluating eqs. (\ref{weyl-split}) to
first order in $v/c$ (Bertschinger \& Hamilton 1994) and it is analogous
to the Lorentz transformation of electric fields into magnetic fields in
a moving frame.  Both terms in eq. (\ref{hij-new}) are of order
$G\rho\vec v$.  They can not be neglected in the Newtonian limit.

The implication of this result is that Lagrangian evolution of matter
and gravity is not purely local except under severe restrictions
such as spherical or plane symmetry.  There exist, of course, local
approximations to evolution such as the Zel'dovich (1970) approximation.
Finding improved local approximations is one of the active areas of
research in large-scale structure theory.  Formulating the problem in
terms of the Lagrangian fluid and field equations not only may suggest
new approaches, it is also likely to clarify the relation between
general relativity and Newtonian dynamics.

\end{document}